\documentclass[12pt,a4paper]{article}            
 \usepackage[skins,theorems]{tcolorbox}
\tcbset{highlight math style={enhanced,
  colframe=red,colback=white,arc=0pt,boxrule=1pt}}
  \usepackage{tikz}
  \usepackage{tikz-3dplot}
 \usetikzlibrary{calc}
 \usetikzlibrary{decorations} %
 \usepackage[UKenglish]{babel}
 \usepackage[toc,page]{appendix}
 \usepackage{amsmath}
 \usepackage{amssymb}
 \usepackage{graphicx}
 \usepackage{hhline}
 \usepackage[bf]{caption}
\usepackage{cite}
\usepackage[vcentermath]{youngtab}
\usepackage{geometry}
\usepackage{slashed}
\usepackage{color}
\usepackage{stackrel}
\usepackage{tikz-cd} 
\usepackage{cancel} 
\usepackage{multirow} 
\usepackage{longtable}
\usepackage[aligntableaux=center]{ytableau}
\RequirePackage{pgfkeys}

\usepackage{tikz}
\usetikzlibrary{shadows}
\usepackage[framemethod=tikz]{mdframed}

\allowdisplaybreaks

\global\mdfdefinestyle{myboxstyle}{%
  shadow=true,
  linecolor=black,
  shadowcolor=black,
  shadowsize=6pt,
  nobreak=false,
  innertopmargin=10pt,
  innerbottommargin=10pt,
  leftmargin=5pt,
  rightmargin=5pt,
  needspace=1cm,
  skipabove=10pt,
  skipbelow=15pt,
  middlelinewidth=1pt,
  afterlastframe={\vspace{5pt}},
  aftersingleframe={\vspace{5pt}},
  tikzsetting={%
draw=black,
very thick} }

\newmdenv[style=myboxstyle]{whitebox} \newmdenv[style=myboxstyle,backgroundcolor=black!20]{graybox}

\definecolor{bluish}{rgb}{0.0, 0.2, 0.4}
\definecolor{darkred}{rgb}{0.7, 0.0, 0.0}
\definecolor{darkgreen}{rgb}{0.0, 0.5, 0.0}

\newmdenv[style=myboxstyle,nobreak=true]{blockwhitebox}
\newmdenv[style=myboxstyle,backgroundcolor=black!20,nobreak=true]{blockgraybox}
\newmdenv[nobreak=true,hidealllines=true]{blockbox}

\def\KN{\mathcal{KN}^{\vec{w}}_{\vec{v}}}
\def\QQ{\mathcal{Q}^{\vec{\boldsymbol{w}}}_{\mathbf{ALE}_n,\vec{\boldsymbol{v}}}}
\def\QQA{\mathcal{Q}^{anom,\vec{\boldsymbol{w}}}_{\vec{\boldsymbol{v}}}}
\usepackage{jheppub}

\newcommand{\bqa}{\begin{eqnarray}}
\newcommand{\eqa}{\end{eqnarray}}

\def\et24{\eta^{24}}
\def\oet24{\frac1{\eta^{24}}}

\hypersetup{
    pdftitle={},
    pdfauthor={},
    pdfsubject={}
}
\numberwithin{equation}{section}
\numberwithin{table}{section}

\makeatletter

\newcommand{\be}{\begin{equation}}
\newcommand{\ee}{\end{equation}}
\newcommand{\beq}{\begin{equation}}
\newcommand{\eeq}{\end{equation}}
\newcommand{\ba}{\begin{aligned}}
\newcommand{\ea}{\end{aligned}}

\newcommand{\bea}{\begin{eqnarray}}
\newcommand{\eea}{\end{eqnarray}}

\newcommand\bi{\begin{itemize}}
\newcommand\ei{\end{itemize}}

\def\Tr{\mathop{\mathrm{Tr}}\nolimits}

\def\unit{{1\kern-.65ex {\rm l}}}
\def\1{{1\kern-.65ex {\rm l}}}

\newcommand{\norm}[1]{\vert\vert#1\vert\vert}

\makeatother

\begin{document}

\title{The ALE Partition Functions of M-String Orbifolds}

\author[\sharp\dagger]{Michele Del Zotto}
\author[\sharp\star\natural]{and Guglielmo Lockhart}

\affiliation[\sharp]{Department of Mathematics, Uppsala University, 75106, Uppsala, Sweden}
\affiliation[\dagger]{Department of Physics and Astronomy, Uppsala University, 75120, Uppsala, Sweden}
\affiliation[\star]{Theory Department, CERN,
 Geneva 23, CH-1211, Switzerland}
\affiliation[\natural]{Bethe Center for Theoretical Physics, Universit\"at Bonn, D-53115, Germany}

\emailAdd{michele.delzotto@math.uu.se}
\emailAdd{glockhar@uni-bonn.de}

\abstract{
The ALE partition functions of a 6d (1,0) SCFT are interesting observables which are able to detect the global structure of the SCFT. They are defined to be the equivariant partition functions of the SCFT on a background with the topology of a two-dimensional torus times an ALE singularity. In this work, we compute the ALE partition functions of \emph{M-string orbifold} SCFTs, extending our previous results for the M-string SCFTs. 
Via geometric engineering, our results about ALE partition functions are connected to the theory of higher-rank Donaldson-Thomas invariants for resolutions of elliptic Calabi-Yau threefold singularities. We predict that their generating functions satisfy interesting modular properties. The partition functions receive contributions from BPS strings probing the ALE singularity, whose worldsheet theories we determine via a chain of string dualities. For this class of backgrounds the BPS strings’ worldsheet theories become relative field theories that are sensitive to discrete data generalizing to 6d the familiar choices of flat connections at infinity for instantons on ALE spaces. A novel feature we observe in the case of M-string orbifold SCFTs, which does not arise for the M-string SCFT, is the existence of \emph{frozen} \emph{BPS} \emph{strings} which are pinned at the orbifold singularity and carry fractional instanton charge with respect to the 6d gauge fields.
}
\maketitle

\section{Introduction}

This paper continues the investigation of the (equivariant) ALE partition functions of 6d theories initiated in \cite{DelZotto:2023rct}. Our motivation for studying these quantities is two-fold. On the field theory side, this class of partition functions are natural probes for the global structure of 6d field theories that can be computed exactly, providing natural six-dimensional generalizations of the four-dimensional Vafa-Witten partition functions \cite{Vafa:1994tf}. On the mathematical side, these partition functions are intimately related with generating functions for higher-rank (equivariant) Donaldson-Thomas (DT) invariants \cite{Donaldson:1996kp,Nekrasov:2014nea}. The main result of this paper is an explicit formula for the ALE partition function of the 6d M-string orbifold SCFTs in terms of a BPS string expansion. In this introduction we give an outline of our results in section \ref{sec:outline}, we comment on the relation with higher-rank DT theory in section \ref{sec:DTBS}, we discuss some interesting directions for future study in section \ref{sec:outlook}, and we conclude by describing the organization of this work in \ref{sec:structure}.

\subsection{ALE partition functions: from M-string to M-string orbifolds}\label{sec:outline}

The focus of \cite{DelZotto:2023rct} are the ALE partition functions of the 6d M-string SCFTs. We have computed these quantities in different ways. On the one hand we have proposed an elliptic generalization of the Nekrasov master formula \cite{Nekrasov:2003vi} (see also \cite{Gasparim:2009sns,Bonelli:2012ny}) together with an elliptic generalization of the gauge theoretic results obtained in \cite{Fucito:2004ry,Fucito:2006kn,Bonelli:2011jx,Bonelli:2012ny,Ito:2013kpa,Alfimov:2013cqa,Dey:2013fea,Bruzzo:2013daa,Bruzzo:2014jza,Mekareeya:2015bla}. On the other hand we have proposed a formula which computes the equivariant ALE partition function in terms of an expansion as a sum of contributions from BPS strings sectors, each contributing with an elliptic genus. 

The matching of all these different approaches is an important consistency check for the results of \cite{DelZotto:2023rct}, which in particular, showcases an important feature of the BPS strings. Namely, the dependence on discrete data at infinity of the 6d ALE partition function is encoded on the solitonic worldvolumes thanks to a collection of 2d chiral fermions, which renders the corresponding 2d worldsheet CFTs relative field theories \cite{Freed:2012bs}. A choice of discrete 6d data at infinity corresponds to selecting different linear combinations of the components of the 2d partition vectors. The 6d ALE partition functions for given 6d data can be reconstructed from the corresponding component of the 2d partition vector for different bound states of BPS strings.

M-strings are among the simplest examples of 6d (1,0) theories, because these models do not have any 6d gauge degrees of freedom.\footnote{\ The only other family of 6d theories with this characteristic are the 6d (1,0) E-strings whose ALE partition functions are analyzed in a separate note \cite{Lockhart:Estrings}.} This simplifies considerably the structure of the possible ALE boundary conditions for M-strings. The main purpose of this work is to extend the analysis of \cite{DelZotto:2023rct} by considering 6d SCFTs that have dynamical gauge vector multiples, and whose ALE partition functions depend also on choices of flat connections at infinity for the corresponding gauge algebras.

The simplest class of models that exhibits non-trivial vector multiplet gauge degrees of freedom are the 6d (1,0) M-string orbifold SCFTs obtained along the worldvolumes of stacks of $r$ M5 branes probing a Taub-NUT space \cite{Taub:1950ez,Newman:1963yy} with $W$ centers, which we denote $\mathbf{TN}_W$, in the low energy limit where the resulting little string decouples. These orbifold theories have a two-fold interpretation. On the one hand these are 6d SCFTs, while on the other hand they can be interpreted as domain walls for a 7d $U(W)$ gauge theory. In this work we consider the 6d equivariant ALE partition functions of the 6d (1,0) M-string orbifold SCFTs. Of the various approaches to computing this quantity which we have developed in \cite{DelZotto:2023rct}, in this project we focus on the BPS string expansion.

The main qualitative difference that distinguishes the orbifolds of M-strings is that these models possess \textit{fractional BPS strings}. In this paper we explore the interplay of these degrees of freedom and the 6d equivariant ALE partition functions, thus extending and generalizing the results of \cite{DelZotto:2023rct}. As expected in analogy to the 4d case, it is the fractional BPS string instantons that give the additional layer of complexity needed to encode the contributions from gauge bundles to the 6d ALE partition functions. Our main result are explicit formulas for the ALE partition functions of 6d M-string orbifold SCFTs (up to a classical prefactor).

A key component of our approach, which is also one of the main results of this paper, is that we are able to determine the 2d (0,4) QFTs that govern the worldvolume theories of BPS instantonic strings thanks to a chain of string dualities. We can read off explicitly the resulting 2d degrees of freedom from a dual network of membranes in the type IIB superstrings. The latter is a generalization the brane configurations studied by \cite{Hanany:2018hlz} (see also  \cite{Witten:2009xu}). The resulting 2d quiver gauge theories have unitary gauge groups which are anomalous. We find that the gauge anomalies can be cured thanks to the presence of additional chiral fermions. In the Type IIB description, these are supported at the transverse intersection of NS5 branes and $(p,q)$ branes. The consistency of the BPS string worldsheet theory for fractional instantons relies on a non-perturbative effect, which gives rise to an anomaly inflow among systems of intersecting $(p,q)$-fivebrane webs and chiral fermions supported at NS5 intersections, and is a generalization of the anomaly inflow analyzed by \cite{Itzhaki:2005tu,Dijkgraaf:2007sw}. Having determined the worldsheet degrees of freedom, we can explicitly determine the corresponding elliptic genera via localization methods \cite{Benini:2013nda,Benini:2013xpa}.

\subsection{ALE partition functions and higher Donaldson-Thomas theory}\label{sec:DTBS}

This section is meant to clarify the connection between our results and the 5d/7d correspondence of \cite{DelZotto:2021gzy}; it can be safely skipped by readers who are not interested in the mathematical counterpart of our results. In the context of \cite{DelZotto:2021gzy}, the equivariant rank-$n$ DT theory \cite{Nekrasov:2014nea} of a Calabi-Yau singularity $X$ was interpreted in terms of the Taub-NUT partition functions for a 5d SCFT. The idea behind it is the following stringy correspondence:
\be
Z_{DT_n}(X) = Z^{7d}_{\mathcal T_{M/\mathbf{TN}_n}}(S^1 \ltimes X) = \mathcal Z_{M}(S^1 \ltimes \mathbf{TN}_n \ltimes X) = Z^{5d}_{\mathcal T_{M/X}}(S^1 \ltimes \mathbf{TN}_n)
\ee
where $\mathcal T_{M/\mathbf{TN}_n}$ is the 7d $U(n)$ gauge theory underlying the equivariant rank $n$ DT theory, and $\mathcal T_{M/X}$ is a 5d SCFT obtained via the geometric engineering of M-theory on the CY singularity $X$. In the case relevant for our paper, $X$ is an elliptic Calabi-Yau thereefold singularity and we have an F-theory uplift \cite{Vafa:1996xn}. The 5d theory $\mathcal T_{M/X}$ has an interpretation as a 5d KK theory for the 6d SCFT obtained from F-theory on $X$. Schematically, we have the relation
\be
\mathcal T_{M/X} = D_{S^1} \mathcal T_{F/X}\,.
\ee
Plugging this relation in the 5d/7d correspondence of \cite{DelZotto:2021gzy} gives a 6d/7d correspondence:
\be
Z_{DT_n}(X) = Z^{5d}_{\mathcal T_{M/X}}(S^1 \ltimes \mathbf{TN}_n) = Z^{6d}_{\mathcal T_{F/X}}(T^2 \ltimes \mathbf{TN}_n)
\ee
The latter is precisely the Taub-NUT partition function of a 6d (1,0) theory of the type we are considering in this paper. $X$ is the elliptic Calabi-Yau threefold singulary dual to the orbifold of M-string 6d SCFT, which can be easily obtained via geometric engineering methods \cite{Heckman:2013pva,DelZotto:2014hpa,Heckman:2015bfa}. Our results therefore have an interpretation as generating functions for higher-rank (equivariant) Donaldson-Thomas invariants for the resolutions of specific elliptic Calabi-Yau singularities \cite{Donaldson:1996kp,Nekrasov:2014nea} (see also \cite{Okounkov:2018yjl,Feyzbakhsh:2021nds}). In this paper, we consider the ALE limit of the above correspondence, which coincides with a 7d $SU(n)$ version of the higher-rank DT theory \cite{DelZotto:2021gzy}:\footnote{\ It is interesting to remark that there are similar relations for all the ALE singularities, also those of type $D_k$ and $E_{6,7,8}$. We expect relations of the form
\be
Z^{7d}_{\mathfrak g}(S^1 \ltimes X) = Z^{6d}_{\mathcal T_{F/X}}(T^2 \ltimes \mathbb{C}^2/\Gamma_{\mathfrak g})
\ee
where $\Gamma_\mathfrak{g}$ is a finite subgroup of $SU(2)$, and it would be interesting to compute the corresponding ALE partition functions.}
\be
Z^{7d}_{SU(n)}(S^1 \ltimes X) = Z^{6d}_{\mathcal T_{F/X}}(T^2 \ltimes \mathbf{ALE}_n).
\ee
One expects the structure of the partition function to simplify in the ALE limit. Namely, turning on a chemical potential 

\subsection{Future directions}\label{sec:outlook}
Before concluding this introduction with the organization of the paper, we would like to mention a few future directions. There are several obvious extensions of our results in this paper, but here we mention three lines of research which we find worth exploring in the near future:
\begin{itemize}
\item \textbf{Computation of ALE partition functions with other methods.} 
The $T^2\times C^2$ partition functions of 6d (1,0) SCFTs can be computed by topological string techniques \cite{Haghighat:2013gba,Haghighat:2013tka,Hohenegger:2013ala,Haghighat:2014vxa,Hayashi:2015zka,Hayashi:2016abm,Gu:2017ccq,Hayashi:2017jze,DelZotto:2017mee,Duan:2020imo}. In particular, the most powerful computational tool towards 6d equivariant partition functions on $T^2 \ltimes \mathbb C^2$ are the 6d generalization of Nakashima-Yoshioka blow up equations. A plethora of exact results have been obtained in the literature about 6d theories exploiting this technique \cite{Gu:2018gmy,Gu:2019dan,Gu:2019pqj,Gu:2020fem}. It would be interesting to generalize that approach to ALE partition functions.
\item \textbf{Non-lagrangian BPS strings.} In the case considered in this paper all the BPS string worldsheet were Lagrangian 2d (0,4) theories. It is well-known that for most 6d SCFTs this is not true, and the BPS string worldsheet do not have a conventional Lagrangian description (which ultimately is related to the lack of an ADHM construction for instantons of exceptional Lie groups). Nevertheless, in the case of strings on $T^2\times\mathbb{C}^2$ it is possible to study the properties of the strings by a number of techniques \cite{Haghighat:2013gba,Haghighat:2013tka,Hohenegger:2013ala,Haghighat:2014pva,Hosomichi:2014rqa,Kim:2014dza,Cai:2014vka,Haghighat:2014vxa,Honda:2015yha,Gadde:2015tra,Kim:2015fxa,Yun:2016yzw,Kim:2016foj,DelZotto:2016pvm,Kim:2018gak,Kim:2018gjo,Duque:2022tub,Lee:2022uiq}. Extending the results of this paper to this larger class of models requires genuinely new techniques with respect to the ones we have developed so far in the ALE case.
\item \textbf{Exploring the modular properties of higher rank DT invariants.} One of the consequences of the mathematical interpretation of our results in the context of the 6d/7d correspondence we have outlined above, is that we expect the higher rank DT invariants must inherit some modularity properties, as they arise as coefficients of meromorphic, vector valued Jacobi forms. It would be interesting to explore these features, in terms of modular actions on the corresponding category of coherent sheaves along the lines of \cite{Schimannek:2019ijf,Cota:2019cjx}.
\end{itemize}

\subsection{Organization of the paper}\label{sec:structure}

This paper is organized as follows. In section \ref{sec:branes} we review basic features of the $\mathcal{T}^{6d}_{r,W}$ SCFTs. In section \ref{sec:data} we review the data that is required to fully specify a background for the $\mathcal{T}^{6d}_{r,W}$ on $T^2\times\mathbb{C}^2/\mathbb{Z}_n$. In section \ref{sec:IIBd} we discuss the brane configurations that gives rise to the $\mathcal{T}^{6d}_{r,W}$ on $T^2\times \mathbf{TN}_n$. In this context, we discuss how the instanton strings of the 6d SCFT can be realized in terms of D3 branes suspended on plaquettes in a brane web, and discuss constraints on the allowed instanton charges. In section \ref{sec:t2c2pf} we review the partition function of $\mathcal{T}^{6d}_{r,W}$ on $T^2\times \mathbb{C}^2$ by carefully examining in section \ref{sec:zper} the contributions arising from NS5 branes and in section \ref{sec:t2c2inst} the contributions from instanton strings. Building on this, in section \ref{sec:zpertc2zn} we determine the contributions of NS5 branes to the $T^2\times\mathbb{C}^2/\mathbb{Z}_{n}$ partition function and show that every NS5 brane supports an $\mathfrak{su}(n)_1$ current algebra. In section \ref{sec:d3z} we turn to the instanton strings and determine their contributions to the $T^2\times\mathbb{C}^2/\mathbb{Z}_{n}$ partition function. We arrive at first in section \ref{sec:d3} at a quiver gauge theory $\QQA$ that captures the degrees of freedom supported on the D3 brane, and by itself suffers from gauge anomalies. In section \ref{sec:anom} we show how the gauge anomalies can be cured by coupling to the quiver gauge theory $\QQA$ the current algebras supported on the NS5 branes. This gives rise to a fully consistent 2d $\mathcal{N}=(0,4)$ worldsheet theory $\QQ$ for the strings. This is a \emph{relative} supersymmetric QFT which depends on boundary conditions for the two-form fields of $\mathcal{T}^{6d}_{r,W}$ at the boundary of $\mathbb{C}^2/\mathbb{Z}_{n}$ through its coupling to the $\mathfrak{su}(n)_1$ current algebras. In section \ref{sec:quivprop} we determine a combinatorial formula for the elliptic genus of arbitrary bound states of instanton strings. In section \ref{sec:zfullc2zn} we combine the NS5 brane and D3 brane contributions into the full $T^2\times\mathbb{C}^2/\mathbb{Z}_n$ partition function of the theories $\mathcal{T}^{6d}_{r,W}$ and discuss its properties. Finally, in section \ref{sec:mod} we determine the central charges and global symmetry levels of $\QQ$ and study the modular properties of its elliptic genus, while in section \ref{sec:examples} we look at concrete examples of elliptic genera that arise in various examples of theories $\mathcal{T}^{6d}_{r,W}$. Additional results are relegated to the appendices: in appendix \ref{sec:TXapp} we derive a bound that restricts the possible ranks of the 2d quiver gauge nodes of instanton strings; in appendix \ref{sec:bps} we determine the partition function of a 5d BPS particle on the background $S^1\times\mathbb{C}^2/\mathbb{Z}_n$; and in appendix \ref{sec:5dabapp} we compute the partition functions of certain 5d abelian quiver gauge theories on $S^1\times\mathbb{C}^2$ and on $S^1\times\mathbb{C}^2/\mathbb{Z}_n$.\newline

\subsection*{Comments on notation}

For the sake of readibility, in the text we employ covariant notation when appropriate. The geometric background $T^2\ltimes\mathbb{C}^2/\mathbb{Z}_n$ depends on additional integer $n$, and we denote the corresponding index by a lowercase $j$. We denote vectors that carries a $j$-index as $\vec{\zeta} = (\zeta_0,\dots,\zeta_{n-1})$.  Additionally, the family of 6d SCFTs we consider depends on two indices: the number of tensor fields $r$ and rank of the gauge algebra $W$. We denote the corresponding indices by a superscript $(a)$ and a subscript $A$ respectively. Vectors carrying such indices are denoted respectively by the notation $\boldsymbol{w} = (w^{(1)},\dots,w^{(r)})$ and $\underline{x} = (x_0,\dots, x_{W-1})$.

\section{The M-string orbifold SCFTs $\mathcal{T}^{6d}_{r,W}$}
\subsection{Review of basic properties}
\label{sec:branes} 
In this section, we will give a brief review of the six-dimensional $\mathcal{N}=(1,0)$ SCFTs of interest in this paper: the M-string orbifold SCFTs. These arise as the worldvolume theory of $r$ M5 branes (where $r$ denotes the \emph{rank} of the SCFT), extended along directions $x_0,\dots,x_5$, probing a Taub-NUT singularity $TN_W$ of charge $W$ along the directions $x_7,\dots,x_{10}$. We denote this six-dimensional theory by 
\be
\mathcal{T}^{6d}_{r,W}.
\ee
We take the first two directions of its worldvolume, $x_0$ and $x_1$, to parametrize a torus $T^2$ of complex modulus $\tau$. For the moment we take $x_2,\dots,x_5$ to parametrize $\mathbb{C}^2$, but in later sections we will generalize this to orbifold singularities of type $\mathbb{C}^2/\mathbb{Z}_{n}.$

The theory has a tensor branch that arises by spacing the $r$ M5 branes along direction $x_6$. The center of mass degrees of freedom correspond to the theory of a single M5 brane probing the $TN_W$ space. Additionally, one has an interacting theory which admits an effective low energy description in terms of $r-1$ tensor multiplets $\widetilde{T}^{(a)}=T^{(a+1)}-T^{(a)},$ for $a=1,\dots, r-1$, where $T^{(a)}$ is the tensor multiplet associated to an individual M5 brane. The bosonic components of a tensor multiplet $T^{(a)}$ are a two-form field $B^{(a)}$ and a real scalar $\phi^{(a)}$, whose vev parametrizes the position of the $a$-th M5 brane along $x_6$. 

On top of this one needs to account for the degrees of freedom arising from the presence of the $TN_W$ singularity. Naively, along each interval separating the $a$-th and $(a+1)$-th M5 brane the singularity gives rise to a 6d gauge algebra $\mathfrak{g}^{(a)} = \mathfrak{u}(W)^{(a)}$, where the superscript in parentheses keeps track of the index. Additionally one expects a 6d global symmetry $\mathfrak{g}^{(0)}\times \mathfrak{g}^{(r)} = \mathfrak{u}(W)^{(a)}\times \mathfrak{u}(W)^{(a)}$ arising from the infinite segments that end on the left- and rightmost M5 branes. However,  $r$ of the $r+1$ abelian factors in $\mathfrak{g}^{(0)}\times\mathfrak{g}^{(1)}\times\dots\times \mathfrak{g}^{(r)}$ receive a mass by the Stuckelberg mechanism \cite{Berkooz:1996iz,Douglas:1996sw,Hanany:1997gh}, so that eventually the gauge algebra of the system is 
\be
\mathfrak{g}
=
\prod_{a=1}^{r-1}\widetilde{\mathfrak{g}}^{(a)}
=
\prod_{a=1}^{r-1}\mathfrak{su}(W)^{(a)},
\ee
and to each factor $\widetilde{\mathfrak{g}}^{(a)}= \mathfrak{su}(W)^{(a)}$ corresponds a 6d (1,0) adjoint vector multiplet in the low energy effective description. The gauge couplings $g^{(a)}_{YM}$ are related to the separations between neighboring M5 branes, which parametrize the tensor branch of the theory:
 \be
 \label{eq:stringtension}
\varphi^{(a)} = \phi^{(a+1)}-\phi^{(a)} = \frac{4\pi^2}{g_{YM}^2}\,,
\ee  
here the $\phi^{(a)}$ are the real scalar components of 6d (1,0) tensormultiplets on the M5 worldvolumes, while $\varphi^{(a)}$ are the combinations that couple to the 6d (1,0) $\mathfrak{su}(W)^{(a)}$ vector multiplet for gauge anomaly cancellation via the Green-Schwarz-Sagnotti mechanism \cite{Green:1984bx,Sagnotti:1992qw}. 

The flavor symmetry has rank $2W - 1 $ and is given by
\be
\mathfrak{f}
=
\widetilde{\mathfrak{g}}^{(0)}\times \widetilde{\mathfrak{g}}^{(r)}\times \mathfrak{u}(1)^{diag}
=
\mathfrak{su}(W)^{(0)}\times \mathfrak{su}(W)^{(r)}\times \mathfrak{u}(1)^{diag},
\ee
where $\mathfrak{u}(1)^{diag}$ is the diagonal subalgebra of the abelian factors of $\mathfrak{g}^{(0)}\times\mathfrak{g}^{(1)}\times\dots\times \mathfrak{g}^{(r)}$, which is not removed by the Stuckelberg mechanism. The remaining low energy degrees of freedom of $\mathcal{T}^{6d}_{r,W}$ consist of a collection of hypermultiplets $H_1,\dots,H_r$, where $H_a$ is in the bifundamental representation of $\mathfrak{su}(W)^{(a-1)}\times \mathfrak{su}(W)^{(a)}$.

The 6d SCFT possesses two dimensional BPS objects that arise to bound states of M2 branes stretched between parallel M5 branes. Due to the Green-Schwarz couplings
\be
\int \widetilde{B}^{(a)} \wedge \Tr F_{\widetilde{\mathfrak{g}}^{(a)}}\wedge F_{\widetilde{\mathfrak{g}}^{(a)}}
\ee
between the tensor fields
\be
\widetilde{B}^{(a)} = B^{(a+1)}-B^{(a)}
\ee
 and the field strength $F_{\widetilde{\mathfrak{g}}^{(a)}}$ of $\widetilde{\mathfrak{g}}^{(a)}$, these BPS configurations carry instanton charge with respect to the gauge algebra $\mathfrak{g}$, which justifies their name of \emph{instanton strings}. 
 
 \medskip
 
 We conclude this review by describing the reduction of this system on a circle of finite size, which we denote $D_{S^1} \mathcal T^{6d}_{r,W}$. \\

\begin{figure}
    \centering
    \includegraphics[scale=0.5]{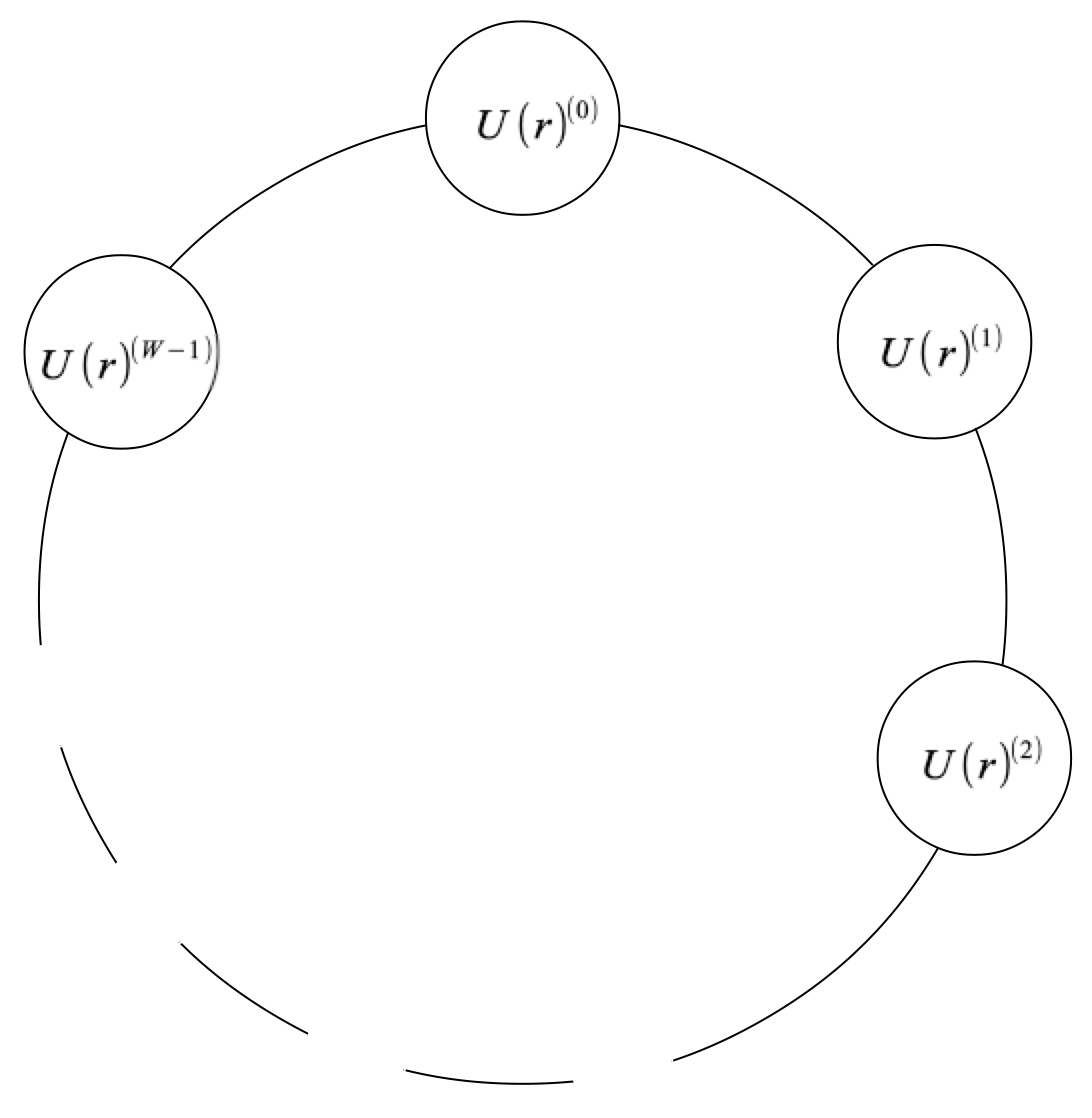}
    \caption{The 5d KK theory describing $r$ D4 branes probing $TN_W$.}
    \label{fig:5dKK}
\end{figure}
\noindent \textbf{5d KK theory}. Let us compactify the M-theory setup along direction $x_1$. This leads to a stack of $r$ D4-branes probing the $TN_W$ singularity in Type IIA. Their worldvolume is described by the 5d quiver gauge theory of figure \ref{fig:5dKK} \cite{Douglas:1996sw}, which we denote by $\mathcal{T}^{KK}_{r,W}$. At low energy, the deformation that corresponds to the tensor branch of the six-dimensional SCFT arises by spacing the $r$ D4-branes along direction $x_6$. In fact, it can be shown that this SCFT has the expected flavor symmetry \cite{Tachikawa:2015mha,Yonekura:2015ksa}
\be
F = SU(W)\times SU(W)\times U(1),
\ee
which is broken to $U(1)^{2W-1}$ away from the superconformal fixed point. To summarize, the theory has a $r\, W$ dimensional Coulomb branch and a $2W-1$ dimensional flavor symmetry group. Considering the description of the 6d theory above, we see that from the circle reduction of the tensor branch low energy effective theory we obtain a 5d Coulomb branch of rank $(W-1) \times r$ from the $r$ gauge sectors $\mathfrak{su}(W)^{(a)}$ together with $r-1$ additional vectors from the tensormultiplets parametrising the tensor branch, in total we obtain:
\be
\text{dim} \, \mathcal M_{Coulomb}(D_{S^1}\mathcal{T}^{6d}_{r,W})^{naive} = (W-1) \cdot (r-1) + r-1 =  W (r - 1)
\ee
however here we are neglecting the center of mass degrees of freedom of the stack of M5 branes probing the $TN_W$ space. In order to understand the latter, we remark that these correspond to a single M5 brane probing the $TN_W$. A circle reduction shows and duality with IIA implies that these degrees of freedom amount to the 6d UV completion of the 5d theory with quiver $\mathcal{T}^{KK}_{1,W}$. The latter has indeed a Coulomb branch of rank W, from which we conclude that
\be
\begin{aligned}
\text{dim} \, \mathcal M_{Coulomb}(D_{S^1}\mathcal{T}^{6d}_{r,W}) &= \text{dim} \, \mathcal M_{Coulomb}(D_{S^1}\mathcal{T}^{6d}_{r,W})^{naive}+W \\
&=  \text{dim} \, \mathcal M_{Coulomb}(\mathcal T^{KK}_{r,W})
\end{aligned}
\ee
thus reconciling the two pictures in terms of the 5d duality
\be
D_{S^1}\mathcal{T}^{6d}_{r,W} \cong \mathcal T^{KK}_{r,W}.
\ee

\subsection{SCFT data on $T^2\times \mathbf{ALE}_n$}
\label{sec:data}
Having reviewed basic aspects of the theories $\mathcal{T}^{6d}_{r,W}$, let us now discuss the data needed to specify a background for its fields on the spacetime $T^2\times \mathbf{ALE}_n$.\footnote{ \  We refer to Section 3 of our previous paper on the topic of ALE partition functions of 6d theories \cite{DelZotto:2023rct} for a description of the relevant background geometry (as well as for our notations and conventions).} There are three collections of discrete data that we can specify: the fluxes and monodromies at infinity for the two-form fields in the $r$ tensor multiplets, the flat connections at infinity for the gauge groups of the models, and the prescriptions of background flavor symmetry Wilson lines. In this section we describe these discrete data in detail.

\paragraph{Fluxes and monodromies at infinity for the two-form fields.} First of all, as in the case of M-strings \cite{DelZotto:2023rct}, we need to take into account possible fluxes and monodromies at infinity for the two-form fields in the $r$ tensor multiplets. To see this, recall that under compactification from six to five dimensions, the $r$ tensor fields $B^{(a)}$  give rise to abelian one-form connections $A^{KK,(a)}$ associated to a collection of $U(1)$ gauge bundles $\mathcal{V}^{(a)}$. The monodromy at infinity for a given gauge connection is specified in terms of an irreducible representation of 
\be
\text{Hom}(\pi_1(S^3/\mathbb{Z}_n),U(1))\simeq \mathbb{Z}_n,
\ee
where 
\be
\pi_1(S^3/\mathbb{Z}_n)\simeq \mathbb{Z}_n
\ee
is the first homotopy group of the asymptotic boundary of $\mathbf{ALE}_n \cong \mathbb{C}^2/\mathbb{Z}_n$. Equivalently, we may label the connection in terms of a phase
\be
e^{2\pi i\frac{\omega^{KK,(a)}}{n}},\qquad \omega^{KK,(a)}\in \mathbb{Z}_n.
\ee
The fluxes of the gauge group $U(1)^{KK,(a)}$ along the compact two cycles of (resolved) $\mathbb{C}^2/\mathbb{Z}_n$ are captured by an $(n-1)$-tuple of integers
\be
u_j^{KK,(a)} = \int_{\Sigma_j}c_1(\mathcal{V}^{(a)}),\qquad j=1,\dots,n-1,
\ee
which are subject to the monodromy constraint
\be\label{eq:omegaKKdef}
\sum_{j=1}^{n-1} j u_j^{KK,(a)}=\omega_j^{KK,(a)}
\ee
and characterize the way in which the bundle $\mathcal{V}^{(a)}$ decomposes in terms of the line bundles $\mathcal{R}_j$ that are described in \cite{DelZotto:2023rct}:
\be
\mathcal{V}^{(a)}=\otimes_{j=1}^{n-1}\mathcal{R}_j^{u_j^{KK,(a)}}.
\ee
Finally, the four-form component of the Chern character of $\mathcal{V}^{(a)}$ encodes the $U(1)^{KK,(a)}$ instanton number
\be
n^{KK,(a)} = \int_{\mathbf{ALE}_n}ch_2(\mathcal{V}^{(a)}).
\ee
\paragraph{Flat connections at infinity for gauge fields.} In addition to the two-form field data, for $r \geq 2$ the monodromy of 6d one-form fields also encodes physically inequivalent field configurations. Specifically, for each factor $\mathfrak{g}^{(a)}=\mathfrak{u}(W)^{(a)}$ of the gauge algebra the gauge field can approach a nontrivial flat connection at infinity. This is labeled by an element
\be
\rho^{(a)} \in \text{Hom}(\pi_1(S^3/\mathbb{Z}_n),\mathfrak{u}(W)^{(a)}),
\ee
which can be characterized in terms of a partition $\vec{w}^{(a)}=(w^{(a)}_0,\dots,w^{(a)}_{n-1})\in \mathbb{Z}_{\geq 0}^n$ of $W$:
\be
w^{(a)}_0 + w^{(a)}_1 + ... + w^{(a)}_{n-1} = W.
\ee
\paragraph{Flavor symmetry backgrounds.} Analogously, one is also allowed to turn on a nontrivial background for the flavor symmetry $\mathfrak{g}^{(0)}\times\mathfrak{g}^{(r)} = \mathfrak{u}(W)^{(0)}\times \mathfrak{u}(W)^{(r)}$, which is characterized by a pair of partitions $\vec{w}^{(0)}$ and $\vec{w}^{(r)}$ of $W$. We may also consider turning on magnetic fluxes for the each $\mathfrak{u}(W)^{(a)}$ factor of the 6d gauge algebra and flavor symmetry. The fluxes are again encoded in the the first Chern class of the corresponding $r+1$ $\mathfrak{u}(W)$ gauge bundles $\mathcal{V}_{W}^{(a)}$. Note that the first Chern class is completely determined by the abelian factor of $\mathfrak{u}(W)\simeq \mathfrak{u}(1)\oplus \mathfrak{su}(W)$. Since the Stuckelberg mechanism gives mass to all of the abelian factors except for the diagonal $\mathfrak{u}(1)^{diag}$, in fact one has:
\be
\label{eq:c1con}
c_1(\mathcal{V}_{W}^{(0)}) = c_1(\mathcal{V}_{W}^{(1)})=\dots=c_1(\mathcal{V}_{W}^{(r)}).
\ee
As for the two-form fields, we may expand the first Chern classes of the one-form connections as follows:
\be\label{eq:bananarepublic}
c_1(\mathcal{V}_{W}^{(a)}) = \sum_{j=1}^{n-1}u^{(a)}_j c_1(\mathcal{R}_j),
\ee
where now the requirement that the $\mathfrak{u}(W)^{(a)}$ bundles have the correct monodromy gives rise to the constraints
\be
\label{eq:ucco}
\sum_{j=1}^{n-1}j\, u_j^{(a)} = \sum_{j=1}^{n-1} j w_j^{(a)} \qquad \text{mod }n.
\ee
Notice in particular that \eqref{eq:c1con} combined with \eqref{eq:bananarepublic} implies that
\be
\label{eq:constro}
\vec{u}^{(a)} = \vec{u}^{(b)}\qquad \forall \,a,b.
\ee
Finally the four-form component of the Chern character determines the instanton charge of $\mathfrak{u}(W)^{(a)}$:
\be
N^{(a)}:=\int_{\mathbf{ALE}_n} \Tr F_{\mathfrak{u}(W)^{(a)}}\wedge F_{\mathfrak{u}(W)^{(a)}} = \int_{\mathbf{ALE}_n} ch_2(\mathcal{V}^{(a)}_W).
\ee
The aim of this paper is to consider supersymmetric configurations which contribute to the 6d BPS partition function. In section \ref{sec:tdua} we will see that this gives rise to very strong constraints on the Chern classes of $\mathcal{V}^{(a)}_W$, or in other words on the $\vec{u}^{(a)}$ and $N^{(a)}$.

\section{Type IIB description}
\label{sec:IIBd}
In this section we study the Type IIB brane configuration which is obtained by replacing the ALE singularity with a Taub-NUT space of charge $n$ and performing T-duality along the Taub-NUT circle. As in the case of M-strings \cite{DelZotto:2023rct}, this very useful dual description of $\mathcal{T}^{6d}_{r,W}$ will lead us in later sections to a clear understanding of the BPS contributions to the partition function of the 6d SCFTs $\mathcal{T}^{6d}_{r,W}$ on $T^2\times \mathbf{ALE}_n$. We begin in section \ref{sec:7dT} by looking at a simpler class of theories: 7d $U(W)$ super-Yang-Mills on $\mathbb{R}\times T^2\times \mathbf{TN}_n$. In section \ref{sec:tdua} we then couple this configuration to the 6d $\mathfrak{u}(r)$ $\mathcal{N}=(2,0)$ theory describing a stack of parallel NS5 branes, leading to the sought-after Type IIB description of the $\mathcal{T}^{6d}_{r,W}$ SCFTs. As already implicitly manifest in the discussion of the previous section, the 6d theories $\mathcal{T}^{6d}_{r,W}$ have a dual interpretation in terms of domain walls for this 7d gauge theory \cite{Gaiotto:2014lca}: the discussion in this second part of this section can be reinterpreted also as a description of the interplay between these domain walls and the Taub-NUT background.

\subsection{Type IIA brane setup and T-dual of 7d $U(W)$ SYM}
\label{sec:7dT}

\begin{figure}
    \centering
    \includegraphics[width=\textwidth]{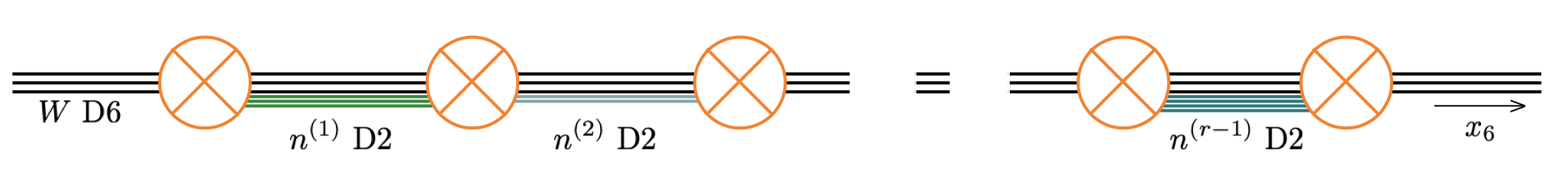}
    \caption{Type IIA brane setup realizing the $\mathcal{T}^{6d}_{r,W}$ SCFT. NS5 branes are displayed in orange. Shown here is a configuration corresponding to a bound state of $n^{(1)}$ instanton strings for the first gauge group, $n^{(2)}$ instanton strings for the second, and so on.}
    \label{fig:IIAframe}
\end{figure}

The starting point to reach the IIB dual frame is to consider a different compactification of the M-theory setup, obtained by compactifying M-theory to Type IIA along the circle fiber of the $\mathbf{TN}_W$ space. The IIA dual of the Taub-NUT geometry is a stack of $W$ D6 branes \cite{Sen:1997js,Sen:1997kz}. Moreover, the $r$ M5 branes dualize to a set of $r$ NS5 branes (whose positions along direction $x_6$ parametrize the tensor branch of the 6d model). This setup is summarized in table \ref{tab:IIA} and displayed in figure \ref{fig:IIAframe}.
\be
\label{tab:IIA}
\begin{tabular}{c|cccccc|cccc}
  & 0 & 1 & 2 & 3 & 4 & 5 & 6& 7 & 8 & 9 \\
\hline
$r$ NS5 & $\times$ & $\times$ & $\times$ & $\times$ & $\times$ & $\times$ &  &  &  & \\
$W$ D6 & $\times$ & $\times$ & $\times$ & $\times$ & $\times$ & $\times$ & $\times$ &  &  & \\   
D2 & $\times$ & $\times$ &  &  &  &  & $\times$ &  &  & \\   
\end{tabular}
\ee
Instanton strings in this duality frame are simply realized as bound states of D2-branes stretching between neighboring NS5-branes. For the purpose of our computation, we take the D2 branes along directions $x_0,x_1$ and $x_6$. Now we are intersted in coupling the 6d theory to a background with topology $T^2 \times \mathbf{ALE}_n$. As explained above, a dual frame is obtained by replacing $\mathbf{ALE}_n$ with $\mathbf{TN}_n$ and then sending the Taub-NUT radius to infinity. Consider the case $r=0$ which corresponds to the 7d $U(W)$ SYM theory describing the low energy worldvolume of the stack of $W$ D6 branes on $T^2\times \mathbf{TN}_n \times \mathbb{R}$ in Type IIA string theory. Now exploit the Taub-NUT circle fibration of $\textbf{TN}_n$ to dualize to Type IIB. Under T-duality, the Taub-NUT space is replaced by a collection of $n$ NS5 branes on $\mathbb{R}^3\times \tilde{S}^1$, where $\tilde{S}^1$ is the dual to the Taub-NUT circle and has radius $1/R$, where $R$ is the Taub-NUT radius of $\mathbf{TN}_n$ (ie. the size of its asymptotic circle). The positions of the NS5 branes along $\mathbb{R}^3$ coincide with the positions $\vec{x}_i$ of the Taub-NUT centers on the Type IIA side. On the other hand, the angular positions of the NS5 branes on $\tilde{S}^1$ coincide with the parameters
\be
\theta_i = \int_{\mathcal{C}_i}B^{NS}
\ee
in Type IIA, where $B^{NS}$ is the two-form B-field. By performing a gauge transformation on $B^{NS}$ one may shift all the angular variables $\theta_i$ by the same amount, which in the Type IIB frame corresponds to a rotation of $\tilde{S}^1$.
 In what follows, we keep the $\theta_i$ to be distinct but set all $\vec{x}_i\to 0$. We arrive at the following Type IIB brane configuration:
\be
\label{tab:7dSYMIIB}
\begin{tabular}{c|ccccc|cc|ccc}
  & 0 & 1 & 2 & 3 & 4 & 5 & 6& 7 & 8 & 9 \\
\hline
$W$ D5 & $\times$ & $\times$ & $\times$ & $\times$ & $\times$ &  & $\times$  &  &  & \\
$n$ NS5 & $\times$ & $\times$ & & & &  & $\times$ &$\times$ & $\times$ & $\times$ \\   
\end{tabular}
\ee
where we take directions $x_0$ and $x_1$ to parametrize the $T^2$, while directions $x_2,x_3,x_4$ and $x_5$ parametrize respectively the $\mathbb{R}^3$ and $\tilde{S}^1$ arising from Taub-NUT, and the remaining four coordinates to parametrize flat space.
The NS5 and D5 branes are both spaced along the $\widetilde{x}_5$ circle, and we assume that no two branes are coincident. In particular this means that the 7d gauge group $U(W)$ is broken to $U(1)^W$. For $j=0,\dots,n-1$, let $w_j$ denote the number of D5 branes between the $j+1$ and $j$-th NS5 brane (where $j$ is taken to be periodic modulo $n$). The positions $s_1,\dots,s_W$ of the D5 branes along $\widetilde{x}_5$ determine the topological class of the bundle corresponding to the gauge fields of 7d $U(1)^W$ SYM on $\mathbf{TN}_n$, as in \cite{Witten:2009xu}. Let us review how this comes about. Since we are on the Coulomb branch, the gauge bundle is simply given by a sum of line bundles 
\be
\label{eq:lbsum}
\mathcal{W}=\oplus_{i=1}^W\mathcal{R}_{s_i}.
\ee
Recall \cite{Witten:2009xu} that a line bundle on $\mathbf{TN}_n$ can alwyas be presented in the form 
\be
\mathcal{R}_s = \mathcal{L}_*^{t}\bigotimes\left(\otimes_{j=1}^n\mathcal{L}_j^{p_j}\right)
\ee
where $(t,p_1,\dots,p_n)\in \mathbb{C}^*\times\mathbb{Z}^n$ are determined up to the equivalence
\be
(t,p_1,\dots,p_n)\sim (t+1,p_1-1,\dots,p_n-1).
\ee
The line bundle corresponding to a brane located at a position $s$ between the $j$-th and $j+1$-st NS5 brane is given by
\be
\mathcal{R}_s = \mathcal{L}_*^{s/2\pi R}\otimes\mathcal{R}_j
\ee
where
\be
\mathcal{R}_j = \bigotimes_{\ell=1}^j\mathcal{L}_\ell^{-1}.
\ee
The topological class of the line bundle jumps if one changes the relative position of the D5-brane to the NS5-branes, since this is not a smooth process. On the other hand, smoothly moving a D5-brane past an NS5-brane, which is not what we do here, would result in the creation of a D3 brane suspended between the two \cite{Hanany:1996ie}. As expected, in the ALE limit $R\to \infty$ the precise position $s$ of the D5 brane becomes irrelevant and the topological class of the corresponding line bundle $\mathcal{R}_j$ depends only on the value of $j$.

Let us now consider the gauge bundle \eqref{eq:lbsum}. The collection of two-forms $c_1(\mathcal{R}_1),$ $\dots,$ $c_1(\mathcal{R}_{n-1})$ forms a basis of $H^2(\mathbf{ALE}_n,\mathbb{Z})$ which is dual to the basis $\Sigma_1,\dots\Sigma_{n-1}$ of $H_2(\mathbf{ALE}_n,\mathbb{Z})$ \cite{Douglas:1996sw}. Written in terms of this basis, the first Chern class of $\mathcal{W}$ is given by
\be
\sum_{j=1}^{n-1}w_j c_1(\mathcal{R}_j),
\ee
where $u_j = w_j$ in the absence of $D3$ branes.

\begin{figure}
    \centering
    \includegraphics[scale=0.5]{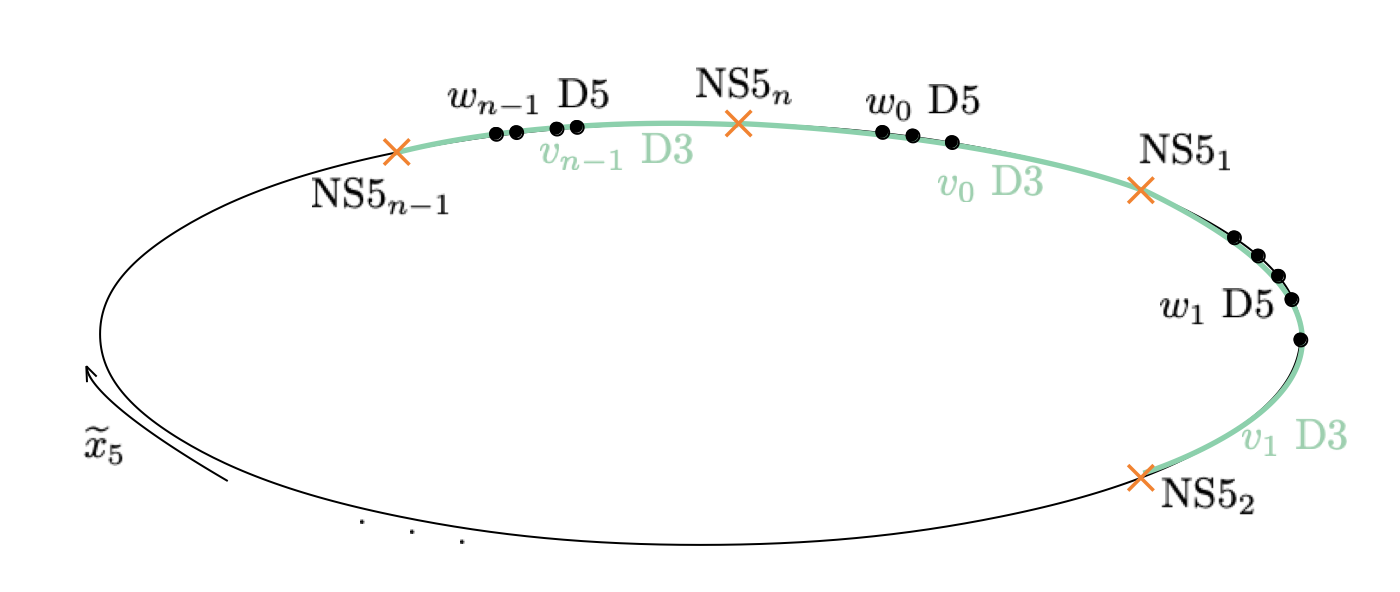}
    \caption{Type IIB brane setup dual to 7d SYM  on $\mathbf{TN}_n$.}
    \label{fig:7dIIB}
\end{figure}

The instanton number, which is generally fractional \cite{Douglas:1996sw}, is also determined in terms of $w_1,\dots,w_{n-1}$:
\be
\int_{\mathbf{ALE}_n}ch_2(\mathcal{W}) =\sum_{j=1}^{n-1}\frac{j(n-j)}{2n}w_j.
\ee
This is however not the entire story: one may consider more general brane configurations that, beside the $W$ D5 and $n$ NS5 branes, also involve D3 branes arranged according to the following table:
\be
\begin{tabular}{r|ccccc|cc|ccc}
  & 0 & 1 & 2 & 3 & 4 & 5 & 6& 7 & 8 & 9 \\
\hline
$W$ D5 & $\times$ & $\times$ & $\times$ & $\times$ & $\times$ &  & $\times$  &  &  & \\
$n$ NS5 & $\times$ & $\times$ & & & &  & $\times$ &$\times$ & $\times$ & $\times$ \\
D3 & $\times$ & $\times$ & & & & $\times$ & $\times$ & & & \\
\end{tabular}
\ee
The D3 branes may be taken to wrap around the $\widetilde{x}^5$ circle, but may also be allowed to end on NS5 or D5 branes. Generic configurations of D3 branes suspended between NS5 or D5 branes are relevant to the case of gauge theory on Taub-NUT spaces; however, in the ALE limit only configurations with D3 branes ending on NS5 branes are relevant.\footnote{ \  This is a generalization of the fact that 4d SYM on $\mathbb{R}^3\times S^1$ admits {caloron} solutions in addition to instantons \cite{Lee:1997vp,Kraan:1998pm}, which can be mapped to D3 branes ending on D5 branes and are not present in the case of gauge theory on $\mathbb{R}^4$.} We are thus led to consider the brane configurations depicted in figure \ref{fig:7dIIB}.
For $j=0,\dots,n-1$, let us denote by $v_j$ the number of D3 branes suspended between the $j+1$-st and $j$-th NS5 brane. Including D3 gives rise to a different bundle which we denote by $\widetilde{\mathcal W}$. The D3 branes do not change the monodromy of the gauge bundle \cite{Witten:2009xu}, but they do affect both the instanton number, which is now given by
\be\label{eq:muahahaha}
N = \int_{ALE_n}ch_2(\widetilde{\mathcal W}) = v_0+\sum_{j=1}^{n-1}\frac{j(n-j)}{2n}w_j,
\ee
as well as the first Chern class, which now has coefficients
\be
\label{eq:uvsj}
u_j = w_j - ({C}^{\widehat{A}_{n-1}}\cdot \vec{v})_j,\quad j=1,\dots,n-1,
\ee
constrained by the relation \eqref{eq:ucco}. Here $C^{\widehat{A}_{n-1}}$ denotes the Cartan matrix of affine $SU(n)$. It is in this sense that the D3 branes in the IIB dual correspond to fractionalizing the instantons which, in the IIA setup, were given by D2 branes.

It is convenient to extend the definition of equation \eqref{eq:uvsj} to the case $j=0$ and to define $\vec{u} = (u_0,u_1,\dots,u_{n-1})$. However, note that $u_0$ is not an independent parameter, since summing over $j=0,\dots,n-1$ in equation \eqref{eq:uvsj} gives:
\be
u_0 = W - \sum_{j=1}^{r-1} u_j.
\ee

\begin{figure}
    \centering
    \includegraphics[scale=0.42]{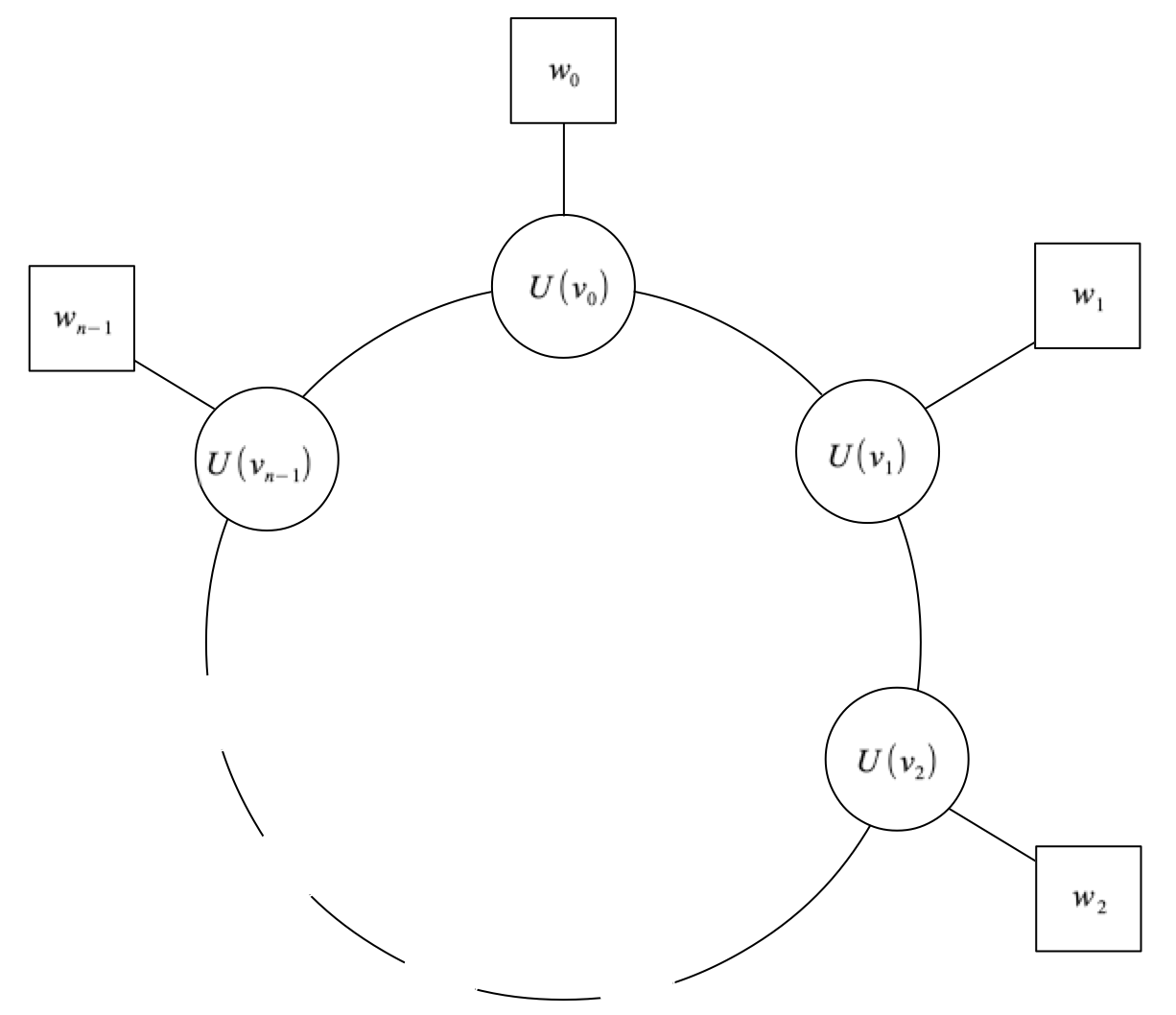}
    \caption{3d Kronheimer-Nakajima quiver gauge theory $\KN$.}
    \label{fig:3dKN}
\end{figure}

The configuration of D3 branes described above is described in terms of the celebrated Kronheimer-Nakajima 3d $\mathcal{N}=4$ quiver gauge theory $\KN$ depicted in figure \ref{fig:3dKN}. Its moduli space of vacua coincides with the moduli space $\mathcal{M}_{\vec{w},\vec{v}}$ of $U(W)$ instantons on $\mathbb{C}^2/\mathbb{Z}_n$, whose dimension is given by:
\be
\dim_{\mathbb{R}} \mathcal{M}_{\vec w,\vec v}  = 4\vec{v}\cdot \vec{w} - 2\vec{v}\cdot C^{\widehat{A}_{n-1}}\cdot \vec{v} = 2\vec{v}\cdot(\vec{w}+\vec{u}).
\ee
These 3d $\mathcal N=4$ theories govern the worldvolume theories of solitonic fractional instantons for the 7d $U(W)$ SYM theory on this backgound. It is interesting to remark that these configurations typically also carry a non-trivial magnetic charge, as per equation \eqref{eq:uvsj}.

\subsection{Type IIB brane setup}
\label{sec:tdua}
We are now ready to discuss the Type IIB setup that corresponds to the 6d SCFTs $\mathcal{T}^{6d}_{r,W}$ on $T^2\times \mathbf{TN}_n$. As per our discussion above, this is obtained by introducing an additional stack of $r$ NS5 branes in Type IIA string theory which are extended along directions $x_0,\dots,x_5$, and spaced along direction $x_6$. The dual Type IIB setup involves, to first approximation, a stack of $r$ NS5 branes, a second stack of $n$ NS5 branes, and a stack of $W$ D5 branes, which are arranged according to the following table:
\be
\begin{tabular}{c|ccccc|cc|ccc}
  & 0 & 1 & 2 & 3 & 4 & 5 & 6& 7 & 8 & 9 \\
\hline
$r$ NS5 & $\times$ & $\times$ & $\times$ & $\times$ & $\times$ & $\times$ & & & & \\
$W$ D5 & $\times$ & $\times$ & $\times$ & $\times$ & $\times$ &  & $\times$  &  &  & \\
$n$ NS5 & $\times$ & $\times$ & & & &  & $\times$ &$\times$ & $\times$ & $\times$ \\
\end{tabular}
\ee
One also needs to account for the possibility of resolving the intersections between the $r$ NS5 branes and $W$ D5 branes in terms of $(p,q)$-fivebranes on the (56) plane \cite{Aharony:1997bh}: we do not do so for the moment but will return to this point below.

Introducing the $r$ NS5 branes gives rise to several new field configurations: we describe them here schematically. Our aim here is to determine the quantum numbers of those BPS configurations that can contribute to the $\textbf{ALE}_n$ partition function. A more detailed description of the corresponding worldvolume theories is presented in Section \ref{sec:zzc2zn}.

\paragraph{Chiral fermions from intersecting NS5s.} The intersections between the two stacks of $r$ and $n$ NS5 branes support chiral 2d fermionic degrees of freedom extended along the directions $x_0,x_1$ \cite{Itzhaki:2005tu}. At the superconformal fixed point these degrees of freedom give rise to a $\mathfrak{u}(n\, r)_1$ current algebra, which on the tensor branch reduces to
\be
    \mathfrak{u}(n)_1^r
    \subset
    \mathfrak{u}(n\, r)_1.
\ee
In taking the ALE limit, each factors of $ \mathfrak{u}(n)_1 $ is replaced by $\mathcal{H}\times\mathfrak{su}(n)_1$, where $\mathcal{H}$ denotes a factor corresponding to a Heisenberg algebra associated to a non-compact free
boson. The resulting system of chiral fermions is described by a bulk-boundary system which is sensitive to the choice of discrete data $\omega^{KK,(a)}$ for the two-form fields $B^{(a)}$ --- see equation \eqref{eq:omegaKKdef}. This translates by the McKay correspondence \cite{Dijkgraaf:2007sw} to a choice of integrable highest weight representation for $\mathfrak{su}(n)_1^r$ labeled by $\boldsymbol{\omega}^{KK}=(\omega^{KK,(1)},\dots,\omega^{KK,(1)})$. As discussed in greater detail in \cite{DelZotto:2023rct}, these 2d chiral fermions indeed give rise to a relative field theory \cite{Freed:2012bs}, and do not possess a partition function. Rather, we obtain a system with a vector of conformal blocks necessary to encode the extra dependence on choices of discrete data at infinity for the 6d theory. More details about this interplay are discussed in Section \ref{sec:d3z} below.

\paragraph{Allowed D3s configurations.} On $T^2\times \mathbb{C}^2$, BPS strings arise from D3 branes suspended between the $r$ NS5 branes; on $T^2\times \mathbf{ALE}_n$, the possibility arises of suspending D3 branes on plaquettes which are supported on a rectangle in the $(\tilde{x}_5,x_6)$ plane which is bounded by two NS5 branes extended along directions $x_0,\dots,x_4$ and $\tilde{x}_5$ and two NS5 branes extended along directions $x_0,x_1$ and $x_6,\dots,x_9$. Schematically, we have the configuration
\be
\begin{tabular}{c|ccccc|cc|ccc}
  & 0 & 1 & 2 & 3 & 4 & 5 & 6& 7 & 8 & 9 \\
\hline
$r$ NS5 & $\times$ & $\times$ & $\times$ & $\times$ & $\times$ & $\times$ & & & & \\
$W$ D5 & $\times$ & $\times$ & $\times$ & $\times$ & $\times$ &  & $\times$  &  &  & \\
$n$ NS5 & $\times$ & $\times$ & & & &  & $\times$ &$\times$ & $\times$ & $\times$ \\
D3 & $\times$ & $\times$ & & & & $\times$ & $\times$ & & & \\
\end{tabular}
\ee
This leads to two-dimensional BPS objects wrapped on $T^2$ which preserve $\mathcal{N}=(0,4)$ supersymmetry \cite{Chung:2016pgt,Hanany:2018hlz}. We will call these BPS objects \textit{fractional BPS strings}. In what follows we aim to determine which configurations of fractional BPS strings are allowed. It is convenient for the moment to work with the $\mathfrak{u}(W)^{(a)}$ gauge factors -- we will impose the constraints from the Stuckelberg mechanism momentarily. Let us focus on a single gauge factor $\mathfrak{u}(W)^{(a)}$, and consider allowed gauge field configurations of the corresponding 7d $U(W)^{(a)}$ SYM theory living on the interval separating the $a$-th and $(a+1)$-st NS5 brane, before imposing boundary conditions at the two ends. Here the analysis from the previous section applies. In particular, recall from equation \eqref{eq:uvsj} that the first Chern classes $\vec{u}^{(a)}$ of the gauge bundle are determined by
\be
\label{eq:e2}
\vec{u}^{(a)} = \vec{w}^{(a)} -C^{\widehat{A}_{n-1}}\cdot \vec{v}^{(a)},
\ee
here the components $v^{(a)}_j$ of $\vec{v}^{(a)}$ denote the numbers of D3 branes suspended between the $a$-th and $a+1$-st vertical NS5 brane and between the $j$ and $j+1$-st horizontal NS5 brane.

As for the gauge symmetry factors, there is also the possibility to consider fractional BPS strings paired to the flavor symmetry factors which arise from the semi-infinite intervals extending to the left and to the right of the stack of $r$ NS5 branes. Such BPS strings are infinitely massive defects from the point of view of the 6d SCFT, being extended over the same semi-infinite intervals as the D6 branes. These contribution would correspond to studying the ALE partition function in presence of surface defects. In order to give a complete characterization of such observables one should also consider adding surface defects with non-trivial 2-form symmetry charges in the defect group \cite{DelZotto:2015isa}. Since we are not studying the ALE partition functions in presence of surface defects insertions, we exclude all these contributions, including the ones from these ``flavor'' BPS surface defects. As a consequence, we have the additional constraints
\be
\vec{u}^{(0)} = \vec{w}^{(0)} \qquad\qquad\vec{u}^{(r)} = \vec{w}^{(r)}.
\ee
Now we can see the effect of the Stuckelberg mechanism on the allowed gauge configuration and background fields: indeed, from equation \eqref{eq:constro}
\be
\vec{u}^{(a)} = \vec{u}^{(b)} \qquad\forall \, a,b=0,...,r
\ee
it follows that for each of the $\mathfrak{u}(W)^{(a)}$, the first Chern classes of all the corresponding bundles are completely determined by the choice of monodromy $\rho^{(0)}$ for the background gauge field associated to the flavor symmetry $\mathfrak{u}(W)^{(0)}$:
\be
\label{eq:e1}
\vec{u}^{(a)} = \vec{w}^{(0)} \qquad \forall\, a.
\ee
In view of the above constraints, we can easily determine for fixed $W, n$ and $r$ what are the allowed possible configurations. Indeed, since for all $a$, $W\geq w^{(a)}_j \geq 0$, combining equations \eqref{eq:e1} and \eqref{eq:e2} gives rise to the following constraint on the allowed fractional BPS string numbers $v^{(a)})_j$:
\be
\label{eq:ineq}
W\geq w^{(0)}_j+(C^{\widehat{A}_{n-1}} \cdot v^{(a)})_j \geq 0\qquad \forall\text{ } j.
\ee
We can exploit the inequality \eqref{eq:ineq} to determine the allowed monodromies for any of the gauge factors $\mathfrak{u}(W)^{(a)}$, and the corresponding configuration of fractional BPS strings. We arrive at the following result, whose derivation is presented in Appendix \ref{sec:TXapp}: Let us denote $\delta^{(a)}_j$ the mismatch between the numbers of D3 branes suspended between the $j$-th and the $j+1$ plaquette
\be
\delta^{(a)}_j = v^{(a)}_{j+1}-v^{(a)}_{j}\in\mathbb{Z}\,
\ee
where the index $j$ is valued in the integers modulo $n$. Borrowing from the Young-tableaux notation, we define
\be
|\vec{v}| = \sum_j v_j\, \qquad\text{and}\qquad \norm{\vec{v}}^2 = \sum_j (v_j)^2
\ee
Consider the finite set $\mathfrak{D}_{\vec{w}^{(0)}}$ of $n$-tuples $\vec{\delta}^{(a)} = (\delta^{(a)}_0,\dots,\delta^{(a)}_{n-1})\in \mathbb{Z}^{n}$ (with $\vert\vec{\delta}^{(a)}\vert=0$) that satisfy the bound
\be
\label{eq:deltacondition}
\norm{\vec{\delta}^{(a)}}^2 \leq \frac{1}{2} \sum_{j=1}^{n-1}j(n-j)\max\left((W-w^{(0)}_j)^2,(w^{(0)}_j)^2\right).
\ee
Let us write 
\be
u^{(a)}_{j} = w^{(a)}_j+\delta^{(a)}_j-\delta^{(a)}_{j-1}= u^{(0)}_j.
\label{eq:wd}
\ee
and denote by 
\be\label{eq:Wset}
\mathfrak{W}_{\vec{w}^{(0)}}
\ee
the set of all partitions $\vec{w}^{(a)}$ of $W$ satisfying equation \eqref{eq:wd} for a certain $\vec{\delta}^{(a)}=\vec{\delta}^{(a)}_{\vec{w}^{(a)}}\in\mathfrak{D}_{w^{(0)}}$.
Any element of $\mathfrak{W}_{\vec{w}^{(0)}} $ corresponds to an allowed choice of monodromy for any of the gauge factors $\mathfrak{u}(W)^{(a)}$. For this choice of monodromy, there exists a one-parameter family of instantons that arise from bound states of D3 branes as follows: let
\be
n^\star_{\vec w^{(a)}} = -\min(0,\delta^{(a)}_{\vec{w}^{(a)},0},\delta^{(a)}_{\vec{w}^{(a)},0}+\delta^{(a)}_{\vec{w}^{(a)},1},\dots,\delta^{(a)}_{\vec{w}^{(a)},0}+\delta^{(a)}_{\vec{w}^{(a)},1}+\dots+\delta^{(a)}_{\vec{w}^{(a)},n-2}).
\ee
Then for each $\kappa^{(a)}\in\mathbb{Z}_{\geq n^\star_{\vec{w}^{(a)}}}$ the bound state of D3 branes described by the $n$-tuple of nonnegative integers
\be
\vec{v}_{\kappa^{(a)},\vec{w}^{(a)}}=(v^{(a)}_0,v^{(a)}_1,\dots,v^{(a)}_{n-1})\label{eq:vsdeltas}
\ee
where
\be
\begin{aligned}
v^{(a)}_0\, &= \kappa^{(a)},\\
v^{(a)}_1\, &= \kappa^{(a)}+\delta^{(a)}_{\vec{w}^{(a)},0},\\
v^{(a)}_2\, &= \kappa^{(a)}+\delta^{(a)}_{\vec{w}^{(a)},0}+\delta^{(a)}_{\vec{w}^{(a)},1},\\
&\vdots\\
v^{(a)}_{n-1}\, &= \kappa^{(a)}+\sum_{j=0}^{n-2}\delta^{(a)}_{\vec{w}^{(a)},j}
\end{aligned}
\ee
gives rise to a valid BPS configuration with instanton number
\be
N_{\kappa^{(a)},\vec{w}^{(a)}} = \kappa^{(a)} +\sum_{j=1}^{n-1}\frac{j(n-j)}{2n}w^{(a)}_j.
\label{eq:instnum}
\ee
The configurations of D3 branes described by the tuples \eqref{eq:vsdeltas} give rise to the valid fractional BPS instanton string configurations of the 6d SCFT for a given choice of monodromies $\boldsymbol{\vec{w}}=(\vec{w}^{(0)},\vec{w}^{(1)},\dots,\vec{w}^{(r)})$.

\medskip

\begin{figure}
    \centering
    \includegraphics[width=\textwidth]{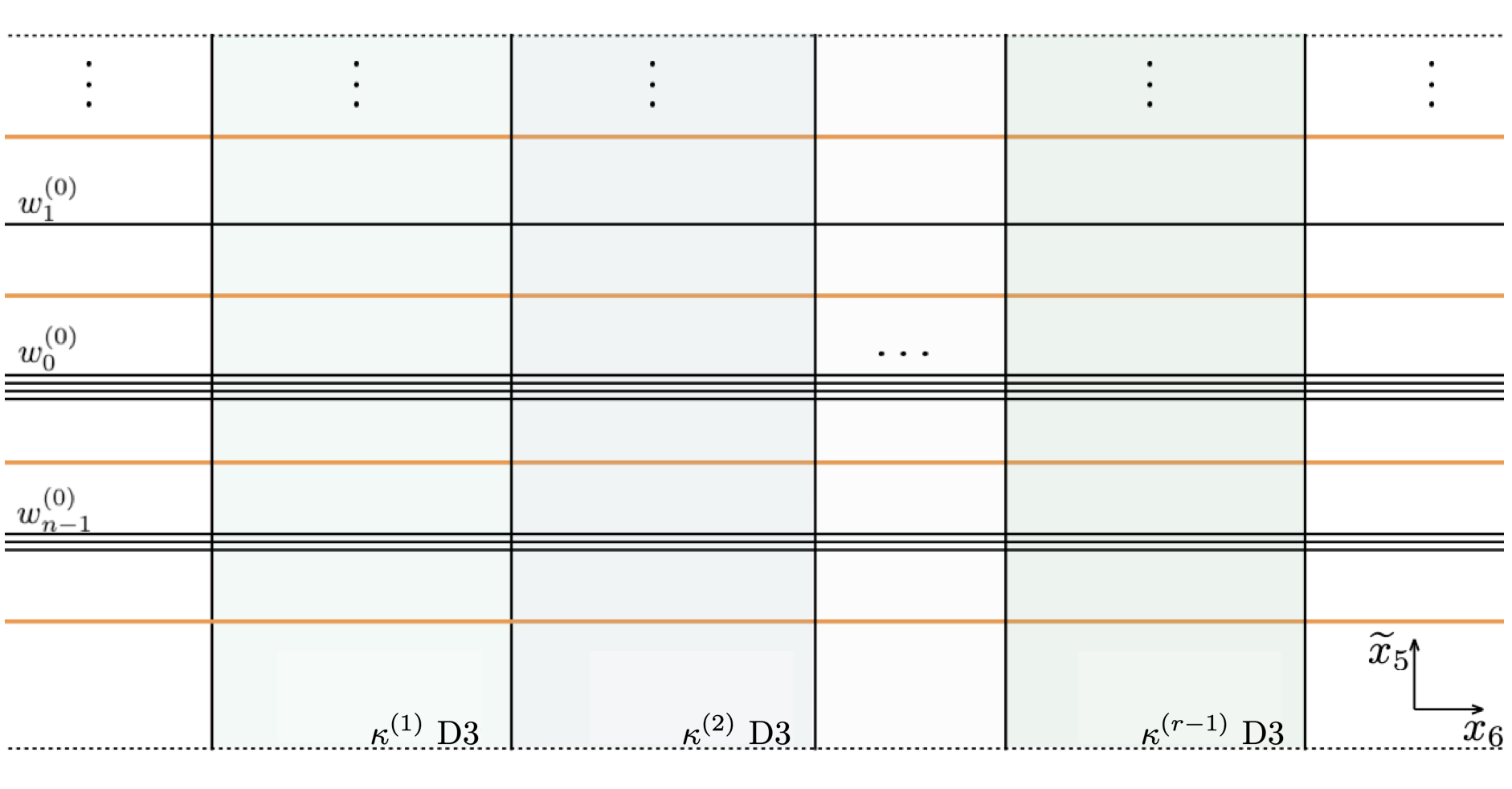}
    \caption{Brane diagram corresponding to an instanton configuration of the theory $\mathcal{T}^{6d}_{r,W}$ in the case where all monodromies are identical. The vertical direction corresponds to the dual circle parametrized by $\widetilde{x}_5$, while the horizontal direction corresponds to direction $x_6$.}
    \label{fig:simpler}
\end{figure}

Note that the integer partition $\vec{w}^{(0)}$ is always an element of the set $\mathfrak{W}_{\vec{w}^{(0)}}$, which corresponds to $\vec\delta_{\vec{w}^{(0)}} = 0$. The simplest configuration of D3 branes one can consider is to take $\vec w^{(a)} = \vec{w}^{(0)}$ for every gauge factor $\mathfrak{u}(W)^{(a)}$. Instanton configurations for this choice of monodromies are parametrized by an $(r-1)$-tuple of non-negative integers $\boldsymbol{\kappa}=(\kappa^{(1)},\dots,\kappa^{(r-1)})$ such that $\vec{v}^{(a)} = (\kappa^{(a)},\dots,\kappa^{(a)})$. The brane diagram corresponding to such a configuration is depicted in figure \ref{fig:simpler}. There are no fractional D3 branes for any of the 6d gauge algebras. On the other hand, D3 branes are allowed to terminate on vertical NS5 branes, which accounts for the possibility that different numbers of BPS strings may couple to the various tensor multiplets in the 6d theory $\mathcal{T}^{6d}_{r,W}$. This situation is akin to the one encountered in \cite{DelZotto:2023rct}, where it was observed that the M-string SCFTs do not allow for fractional BPS strings.\newline

Situations in which $\vec w^{(a)} \neq \vec{w}^{(0)}$ for some $a$ require a more nontrivial construction, in which the intersections between the D5 and NS5 branes are resolved. In order to illustrate this effect, we discuss a specific example of a model with fractional BPS strings.

\begin{figure}
    \centering
    \includegraphics[width=\textwidth]{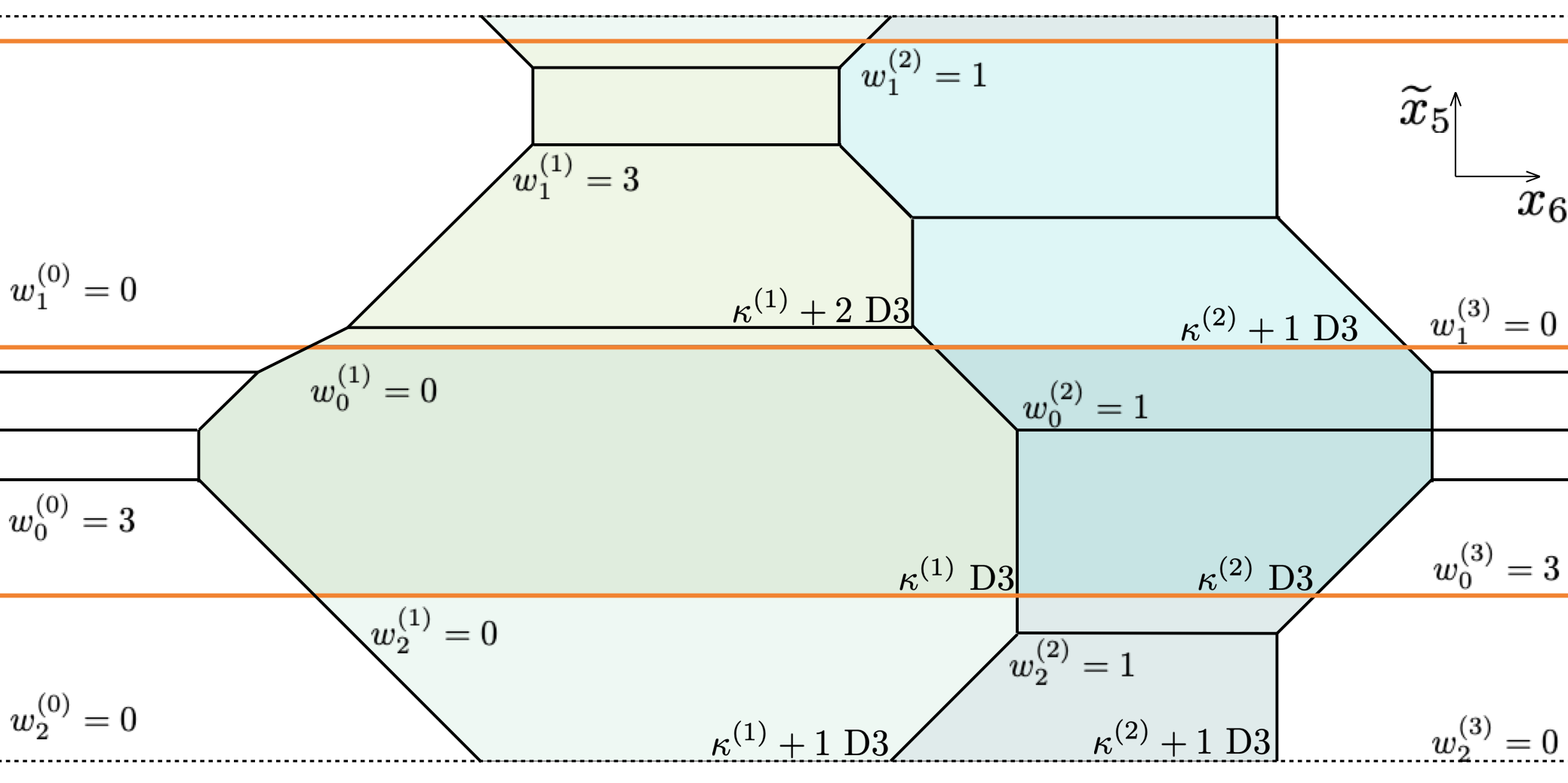}
    \caption{Type IIB brane configuration corresponding to a generic instanton solution for the theory $\mathcal{T}^{6d}_{3,3}$, with choice of monodromies $w^{(0)} = (3,0,0)$, $w^{(1)} = (0,3,0)$, and $w^{(2)} = (1,1,1)$. The different shaded regions indicate different numbers of D3 branes suspended between the horizontal NS5 branes. The vertical direction $\widetilde{x}_5$ is periodically identified along the dashed line.}
    \label{fig:beetles}
\end{figure}

\paragraph{Fractional BPS strings: a specific example.} The case of the theory $\mathcal{T}^{6d}_{3,3}$ on $\mathbf{TN}_3$ is a simple model that can be used to illustrate the complexity arising from fractional BPS string contributions. Take the monodromy for $\mathfrak{u}(W)^{(0)}$ to be given by $\vec{w}^{(0)} = (3,0,0)$. Then, one finds the following set of possible monodromies for the 6d gauge algebra factors:
\be
\mathfrak{W}_{\vec{w}^{(0)}} = \{(3,0,0),(0,3,0),(0,0,3),(1,1,1)\}.
\ee
The corresponding four solutions for $\vec\delta_{\vec w}$ are:
\bea
\vec\delta_{(3,0,0)} &=& (0,0,0);\\
\vec\delta_{(0,3,0)} &=& (2,-1,-1);\\
\vec\delta_{(0,0,3)} &=& (1,1,-2);\\
\vec\delta_{(1,1,1)} &=& (1,0,-1).
\eea
Figure \ref{fig:beetles} illustrates the allowed instanton configurations for the choice of monodromies $\vec{w}^{(1)} = (0,3,0)$ and $\vec{w}^{(2)} = (1,1,1)$. The allowed D3-brane multiplicities are given by $\vec{v}^{(1)} = (\kappa^{(1)},\kappa^{(1)}+2,\kappa^{(1)}+1)$  for $\kappa^{(1)}\geq 0$ and $\vec{v}^{(2)}=(\kappa^{(2)},\kappa^{(2)}+1,\kappa^{(2)}+1)$ for $\kappa^{(2)}\geq 0$, corresponding to instanton numbers $N^{(1)} = \kappa^{(1)}+1$ and $N^{(2)} = \kappa^{(2)}+\frac{2}{3}$. 
In section \ref{sec:anom} we will obtain a simple description of the worldsheet theory describing such generic configurations of fractional BPS strings in the ALE limit.

\section{The $T^2\times \mathbb{C}^2$ partition function revisited}
\label{sec:t2c2pf}
We now turn to the computation of partition functions. We begin in this section by computing the partition function of the theory $\mathcal{T}^{6d}_{r,W}$ on the background $T^2\times\mathbb{C}^2$. We rely largely on \cite{Hohenegger:2013ala,Haghighat:2013tka, Gadde:2015tra}, but some of the details presented here are new. In sections \ref{sec:zpertc2zn} and \ref{sec:d3z} we will discuss how the the results of this section generalize to the background $T^2\times \mathbb{C}^2/\mathbb{Z}_n$.
 The partition function, which we denote by $\mathcal{Z}_{T^2\times\mathbb{C}^2}$, can be be written in the form of a product of classical, perturbative, and instanton contributions with respect to the 6d gauge symmetry algebra:
\be
\mathcal{Z}_{T^2\times\mathbb{C}^2} = \mathcal{Z}_{T^2\times\mathbb{C}^2}^{class}\cdot \mathcal{Z}_{T^2\times\mathbb{C}^2}^{pert}\cdot \mathcal{Z}_{T^2\times\mathbb{C}^2}^{inst}.
\ee
In the following we concentrate on the perturbative and instanton pieces.

\subsection{Perturbative contributions}
\label{sec:zper}

\begin{figure}[t]
    \centering
    \includegraphics[scale=0.52]{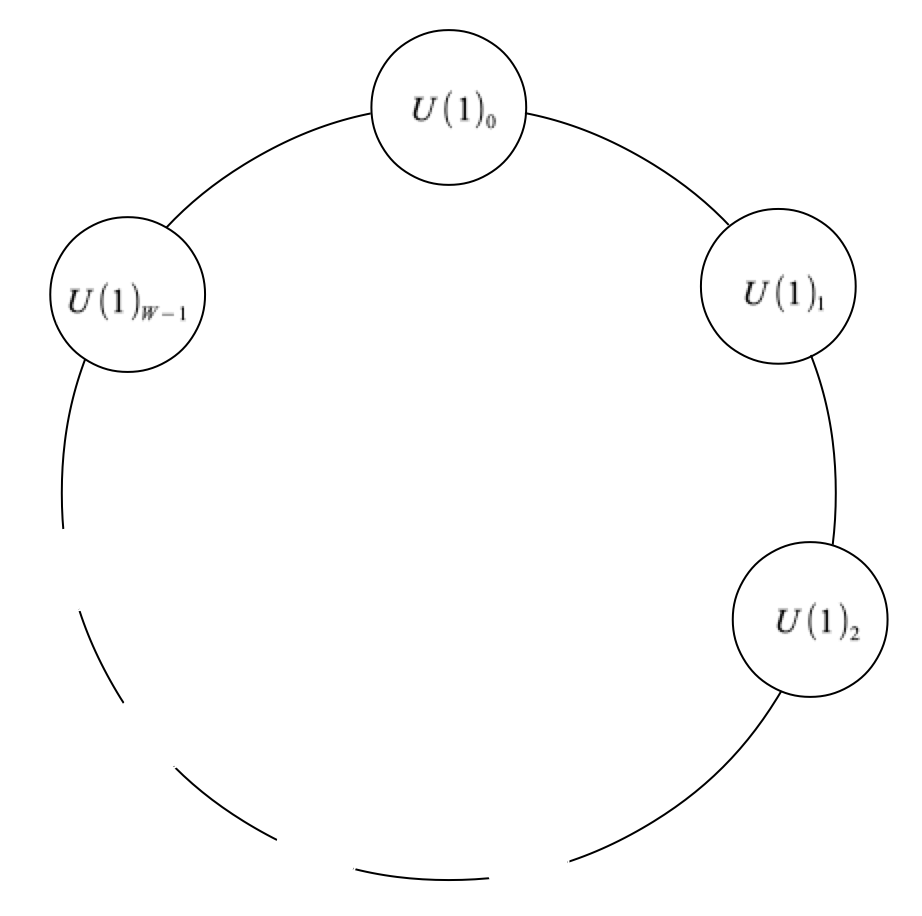}
    \caption{Five-dimensional quiver corresponding a single M5 brane probing a $\mathbb{C}^2/\mathbb{Z}_W$ singularity.}
    \label{fig:5dquiv}
\end{figure}

The perturbative piece of the partition function includes contributions from degrees of freedom localized on the M5 branes:
\be
\mathcal{Z}_{T^2\times\mathbb{C}^2}^{pert} = \prod_{a=1}^r \mathcal{Z}_{T^2\times\mathbb{C}^2}^{M5^{(a)}}.
\ee
To determine these contributions we can assume that we are far on the tensor branch, and for the moment ignore any interactions between M5 branes. The theory that describes a single M5 brane probing a transverse $\mathbf{TN}_W$ space admits a description in terms of the 5d KK theory $\mathcal{T}^{KK,(a)}$ of figure \ref{fig:5dquiv}, which is simply the abelian ($r=1$) case of the quiver of figure \ref{fig:5dKK}. We resort to this dual description since it makes it easier to detect the existence of a 2d chiral boson (which is responsible for a Heisenberg algebra factor) among the degrees of freedom contributing to the partition function.
This quiver arises by considering the dual type IIA description  in terms of a D4 brane probing $\mathbf{TN}_W$. The D4 brane splits into a collection of $W$ fractional D4 branes, each carrying $U(1)$ gauge degrees of freedom. We denote by $F_{U(1)^{(a)}_{A}}$, $A=0,\dots,W-1$, the field strength corresponding to the $A$-th gauge group, and by $\tau^{(a)}_{A}$ the gauge coupling corresponding to the $A$-th $U(1)$ gauge node. The gauge couplings are related to the complex modulus $\tau$ of $T^2$ by:
\be
\sum_{A=0}^{W-1} \tau^{(a)}_{A} = \tau,
\ee
and we also define $q = e^{2\pi i \tau}$ and $q^{(a)} = e^{2\pi i \tau^{(a)}}$. The spacetime $\mathbb{C}^2$ has isometry $SO(4)\sim SU(2)_x\times SU(2)_t$, and we denote respectively by $\epsilon_-$ and $\epsilon_+$ the chemical potentials for $SU(2)_x$ and $SU(2)_t$. We also define
\be
x=e^{2\pi i \epsilon_-},\qquad t = e^{2\pi i \epsilon_+}.
\ee
Denote by $m^{(a)}_A$ the mass of the bifundamental hypermultiplet which is charged under $U(1)^{(a)}_A\times U(1)^{(a)}_{A-1}$, and let
 \be
\mu^{(a)}_{A} = e^{2\pi i (m^{(a)}_{A}+\epsilon_+)}.
 \ee
The partition function depends on the hypermultiplet masses through the following combination:
\be
\tilde\mu^{(a)}_{A} = \mu^{(a)}_{A}\, e^{2\pi i(a^{(a)}_{A}-a^{(a)}_{A-1})} = e^{2\pi i \nu^{(a)}_A},
\label{eq:tildemu}
\ee
where $a^{(a)}_{A}$ is the vacuum expectation value of the scalar in the $A$-th vector multiplet. In appendix \ref{sec:5dc2app} we find that the contributions localized at the $a$-th M5 brane can be written (up to the classical piece) as:
\bea
\nonumber
&&
\resizebox{.4\textwidth}{!}{$\displaystyle{
\mathcal{Z}_{T^2\times\mathbb{C}^2}^{M5^{(a)}}(\epsilon_+,\epsilon_-,\underline{\tilde\mu}^{(a)},\underline{q}^{(a)})
=
q^{\frac{1}{24}}
\widehat{\chi}_{\mathcal{H}}(\tau)
}$}
\\
\nonumber
&&
\resizebox{.62\textwidth}{!}{$\displaystyle{
\times
\prod_{A=0}^{W-1}
\prod_{i,j,k=0}^\infty
\frac{
(1- q^{k}{\tilde\mu}^{(a)}_{A} t^{i-j} x^{i+j+1})
(1- q^{k+1}\frac{1}{\tilde\mu^{(a)}_{A}}t^{i-j} x^{i+j+1})}
{(1- q^{k} t^{i-j+1} x^{i+j+1})
(1- q^{k+1}t^{i-j-1} x^{i+j+1})}
}$}
\\
\nonumber
&&
\resizebox{.98\textwidth}{!}{$\displaystyle{
\times
\prod_{\substack{A,B=0\\B> A}}^{W-1}
\prod_{i,j,k=0}^\infty
\frac{
(1- q^{k} \sqrt{{\tilde\mu}^{(a)}_{A} {\tilde\mu}^{(a)}_{B}}(\prod_{L=A}^{B-1}q^{(a)}_L)t^{i-j} x^{i+j+1})
(1- q^{k+1} \frac{1}{\sqrt{{\tilde\mu}^{(a)}_{A}{\tilde\mu}^{(a)}_{B}}}(\prod_{L=A}^{B-1}\frac{1}{q^{(a)}_L})t^{i-j} x^{i+j+1})}
{(1- q^{k}\sqrt{\frac{{\tilde\mu}^{(a)}_{B}}{{\tilde\mu}^{(a)}_{A}}}(\prod_{L=A}^{B-1}q^{(a)}_L)t^{i-j+1} x^{i+j+1})
(1- q^{k+1}\sqrt{\frac{{\tilde\mu}^{(a)}_{A}}{{\tilde\mu}^{(a)}_{B}}}(\prod_{L=A}^{B-1}\frac{1}{q^{(a)}_L})t^{i-j+1} x^{i+j+1})}
}$}
\\
\nonumber
&&
\resizebox{.98\textwidth}{!}{$\displaystyle{
\times
\prod_{\substack{A,B=0\\B> A}}^{W-1}
\prod_{i,j,k=0}^\infty
\frac{
(1- q^{k}\frac{1}{\sqrt{{\tilde\mu}^{(a)}_{A} {\tilde\mu}^{(a)}_{B}}}(\prod_{L=A}^{B-1}q^{(a)}_L)t^{i-j} x^{i+j+1})
(1- q^{k+1}\sqrt{{\tilde\mu}^{(a)}_{A} {\tilde\mu}^{(a)}_{B}}(\prod_{L=A}^{B-1}\frac{1}{q^{(a)}_L})t^{i-j} x^{i+j+1})}
{(1- q^{k} \sqrt{\frac{{\tilde\mu}^{(a)}_{A}}{{\tilde\mu}^{(a)}_{B}}}(\prod_{L=A}^{B-1}q^{(a)}_L)t^{i-j-1} x^{i+j+1})
(1- q^{k+1}\sqrt{\frac{{\tilde\mu}^{(a)}_{B}}{{\tilde\mu}^{(a)}_{A}}}(\prod_{L=A}^{B-1}\frac{1}{q^{(a)}_L})t^{i-j-1} x^{i+j+1})},
}$}
\\
\label{eq:Zm5a}
\eea
where we denote respectively by $\underline{\tilde\mu}^{(a)}$ and $\underline{q}^{(a)}$ the $W$-tuples $(\tilde\mu^{(a)}_0,\dots,\tilde\mu^{(a)}_{W-1})$ and $(q^{(a)}_0,\dots,q^{(a)}_{W-1})$. The infinite product factors in the partition function are the contributions of KK towers of 5d BPS hypermultiplets and vector multiplets, which arise from 6d BPS vector, tensor, and hypermultiplets upon compactification. In the context of geometric engineering \cite{Katz:1996fh,Lawrence:1997jr,Iqbal:2003ds}, these contributions arise from BPS M2 branes wrapping holomorphic two-cycles of a Calabi-Yau threefold. On the other hand, the factor of
\be
\widehat{\chi}_{\mathcal{H}}(\tau)=\frac{1}{\eta(\tau)}= \frac{1}{q^{\frac{1}{24}}\prod_{k=1}^\infty(1-q^k)}
\ee
is the character of the Heisenberg algebra. Unlike other contributions to the partition function, \textit{it does not arise from M2 branes wrapping holomorphic cycles in a Calabi-Yau threefold}, and indeed it has  does not take the standard form of a BPS particle contribution to the topological string partition function \cite{gopakumar1998mtheory,gopakumar1998mtheory2}.

The combination
\be
\left(\prod_{A=0}^{W-1}\widetilde\mu^{(a)}_A\right)^{1/W} := e^{2\pi i\mathfrak{m}}.
\ee
was identified in \cite{Haghighat:2013tka} with the exponentiated chemical potential for an additional $U(1)_{\mathfrak{m}}$ global symmetry acting on the space transverse to the D6 branes, which is independent of the NS5 brane index $a$.

It is useful to rewrite the partition function of the 5d abelian quiver gauge theory of figure \ref{fig:5dquiv} in terms of the $\mathfrak{su}(W)^{(a)}$ Wilson lines of the dual 6d SCFT. We denote by $\{s^{(a)}_A\}_{A=0,\dots,W-1}$ the Wilson line fugacity for the $U(1)^{(a)}_A$ factors in the Cartan of $\mathfrak{su}(W)^{(a)}$ (we fix a Weyl chamber such that $s^{(a)}_{A}\leq s^{(a)}_{B}$ for $A<B$). In appendix \ref{sec:5dc2app} we find:
\bea
\nonumber
&&
\mathcal{Z}_{T^2\times\mathbb{C}^2}^{M5^{(a)}}
=
q^{\frac{1}{24}}
\widehat{\chi}_{\mathcal{H}}(\tau)\\
\nonumber
&\times&
\prod_{k=0}^{\infty}
\prod_{\substack{A,B=0\\A\leq B}}^{W-1} 
Z^{BPS,U(1)_A^{(a-1)},U(1)_B^{(a-1)}}_{S^1\times{\mathbb{C}^2}}(k\tau,1)
Z^{BPS,U(1)_A^{(a-1)},U(1)_B^{(a-1)}}_{S^1\times{\mathbb{C}^2}}((k+1)\tau,-1)\\
\nonumber
&\times&
\prod_{k=0}^{\infty}
\prod_{\substack{A,B=0\\A < B}}^{W-1}
Z^{BPS,U(1)_A^{(a)},U(1)_B^{(a)}}_{S^1\times{\mathbb{C}^2}}(k\tau,1)
Z^{BPS,U(1)_A^{(a)},U(1)_B^{(a)}}_{S^1\times{\mathbb{C}^2}}((k+1)\tau,-1)\\
\nonumber
&\times&
\prod_{k=0}^{\infty}
\prod_{(A,B)\in\mathfrak{S}^{(a)}}
Z^{BPS,U(1)_A^{(a-1)},U(1)_B^{(a)}}_{S^1\times\mathbb{C}^2}(k\tau,0)
Z^{BPS,U(1)_B^{(a)},U(1)_A^{(a-1)}}_{S^1\times\mathbb{C}^2}((k+1)\tau,0)\\
&\times&
\prod_{k=0}^{\infty}
\prod_{(A,B)\in\overline{\mathfrak{S}}^{(a)}}
Z^{BPS,U(1)_B^{(a)},U(1)_A^{(a-1)}}_{S^1\times\mathbb{C}^2}(k\tau,0)
Z^{BPS,U(1)_A^{(a-1)},U(1)_B^{(a)}}_{S^1\times\mathbb{C}^2}((k+1)\tau,0)
\label{eq:zz22full}
\eea
where we have introduced the notation
\be
Z^{BPS,U(1)_A^{(a)},U(1)_B^{(b)}}_{S^1\times\mathbb{C}^2}(m,k_t)
=
\prod_{i,j=0}^\infty (1-e^{2\pi i(m+s^{(b)}_B-s^{(a)}_A)} t^{i-j+2k_t}x^{i+j+1})^{(-1)^{2k_t}}
\label{eq:zbps}
\ee
for a particle with $SU(2)_x\times SU(2)_t$ charge $(0,k_t)$, and we denote by $\mathfrak{S}^{(a)}$ the set of index pairs $(A,B)$ such that $s^{(a-1)}_A\leq s^{(a)}_B $, and by $\overline{\mathfrak{S}}^{(a)}$ its complement among the set of all possible $(A,B)$.

\subsection{Instanton contributions}
\label{sec:t2c2inst}

\begin{figure}
\begin{center}
\includegraphics[width=\textwidth]{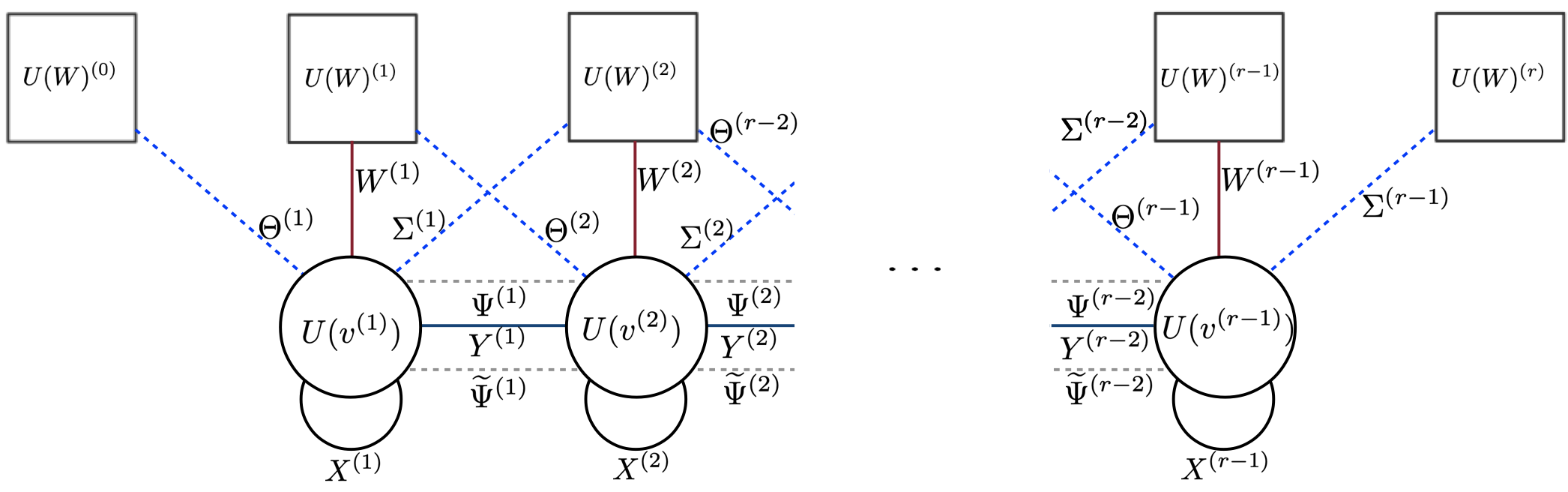}
\caption{Quiver $\mathcal{Q}_{W,\boldsymbol{{v}}}$ describing a bound state of $\boldsymbol{{v}}=({v}^{(1)},{v}^{(2)},\dots,{v}^{(r-1)})$ strings of the 6d SCFT $\mathcal{T}^{6d}_{r,W}$ on $T^2\times\mathbb{C}^2.$}
\label{fig:orbi}
\end{center}
\end{figure}

Let us now turn to the instanton part of the partition function. Consider a bound state of ${v}^{(1)}$ M2-branes suspended between the first and second M5 brane, ${v}^{(2)}$ M2-branes wrapped on $T^2$ suspended between the second and third M5 brane, and so on. Such a bound state carries instanton charges $\boldsymbol{{v}}=({v}^{(1)},\dots,{v}^{(r-1)})$ with respect to $\mathfrak{su}(W)^{(1)}\oplus\dots\oplus\mathfrak{su}(W)^{(r-1)}$. Its worldsheet degrees of freedom are described in terms of the 2d $\mathcal{N}=(0,4)$ quiver gauge theory $\mathcal{Q}_{W,\boldsymbol{{v}}}$ of figure \ref{fig:orbi} \cite{Haghighat:2013tka}.\newline

The field content of (0,4) quiver gauge theories assembles into multiplets of the $(0,4)$ supersymmetry algebra, whose components are listed in table \ref{tab:04mult} along with their charges with respect to the R-symmetry algebra $SU(2)_t\times SU(2)_R$. The 2d $\mathcal{N}=(0,4)$ field content associated to any given gauge node of $\mathcal{Q}_{W,\boldsymbol{v}}$ consists of the following fields:
\begin{itemize}
\item[-]
Vector multiplets ${V}^{(a)}$
in the adjoint representation of $U({v}^{(a)})$;
 \item[-]
Twisted hypermultiplets ${X}^{(a)}$
in the adjoint representation of $U({v}^{(a)})$ and in the doublet representation of $SU(2)_x$;
 \item[-]
Hypermultiplets ${Y}^{(a)}$
in the bifundamental representation of $U({v}^{(a)})\times U({v}^{(a+1)})$;
\item[-] $SU(2)_x$ doublets of Fermi multiplets $({\Psi}^{(a)},\widetilde{{\Psi}^{(a)}})$ in the bifundamental representation of $U({v}^{(a)})\times U({v}^{(a+1)})$;
\item[-] Hypermultiplets ${W}^{(a)}$ in the bifundamental representation of $U({v}^{(a)})\times U(W)^{(a)}$;
\item[-] Fermi multiplets ~${\Sigma}^{(a)}$~
in the bifundamental representation of $\,\, U({v}^{(a)})\,\times\, U(W)^{(a+1)}$;
\item[-] Fermi multiplets ${\Theta}^{(a)}$
in the bifundamental representation of $U(W)^{(a-1)}\times U({v}^{(a)})$.
\end{itemize}
The factor $SU(2)_t$ of the R-symmetry group acts as an isometry of the spacetime $\mathbb{C}^2$ of the 6d SCFT; other factor $SU(2)_x$ of the isometry group is realized as a flavor symmetry of the 2d quiver gauge theory. The gauge symmetry group of the D6 branes gives rise to the additional 2d flavor symmetry
\be
\frac{\prod_{a=0}^{r} U(W)^{(a)}}{U(1)^{r}}
\ee
for the theory $\mathcal{Q}_{W,\boldsymbol{{v}}}$. Let us denote by $J_R, J_t,$ and $J_x$ the Cartan generators for $SU(2)_R\times SU(2)_t\times SU(2)_x$. Also denote by $\underline{J}^{(a)}=(J^{(a)}_0,\dots J^{(a)}_{W-1})$ the Cartan generators for $U(W)^{(a)}$, keeping in mind that the overall abelian factors for different $a$ are ultimately to be identified with each other.
\begin{table}
\centering
\resizebox{\textwidth}{!}{
\begin{tabular}{c|l}
Multiplet & Field content and $SU(2)_t\times SU(2)_R$ transformation \\
\hline
Vector multiplet & Vector $A_\mu$: \textbf{(1,1)} + complex fermions $\lambda^{\dot{\alpha}A}$: \textbf{(2,2)} \\
Fermi multiplet & Complex fermion $\psi$: \textbf{(1,1)} \\
Hypermultiplet & Complex scalars $x_{\dot\alpha}$: \textbf{(1,2)} + complex fermions $\chi^{A}$: \textbf{(2,1)} \\
Twisted hypermultiplet & Complex scalars $y_A$: \textbf{(2,1)} + complex fermions $\,\xi^{\dot{\alpha}}$: \textbf{(1,2)} \\
\end{tabular}}
\caption{Field content and R-symmetry transformation of $(0,4)$ multiplets. The indices $\dot \alpha, A$ label respectively the components of the doublet of $SU(2)_r$ and  $SU(2)_R$.} 

\label{tab:04mult}
\end{table}

The excitations that contribute to the Nekrasov partition function are captured by the elliptic genus
\be
\mathbb{E}_{\boldsymbol{{v}}}(\epsilon_+,\epsilon_-,\underline{\boldsymbol{s}},\tau)
=
\Tr (-1)^F q^{H_L}  \overline{q}^{H_R}e^{2\pi i \epsilon_+(J_R+J_t)}e^{2\pi i \epsilon_- J_x} e^{2\pi i \sum_{a=0}^{r}\underline{s}^{(a)}\cdot \underline{J}^{(a)}}.
\ee
From the perspective of the 2d theory, the Wilson lines $s^{(a)}_A$ are realized as chemical potentials along $T^2$ for the $U(W)^{(a)}$ flavor symmetry. Due to the Stuckelberg mechanism which identifies the abelian factors, the chemical potentials are not all independent. Rather, they obey the following relation \cite{Haghighat:2013tka}:
\be
\sum_{A=0}^{W-1} s^{(a+1)}_A = \sum_{A=0}^{W-1} s^{(a)}_A + W\, \mathfrak{m}. 
\label{eq:stucondition}
\ee
The elliptic genus can be computed by supersymmetric localization \cite{Benini:2013nda,Benini:2013xpa} and is given by the following integral over the 2d gauge group holonomies $\boldsymbol{z}$:
\be
\mathbb{E}_{\boldsymbol{{v}}}(\epsilon_+,\epsilon_-,\underline{\boldsymbol{s}},\tau)
=
\frac{1}{\prod_{a=1}^{r-1}{v}^{(a)}!}\int \bigg(\prod_{a=1}^{r-1}\prod_{k=1}^{{v}^{(a)}}d z^{(a)}\bigg)Z^{1-loop}_{\mathcal{Q}_{\boldsymbol{{v}}}}(\vec{\boldsymbol{z}},\epsilon_+,\epsilon_-,\underline{\boldsymbol{s}},\tau)
\ee
The one-loop partition function in the integrand is given by
\bea
\nonumber
&&
\hskip-.3in
Z^{1-loop}_{\mathcal{Q}_{\boldsymbol{{v}}}}(\vec{\boldsymbol{z}},\epsilon_+,\epsilon_-,\underline{\boldsymbol{s}},\tau)
=
\left[\left(\frac{\eta(\tau)^3\,\theta_1(2\epsilon_+,\tau)}{\theta_1(\epsilon_++\epsilon_-,\tau)\theta_1(-\epsilon_++\epsilon_-,\tau)}\right)^{\sum_{a=1}^{r-1}{v}^{(a)}}\right.
\\
\nonumber
&\times&
\left(\prod_{a=1}^{r-1}\prod_{\substack{k,l=1 \\ k\neq l}}^{{v}^{(a)}}\frac{\theta_1(z^{(a)}_{k}-z^{(a)}_{l},\tau)\theta_1(2\epsilon_+ + z^{(a)}_{k}-z^{(a)}_{l},\tau)}{\theta_1(\epsilon_++\epsilon_-+ z^{(a)}_{k}-z^{(a)}_{l},\tau)\theta_1(-\epsilon_++\epsilon_-+ z^{(a)}_{k}-z^{(a)}_{l},\tau)}\right)\\
\nonumber
&\times&
\left(\prod_{a=1}^{r-2}\prod_{k=1}^{{v}^{(a)}}\prod_{l=1}^{{v}^{(a+1)}}\frac{\theta_1(\epsilon_-+ z^{(a)}_{k}-z^{(a+1)}_{l},\tau)\theta_1(-\epsilon_-+ z^{(a)}_{k}-z^{(a+1)}_{l},\tau)}{\theta_1(\epsilon_++ z^{(a)}_{k}-z^{(a+1)}_{l},\tau)\theta_1(-\epsilon_++z^{(a)}_{k}- z^{(a+1)}_{l},\tau)}\right)\\
&\times&
\left.\left(\prod_{a=1}^{r-1}\prod_{k=1}^{{v}^{(a)}}\prod_{A=0}^{W-1}\frac{\theta_1(s^{(a-1)}_{A}-z^{(a)}_{k},\tau)\theta_1(z^{(a)}_{k}-s^{(a+1)}_{A},\tau)}{\theta_1(\epsilon_++ z^{(a)}_{k}-s^{(a)}_{A},\tau)\theta_1(-\epsilon_++z^{(a)}_{k}-s^{(a)}_{A},\tau)}\right)\right],
\eea
where
\be
\theta_1(z,\tau) = i\, q^{\frac{1}{8}}e^{-\pi i z}\prod_{k=0}^\infty (1-q^{k+1})(1-e^{2\pi i z}q^{k})(1-e^{-2\pi i z}q^{k+1}),
\ee
and the notation $\prod \!\phantom{\vert}^{'}$ indicates that all occurrences of $\theta_1(0,\tau)$ in the product must be replaced by a factor of $\eta(\tau)$. 
The elliptic genus can be expressed as a combinatorial sum over the appropriate set of residues of the one-loop partition function of $\mathcal{Q}_{\boldsymbol{{v}}}$:
\bea
\mathbb{E}_{\boldsymbol{{v}}}(\epsilon_+,\epsilon_-,\underline{\boldsymbol{s}},\tau)
&=&
\sum_{\substack{\underline{Y}^{(1)}\in \mathcal{Y}_{{v}^{(1)}}\\\dots\\\underline{Y}^{(r-1)}\in \mathcal{Y}_{{v}^{(r-1)}}}}
\text{Res}_{\{z^{(a)}_k=z^{*}_{\underline{Y}^{(a)},k}\}_{k=1}^{{v}^{(a)}}}Z^{1-loop}_{\mathcal{Q}_{\boldsymbol{{v}}}}(\epsilon_+,\epsilon_-,\underline{\boldsymbol{s}},\tau).
\eea
Here $\mathcal{Y}_{{v}^{(a)}}$ denotes the set of $W$-tuples of Young diagrams $\underline{Y}^{(a)}=(Y^{(a)}_0,\dots,Y^{(a)}_{W-1})$ of total size  $\sum_{A=0}^{W-1}\vert Y^{(a)}_A\vert = {v}^{(a)}$, whose boxes we denote by $\{\mathfrak{b}^{(a)}_k\}_{k=1}^{{v}^{(a)}}$. We let $A(\mathfrak{b}^{(a)}_k)\in \{0,\dots,W-1\}$ denote the label of the Young diagram to which box $\mathfrak{b}^{(a)}_k$ belongs. The residues are evaluated at the values
\be
z^*_{\underline{Y}^{(a)},k} = s^{(a)}_{A(\mathfrak{b}^{(a)}_k)} + \epsilon_+ + (\epsilon_+ + \epsilon_-)\mathfrak{b}^{(a),1}_{k}+(\epsilon_+-\epsilon_-)\mathfrak{b}^{(a),2}_{k},
\ee
of the $U({v}^{(a)})$ holonomies, where $(\mathfrak{b}^{(a),1}_k,\mathfrak{b}^{(a),2}_k)$ denotes the coordinates of box $\mathfrak{b}^{(a)}_k$.  We refer the reader to \cite{Gadde:2015tra} for further details.\\

The instanton partition function is then obtained by summing over the contributions of all possible bound states of BPS strings:
\be
\mathcal{Z}_{T^2\times\mathbb{C}^2}^{inst} = \sum_{\boldsymbol{{v}}\in\mathbb{Z}^{r-1}_{\geq 0}}e^{-\sum_{a=1}^{r-1}{v}^{(a)} \varphi^{(a)}} \mathbb{E}_{\boldsymbol{{v}}}(\epsilon_+,\epsilon_-,\underline{\boldsymbol{s}},\tau),
\ee
where $\varphi^{(a)}$ denotes the separation between the $a$-th and the $a+1$-st M5 brane as in equation \eqref{eq:stringtension}, which coincides with the tension of a BPS string carrying instanton charge $1$ with respect to $\mathfrak{su}(W)^{(a)}$ and 0 with respect to the remaining gauge group factors.

\section{The $T^2\times \mathbb{C}^2/\mathbb{Z}_n$ Partition Function}
\label{sec:zzc2zn}
We now turn to the computation of the partition function on $T^2\times \mathbb{C}^2/\mathbb{Z}_n$. Generalizing the case of M-strings \cite{DelZotto:2023rct}, the naive expectation that the partition function takes a factorized form with separate contributions from NS5 and D3 branes\footnote{ \  The classical free energy, on general grounds, appears as an overall prefactor that scales as the (equivariant) volume of $\mathbb{C}^2/\mathbb{Z}_n$ which is given by $\frac{1}{n\epsilon_1\epsilon_2}$ \cite{Moore:1997dj}. Since the latter is the logarithm of the partition function, one expects the simple relation
\be
\mathcal{Z}^{class}_{T^2\times\mathbb{C}^2/\mathbb{Z}_n}=  \left(\mathcal{Z}^{class}_{T^2\times\mathbb{C}^2}\right)^{\frac{1}{n}}\,
\ee
which is a strightfoward extension to 6d of the perturbative part of Nekrasov master formula \cite{Nekrasov:2003vi}(see also eg. the main theorem on p.4 of  \cite{Gasparim:2009sns}).
},
\be
\mathcal{Z}^{\mathcal{T}^{6d}_{r,W}}_{T^2\times\mathbb{C}^2/\mathbb{Z}_n} \overset{?}{=} \mathcal{Z}^{class}_{T^2\times\mathbb{C}^2/\mathbb{Z}_n} (\prod_{a=1}^r \mathcal{Z}^{NS5^{(a)}}_{T^2\times\mathbb{C}^2/\mathbb{Z}_n}) \mathcal{Z}_{D3},
\label{eq:zfac}
\ee
is \emph{not} realized due to the existence of an anomaly inflow that couples the degrees of freedom arising from intersecting NS5s with those along the suspended D3s. In section \ref{sec:zpertc2zn} we discuss the contributions to the partition function arising from the degrees of freedom supported on individual NS5 branes. We find that these include KK towers of BPS particles as well as an $\mathfrak{su}(n)_1$ current algebra living on $T^2$. In section \ref{sec:d3z} we then turn to the discussion of the degrees of freedom arising from D3 branes, which due to anomaly inflow couple nontrivially to the $\mathfrak{su}(n)_1$ current algebras supported on NS5 branes. Finally in section \ref{sec:zfullc2zn} we comment on some salient properties of the partition function of the theory $\mathcal{T}^{6d}_{r,W}$ which is obtained by combining the perturbative and instanton contributions.

\subsection{Perturbative Contributions}
\label{sec:zpertc2zn}

Let us consider for the moment the theory of a single M5 (equivalently, NS5) brane $M5^{(a)}$. The latter corresponds to a domain wall for the 7d $U(W)$ theory, as we discussed in section \ref{sec:7dT}. Recall the structure of its $T^2\times\mathbb{C}^2$ partition function from equation \eqref{eq:zz22full}. This can be understood as a product of contributions from KK towers of 5d BPS particles, as well as the contribution from a free 2d chiral noncompact boson on $T^2$. When one takes the spacetime to be $T^2\times \mathbb{C}^2/\mathbb{Z}_n$, due to the nontrivial topology on the asymptotic boundary of $\mathbf{ALE}_n$ there exist topologically inequivalent field configurations that give rise to superselection sectors. The partition function depends on a choice of discrete data both for the $\mathfrak{u}(W)^{(a-1)}\times \mathfrak{u}(W)^{(a)}$ symmetry (which is a global symmetry in the approximation where we take the $a$-th NS5 brane to be infinitely far away from the other $r-1$ NS5 branes), and for the tensor field supported on the NS5 brane. Recall from section \ref{sec:tdua} that the topological data for $\mathfrak{u}(W)^{(a-1)}\times \mathfrak{u}(W)^{(a)}$ consists of a fixed choice of flux $\vec{u}^{(a-1)}=\vec{u}^{(a)}=\vec{w}^{(0)}$ and a choice of monodromies $w^{(a-1)}$ and $w^{(a)}$ chosen from the set $\mathfrak{W}_{\vec{w}^{(0)}}$ defined around equation \eqref{eq:Wset}. On the other hand,  as in section \ref{sec:data}, the topological data for the tensor field involves a fixed choice of monodromy $\omega^{KK,(a)}\in\mathbb{Z}_n$. Within this superselection sector, one must sum over inequivalent choices of flux arising from the the two-form field, which are labeled by a tuple of integers $\vec{u}^{KK,(a)}\in\mathbb{Z}^{n-1}$ subject to the condition
\be
\sum_{j=1}^n u^{KK,(a)}_j = \omega^{KK,(a)}\text{ mod }n.
\ee

In fact, upon closer inspection we find that the $T^2\times\mathbb{C}^2/\mathbb{Z}_n$ partition function takes a particularly simple form. Specifically, note that the tensor field has couplings to BPS strings but not to the BPS particles contributing to the partition function of a single M5 brane. As a consequence, the contribution from the BPS particles does not depend on the topological data for the two-form fields and in particular is the same in all two-form flux sectors. Schematically the domain wall partition function takes the following form:
\be
\mathcal{Z}^{NS5^{(a)},\vec{w}^{(a-1)},\vec{w}^{(a)},\omega^{KK,(a)}}_{T^2\times\mathbb{C}^2/\mathbb{Z}_n}
=
 \mathcal{Z}^{(a),\text{BPS particles},\vec{w}^{(a-1)},\vec{w}^{(a)}}_{T^2\times\mathbb{C}^2/\mathbb{Z}_n}
\sum_{\vec{u}^{KK,(a)}} \mathcal{Z}^{\text{chiral boson},\omega^{KK,(a)}, \vec{u}^{KK,(a)}}_{T^2\times \mathbb{C}^2/\mathbb{Z}_n}.
\ee
We defer the discussion of the dependence of the chiral boson on the 6d gauge algebra data to section \ref{sec:anom}. We proceed to analyze the two contributions separately.
\paragraph{BPS particles.} In order to write down the contribution from the BPS particles explicitly, it is convenient to interpret the $T^2\times\mathbb{C}^2/\mathbb{Z}_n$ background by compactifying along $S^1$ down to $S^1\times\mathbb{C}^2/\mathbb{Z}_n$, so that each 6d BPS particle gives rise to a KK tower of BPS particles in 5d on $S^1\times\mathbb{C}^2/\mathbb{Z}_n$. Because BPS particles are mutually non-interacting, we can compute their contributions to the partition function one by one. Specifically, we take the BPS contributions to the $S^1\times\mathbb{C}^2$ partition function, which appear in equation \eqref{eq:zz22full}, and apply the $\mathbb{Z}_n$ orbifold projection on each 5d BPS particle individually. We find:
\bea
\nonumber
&&
\mathcal{Z}^{(a),\text{BPS particles},\vec{w}^{(a-1)},\vec{w}^{(a)}}_{T^2\times\mathbb{C}^2/\mathbb{Z}_n}(\epsilon_+,\epsilon_-,\tau,\underline{s}^{(a-1)},\underline{s}^{(a)})
=\\
\nonumber
&\times&
\prod_{k=0}^{\infty}
\prod_{\substack{A,B=0\\A\leq B}}^{W-1} 
Z^{BPS,U(1)_A^{(a-1)},U(1)_B^{(a-1)}}_{S^1\times{\mathbb{C}^2/\mathbb{Z}_n}}(k\tau,1)
Z^{BPS,U(1)_A^{(a-1)},U(1)_B^{(a-1)}}_{S^1\times{\mathbb{C}^2/\mathbb{Z}_n}}((k+1)\tau,-1)\\
\nonumber
&\times&
\prod_{k=0}^{\infty}
\prod_{\substack{A,B=0\\A < B}}^{W-1}
Z^{BPS,U(1)_A^{(a)},U(1)_B^{(a)}}_{S^1\times{\mathbb{C}^2/\mathbb{Z}_n}}(k\tau,1)
Z^{BPS,U(1)_A^{(a)},U(1)_B^{(a)}}_{S^1\times{\mathbb{C}^2/\mathbb{Z}_n}}((k+1)\tau,-1)\\
\nonumber
&\times&
\prod_{k=0}^{\infty}
\prod_{(A,B)\in\mathfrak{S}^{(a)}}
Z^{BPS,U(1)_A^{(a-1)},U(1)_B^{(a)}}_{S^1\times\mathbb{C}^2/\mathbb{Z}_n}(k\tau,0)
\prod_{(A,B)\in\overline{\mathfrak{S}}^{(a)}}
Z^{BPS,U(1)_B^{(a)},U(1)_A^{(a-1)}}_{S^1\times\mathbb{C}^2/\mathbb{Z}_n}((k+1)\tau,0),\\
\label{eq:bpsp}
\eea
where the contributions of individual BPS particles on $\mathbb{C}^2/\mathbb{Z}_n$ are derived in appendix \ref{sec:bps} and are given in terms of equation \eqref{eq:zbpsproj}, which we report here for convenience:
\be
Z^{BPS,U(1)_A^{(a)},U(1)_B^{(b)}}_{S^1\times{\mathbb{C}^2/\mathbb{Z}_n}}(m,k_t)
=
\hspace{-.4in}
\prod_{\substack{i,j=0\\i+j+1=\omega_{A}^{(a)}-\omega_{B}^{(b)}\text{ mod }n}}^\infty
\hspace{-.4in}
(1-e^{2\pi i(m+s^{(b)}_B-s^{(a)}_A)} t^{i-j+2k_t}x^{i+j+1})^{(-1)^{2k_t}}.
\label{eq:zbpspro}
\ee
In this expression, $\omega^{(a)}_A$ labels the monodromy associated to the $A$-th $U(1)$ factor in the Cartan subgroup of $U(W)^{(a)}$. In appendix \ref{sec:bps} we also derive an alternative  expression for contribution of an individual BPS particle as a product over fixed points of the equivariant $SU(2)_x\times SU(2)_t$ action on $\mathbb{C}^2$ which arises from localization on on the resolved space $\widetilde{\mathbb{C}^2/\mathbb{Z}_n}$ and is nontrivially equivalent to equation \eqref{eq:zbpspro}. \newline

\paragraph{Chiral boson and 6d/7d correspondence.} Let us now turn to the contribution from the chiral boson. We will assume for the moment that the chiral boson does not couple to the 6d gauge algebra, so we can ignore the presence of D6 branes and exploit known results for the the partition function of the abelian 6d $\mathcal{N}= (2,0)$ theory \cite{Vafa:1994tf,Bruzzo:2013daa}. In section \ref{sec:anom} we will modify this picture to account for the presence of the D6 branes. In passing from ${T}^2\times \mathbb{C}^2$ to ${T}^2\times \mathbb{C}^2/\mathbb{Z}_n$, the Heisenberg algebra $\mathcal{H}$ arising from an individual M5 brane gets replaced by a copy of the $ \mathcal{H}\times\mathfrak{su}(n)_1$ chiral algebra. We now discuss the way in which the current algebra encodes the topological data associated to the two-form field $B^{(a)}$ in the background $T^2\times\mathbb{C}^2/\mathbb{Z}_n$. For the moment, we take the spacetime to be $T^2\times\mathbf{TN}_n$ where the discussion of \cite{Dijkgraaf:2007sw} applies. 
In M-theory we are allowed to turn the following three-form field background:
\be
C_{3} = \sigma \wedge d z + \bar\sigma \wedge d \bar z,
\label{eq:c3}
\ee
where
\be
\sigma= \sum_{j=1}^{n} \sigma_j\alpha_{j}
\ee
and $\alpha_j$ are the $n$ normalizable two-forms on $\mathbf{TN}_n$. 

The field strength $H^{(a)}$ of the two-form field $B^{(a)}$ associated to the $a$-th M5 brane can be expanded in the following way:
\be
H^{(a)} = F^{(a)}_{+}\wedge d\bar{z} +F^{(a)}_{-}\wedge dz,
\label{eq:6dexp}
\ee
where $(z,\bar z)$ are complex coordinates on the $T^2$ and $F^{(a)}_+$, $F^{(a)}_{-}$ denote respectively the selfdual and anti-selfdual components of the field strength of the 4d abelian gauge fields $U(1)^{KK,(a)}$ that arise upon by compactifying the $\mathcal{N}=(2,0)$ 6d theory to 4d $\mathcal{N}=4$ SYM on the Coulomb branch. In the partition function we restrict to self-dual field configurations for which $F_-=0$. The following coupling on the M5-brane worldvolume,
\be
\int_{T^2\times \mathbf{TN}_n} H^{(a)}\wedge C_{3},
\ee
descends to a coupling in 4d $\mathcal{N}=4$ SYM of the form
\be
\int_{\mathbf{TN}_n} F_+^{(a)}\wedge \sigma
\ee
which shows that the $\sigma_j$ serve as chemical potentials for the magnetic flux. Alternatively, one may decompose the three-form background field \eqref{eq:c3} as
\be\label{eq:AZZ}
C_3 = \sum_{j=1}^{n} A^{U(n)}_j \wedge \alpha_j,\qquad A^{U(n)}_j = \sigma_j dz +\bar\sigma_j d\bar{z}
\ee
where the $A^{U(n)}_j$ can be interpreted as the gauge fields localized on the $n$ D6 branes that arise upon compactifying M-theory on the $\mathbf{TN}_n$ Taub-NUT circle. Notice that this gives a third duality frame in addition to the IIA and the IIB frames we discussed in Section \ref{sec:IIBd}, which is the root of the 6d/7d correspondence we discussed in the introduction, generalizing \cite{Nekrasov:2014nea,DelZotto:2021gzy}. In this picture, the $\sigma_j$ give rise to Wilson lines for the 7d $U(n)$ gauge symmetry, and the M5 is dualized to a D4 brane, which intersects the D6 along the $T^2$ directions. The intersections between D4 and D6 branes support a system of chiral fermions that couples to the $U(n)$ symmetry. The latter gives rise to a $\mathfrak{u}(n)_1$ current algebra on the 2d worldsheet. Moreover, from the perspective of these degrees of freedom, the $\sigma_j$ are realized as chemical potentials associated to the Cartan of the current algebra, whose generators we denote by $\vec{J}_{\mathfrak{su}(n)}$. In order to pass to the ALE case, we switch off the coefficient of the two-form $\omega = \sum_j\omega_j $, which becomes non-normalizable in the ALE limit. This corresponds to imposing the tracelessness of the chemical potentials:
\be
\sum_{j=1}^n \sigma_j = 0.
\ee
It is useful to expand the chemical potential in terms of the cohomology basis $c_1(\mathcal{R}_1),\dots,c_1(\mathcal{R}_{n-1})$ dual to the basis $\Sigma_1,\dots,\Sigma_{n-1}$ of $H_2(\widetilde{\mathbb{C}_2/\mathbb{Z}_n})$:
\be
\sigma = \sum_{j=1}^{n-1}\xi_jc_1(\mathcal{R}_j),\qquad \xi_j=\sigma_j-\sigma_{j+1}.
\ee
Writing $F_+^{(a)} = \sum_{j=1}^{n-1} u^{KK,(a)}_j c_1(\mathcal{R}_j)$ and employing the fact that
\be
\int_{\widetilde{\mathbb{C}_2/\mathbb{Z}_n}} c_1(\mathcal{R}_j)\wedge c_1(\mathcal{R}_k)=-(C^{A_{n-1}})^{-1}_{jk},
\ee
so that
\be
\int_{\mathbb{C}^2/\mathbb{Z}_n} F_+^{(a)}\wedge \sigma = \vec{\xi}\cdot (C^{A_{n-1}})^{-1}\cdot\vec{u}^{KK,(a)},
\ee
one finds that the partition function of the chiral boson on $\mathbb{C}^2/\mathbb{Z}_n$ is given by:
\be
\sum_{\vec{u}^{KK,(a)}} \mathcal{Z}^{\text{chiral boson},\omega^{KK,(a)}, \vec{u}^{KK,(a)}}_{T^2\times \mathbb{C}^2/\mathbb{Z}_n}
=
q^{\frac{n}{24}}
\widehat{\chi}_{\mathcal{H}}(\tau)
\widehat{\chi}_{\omega^{KK,(a)}}^{\mathfrak{su}(n)_1}(\vec{\xi},\tau),
\label{eq:chibos}
\ee
where
\bea
\nonumber
\widehat{\chi}^{\mathfrak{su}(n)_1}_{\omega^{KK}}(\vec{\xi},\tau)
&=&
\Tr_{V_{\lambda^{\omega^{KK}}}}\left(q^{H_L}e^{2\pi i \vec{\xi}\cdot (C^{A_{n-1}})^{-1}\cdot\vec{J}^{\mathfrak{su}(n)}}\right)
\\
\nonumber
&=&
\eta(\tau)^{-n+1}
\hspace{-.4in}
\sum_{\substack{\vec{u}^{KK}\in \mathbb{Z}^{n-1}\\\sum_{j=1}^{n-1}ju_j =\omega^{KK}\mod n}}
\hspace{-.4in}
q^{\frac{1}{2}\vec{u}^{KK} \cdot (C^{A_{n-1}})^{-1}\cdot \vec{u}^{KK}}
e^{2\pi i \vec\xi\cdot(C^{A_{n-1}})^{-1}\cdot\vec{u}^{KK}}\\
\label{eq:sunchar}
\eea
is the character of the $\mathfrak{su}(n)$ level-1 integrable highest weight module $V_{\lambda^{\omega_{KK}}}$ whose Dynkin label has entries $\lambda^{\omega^{KK}}_j = \delta_{j,\omega^{KK}}$ for $j=1,\dots,n-1$. Here, $H_{\omega^{KK,(a)}}$ denotes the Hilbert space of states in the integrable highest weight representation associated to $\omega^{KK,(a)}$, and $\vec{J}^{\mathfrak{su}(n)}$ denote currents for the Cartan of $\mathfrak{su}(n)$. Here, $(C^{\widehat{A}_{n-1}})^{-1}$ is the inverse of the $A_{n-1}$ Cartan matrix, which has entries
 \be
(C^{\widehat{A}_{n-1}})^{-1}_{i j} = \min(i,j) - \frac{i j}{n},\qquad i,j=1,\dots, n-1.
 \ee
\newline
Note that from the perspective of the chiral boson the dependence on the three-form field parameters $\vec{\xi}$ is encoded in the existence of a coupling
\be
\int_{\Sigma} \vec{\mathcal{J}}^{\mathfrak{su}(n),(a)}\cdot\vec{\xi}
\label{eq:c3cplg}
\ee
of the theory on the worldsheet $\Sigma$.
\newline
\paragraph{Single NS5-brane contribution.} Combining terms arising from the chiral boson and BPS particles, we find the following contribution to the partition function from the $a$-th NS5 brane:
\be
\mathcal{Z}^{NS5^{(a)},\vec{w}^{(a-1)},\vec{w}^{(a)},\omega^{KK,(a)}}_{T^2\times\mathbb{C}^2/\mathbb{Z}_n}
=
q^{\frac{n}{24}}
\widehat{\chi}_{\mathcal{H}}(\tau)
\widehat{\chi}_{\omega^{KK,(a)}}^{\mathfrak{su}(n)_1}(\vec{\xi},\tau)
\mathcal{Z}^{(a),\text{BPS particles},\vec{w}^{(a-1)},\vec{w}^{(a)}}_{T^2\times\mathbb{C}^2/\mathbb{Z}_n},
\label{eq:zns5full}
\ee
with $\mathcal{Z}^{(a),\text{BPS particles},\vec{w}^{(a-1)},\vec{w}^{(a)}}_{T^2\times\mathbb{C}^2/\mathbb{Z}_n}$ given by equation \eqref{eq:bpsp}. We stress again that for the moment we have neglected effects on the chiral boson arising from the presence of the $W$ D6 branes, and likewise we have not yet taken into account effects due to the presence of D2 branes. We will incorporate these effects in section \ref{sec:anom}. It would be nice to have a purely gauge-theoretic derivation of equation \eqref{eq:zns5full}. In appendix \ref{sec:5dpf}, by restricting to the case $\vec{w}^{(0)} = \vec{w}^{(a-1)} = \vec{w}^{(a)} = 0$, we are able to derive this expression from a gauge theory perspective by exploiting the dual 5d quiver gauge theory of figure \ref{fig:5dquiv}. The restriction to trivial $U(W)$ monodromies is required in order to be able to compare with the existing results in the gauge theory literature. It would be very interesting to extend the formalism of \cite{Bruzzo:2014jza} to encompass the more general choices of monodromy that are relevant to the 6d ALE partition function of $\mathcal{T}^{6d}_{r,W}$.

\subsection{The $T^2\times \mathbb{C}^2/\mathbb{Z}_n$ Partition Function: Instanton Contributions}
\label{sec:d3z}
We now turn to the discussion of the BPS degrees of freedom of the 6d SCFT $\mathcal{T}^{6d}_{r,W}$ that carry instanton charge with respect to the 6d gauge algebra. These objects are BPS strings which arise in the duality frame of section \ref{sec:tdua} from bound states of D3 branes supported on plaquettes bounded by NS5 branes. In section \ref{sec:d3} we derive a 2d quiver gauge theory describing the degrees of freedom of the string that arise from the D3 branes. In section \ref{sec:anom} we study the gauge anomalies of the quiver obtained from the D3 branes, and find that cancelation of the abelian anomalies requires gauging the $\mathfrak{su}(n)_1$ current algebras and coupling them to the gauge group on the worldsheet of the instanton strings. In section \ref{sec:quivprop} we determine the elliptic genus of the quiver gauge theories $ \QQ $ thus obtained, which describe the instanton degrees of freedom of theory $\mathcal{T}^{6d}_{r,W}$.

\subsubsection{Anomalous quiver from D3 branes}
\label{sec:d3}
In this section we consider the contributions to the partition function that arise from D3 brane tilings like the one in figure \ref{fig:beetles}. The degrees of freedom that arise from the worldvolume of D3 branes are captured by a 2d $(0,4)$ quiver gauge theory, which we will denote by $ \QQA $ to stress that, by itself, it suffers from gauge anomalies and as such is ill-defined. In section \ref{sec:anom} we will see how anomaly inflow which couples the worldsheet degrees of freedom to the chiral fermions provides a resolution of this issue.\\

For simplicity, we begin by considering the case in which the monodromies for $\mathfrak{u}(W)^{(0)},\mathfrak{u}(W)^{(1)},\dots,\mathfrak{u}(W)^{(r)}$ are all identical:
\be
\vec{w}^{(0)}=\vec{w}^{(1)}=\dots=\vec{w}^{(r)}.
\ee
This corresponds to the configurations depicted in figure \ref{fig:simpler}. This setup belongs to a class of brane arrangements that were considered in \cite{Hanany:2018hlz}; the analysis performed there will lead us to a worldsheet description of the strings.

The D3 branes are supported on plaquettes that have boundaries on the vertical and horizontal NS5 branes. To the plaquette which is delimited by the $a$-th and $(a+1)$-st vertical brane and by the $j$-th and $(j+1)$-st horizontal brane corresponds a gauge group $U(v^{(a)}_j)$ in a 2d $\mathcal{N}=(0,4)$ quiver gauge theory. Due to the condition \eqref{eq:vsdeltas}, the gauge group is the same for each of the $n$ plaquettes corresponding to a given $a$:
\be
v^{(a)}_0 = \dots = v^{(a)}_{n-1} := \kappa^{(a)}\geq 0.
\ee
We impose $\kappa^{(0)}=\kappa^{(r)}=0$, corresponding to the fact that we do not allow for D3 branes which are infinitely extended along $x_6$.

Using the methods of \cite{Hanany:2018hlz} it is straightforward to determine the multiplet content of the theory $ \QQA$. From D3-D3 strings one obtains:
\begin{itemize}
\item[-]
 $(0,4)$ vector multiplets $V^{(a)}_j$;
 \item[-]
Twisted hypermultiplets $X^{(a)}_{j}$ in the bifundamental representation of $U(v^{(a)}_{j})\times U(v^{(a)}_{j+1})$;
 \item[-]
Hypermultiplets $Y^{(a)}_{j}$ in the bifundamental representation of $U(v^{(a)}_{j})\times U(v^{(a+1)}_{j})$
\item[-] Pairs of Fermi multiplets $\Psi^{(a)}_{j},\widetilde\Psi^{(a)}_{j},$ respectively in the bifundamental representation of $U(v^{(a)}_{j})\times U(v^{(a+1)}_{j + 1})$ and of $U(v^{(a)}_{j})\times U(v^{(a+1)}_{j - 1})$.
\end{itemize}
From D3-D5 strings one obtains:
\begin{itemize}
\item[-] Hypermultiplets $W^{(a)}_j$ in the bifundamental of $U(v^{(a)}_j)\times U(w^{(a)}_j)$;
\item[-] Fermi multiplets $\Sigma^{(a)}_j$ in the bifundamental of $U(v^{(a)}_j)\times U(w^{(a+1)}_j)$;
\item[-] Fermi multiplets $\Theta^{(a)}_j$ in the bifundamental  of $U(v^{(a-1)}_j)\times U(w^{(a)}_j)$.
\end{itemize}

\begin{figure}
\begin{center}
\includegraphics[width=\textwidth]{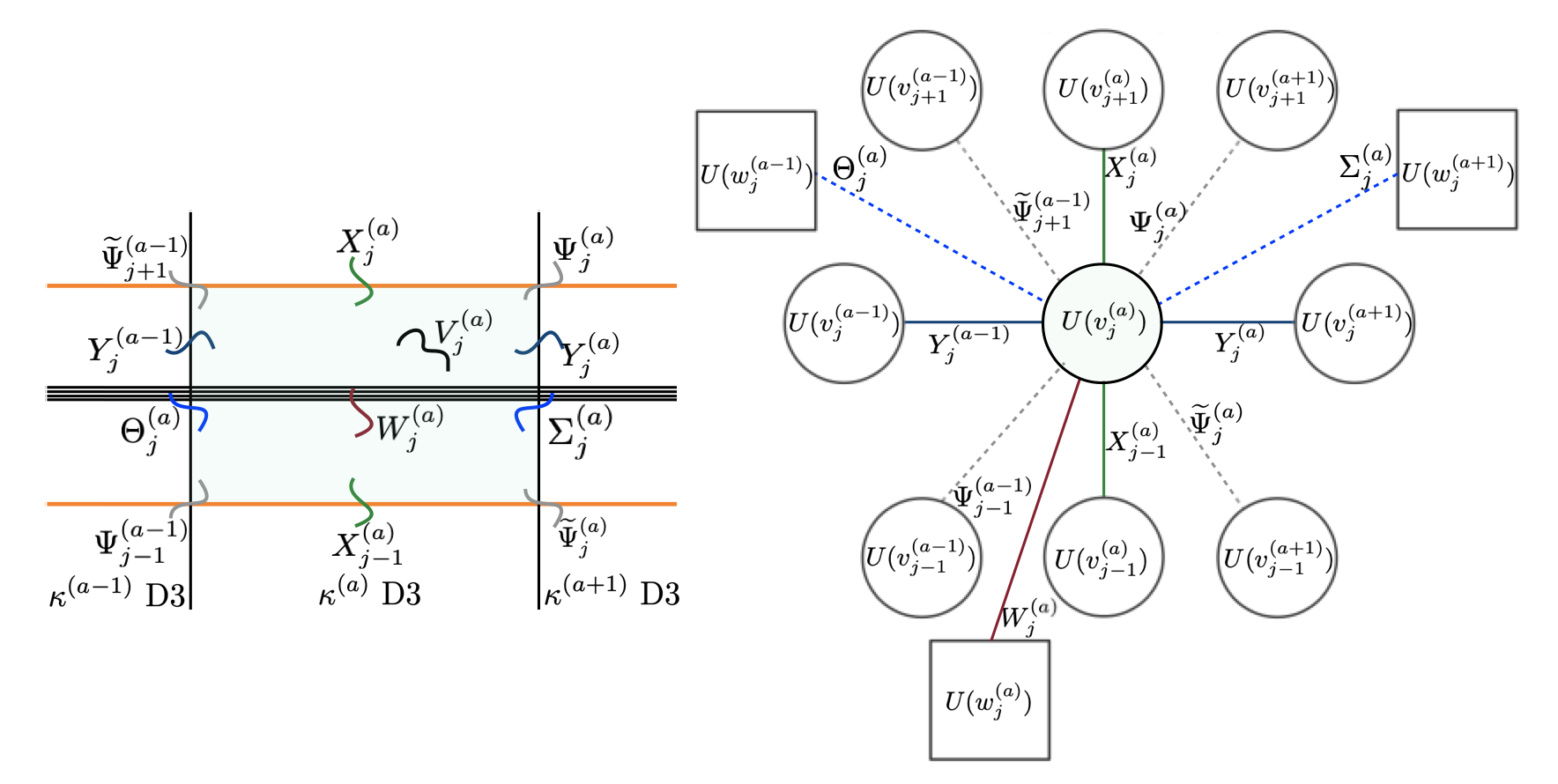}
\caption{Fields that couple to the $(a,j)$ D3 brane (highlighted in light green) in the Type IIB brane setup (where $v^{(a)}_0=\dots=v^{(a)}_{n-1}=\kappa^{(a)}$). On the left figure, the strings ending on the $(a,j)$ D3 brane are shown schematically. On the right figure, the gauge node that corresponds to the $(a,j)$ D3 brane, as well as the fields that couple to it and the neighboring nodes are shown. Hyper- and Fermi multiplets are depicted respectively as continuous and dashed lines. The resulting quiver, $\QQA$, suffers from abelian gauge anomalies that will be canceled by coupling to additional degrees of freedom in section \ref{sec:anom}. The ranks of different nodes in the quiver of the figure on the right can be taken to be more general, in which case the quiver describes the worldsheet degrees of freedom of instanton strings corresponding to more general brane configurations.}
\label{fig:plaqanom}
\end{center}
\end{figure}

The fields that couple to the gauge group arising from a specific plaquette are displayed in figure \ref{fig:plaqanom}.
The resulting 2d quiver gauge theory, however, is not well-defined by itself as it suffers from an abelian gauge anomaly as already observed in \cite{Hanany:2018hlz}. This anomaly can be canceled by coupling the quiver gauge theory to the chiral fermions that are supported at the intersections between the two sets of NS5 branes. This point is addressed in section \ref{sec:anom}.\newline

Let us turn now to the case where the flavor ranks $\vec{w}^{(a)}$ are not necessarily all identical. In order to do this, it is convenient to re-derive the 2d quiver from a different approach. Namely, we will begin by considering the 2d $(0,4)$ quiver gauge theories $\mathcal{Q}_{W,\boldsymbol{v}}$ of figure \ref{fig:orbi}, which describe the strings of the 6d SCFT $\mathcal{T}^{6d}_{r,W}$ on $T^2\times\mathbb{C}^2$ \cite{Haghighat:2013tka,Gadde:2015tra}, and implement the $\mathbb{Z}_n$ orbifold along the lines of \cite{Douglas:1996sw,Okuyama:2005gq,Haghighat:2013tka} to obtain the quivers describing strings on $T^2\times \mathbb{C}^2/\mathbb{Z}_n$. Recall that the global symmetries of theory $\mathcal{Q}_{W,{\boldsymbol{v}}}$ consist of the R-symmetry $ SU(2)_R\times SU(2)_t $ and of the flavor symmetry
\be
\left(\frac{\prod_{a=0}^{r-1}U(W)^{(a)}}{U(1)^{r}}\right)\times SU(2)_x.
\ee
The field content of this class of theories was reviewed in section \ref{sec:t2c2inst}.  We now perform the orbifold with respect to $\mathbb{Z}_n \in SU(2)_x$. First of all, we need to discuss the symmetries of the orbifolded quiver. Th 6d gauge symmetry $U(W)^{(a)}$ breaks to
\be
\prod_{j=0}^{n-1} U(w^{(a)}_j), \qquad \vert \vec{w}^{(a)}\vert = W,
\ee
where we impose the restrictions of section \ref{sec:tdua} on the ranks, while the 2d gauge group $U(\boldsymbol{v}^{(a)})$ is projected to
\be
\prod_{j=0}^{n-1} U(v^{(a)}_j),
\ee
where $\sum_{j=0}^{n-1}v^{(a)}_j = \boldsymbol{v}^{(a)}$. We also need to discuss the effect of the orbifold projection on the matter fields of the quiver. These consist of:
\begin{itemize}
\item[-] multiplets $(W^{(a)},\Sigma^{(a)},\Theta^{(a)})$ in the bifundamental representation of $U({v}^{(a)})\times U(W)$, which are neutral under $SU(2)_x$ and give rise in the orbifolded quiver to a collection of multiplets $(\vec{W}^{(a)},\vec{\Sigma}^{(a)},\vec{\Theta}^{(a)})$ whose $j$-th components transform respectively  in the bifundamental of $U({v}^{(a)}_j)\times U(w^{(a)}_j)$, of $U({v}^{(a)}_j)\times U(w^{(a+1)}_j)$, and of $U({v}^{(a)}_j)\times U(w^{(a-1)}_j)$;
\item[-] multiplets $X^{(a)}$ in the adjoint representation of $U({v}^{(a)})$ which carry charge $+1$ with respect to the Cartan of $SU(2)_x$, which give rise to a collection of multiplets $\vec{X}^{(a)}=(X^{(a)}_0,\dots,{X}^{(a)}_{n-1})$ where ${X}^{(a)}_j$ transforms in the bifundamental of $U(v^{(a)}_j)\times U(v^{(a)}_{j+1})$;
\item[-] multiplets $(Y^{(a)},\Psi^{(a)},\widetilde{\Psi}^{(a)})$ in the bifundamental representation of $U({v}^{(a)})\times U({v}^{(a+1)})$ which carry respectively charge $0,+1$, and $-1$ with respect to the Cartan of $SU(2)_x$ and give rise to a collection of multiplets $(\vec{Y}^{(a)},\vec{\Psi}^{(a)},\vec{\widetilde{\Psi}}{}^{(a)})$ whose $j$-th components transform respectively in the bifundamental of $U(v^{(a)}_j)\times U(v^{(a+1)}_{j})$,$U(v^{(a)}_j)\times U(v^{(a+1)}_{j+1})$, and $U(v^{(a)}_j)\times U(v^{(a+1)}_{j-1})$.
\end{itemize}
This way, one recovers all the fields depicted in the quiver of figure \ref{fig:plaqanom}, now for more general choices of monodromies $\vec{\boldsymbol{w}}$. We will rederive in section \ref{sec:anom} the conditions on the ranks $\vec{\boldsymbol{v}}$ by imposing cancelation of non-abelian gauge anomalies of $\QQA$.\newline

\begin{figure}
    \centering
    \includegraphics[scale=0.45]{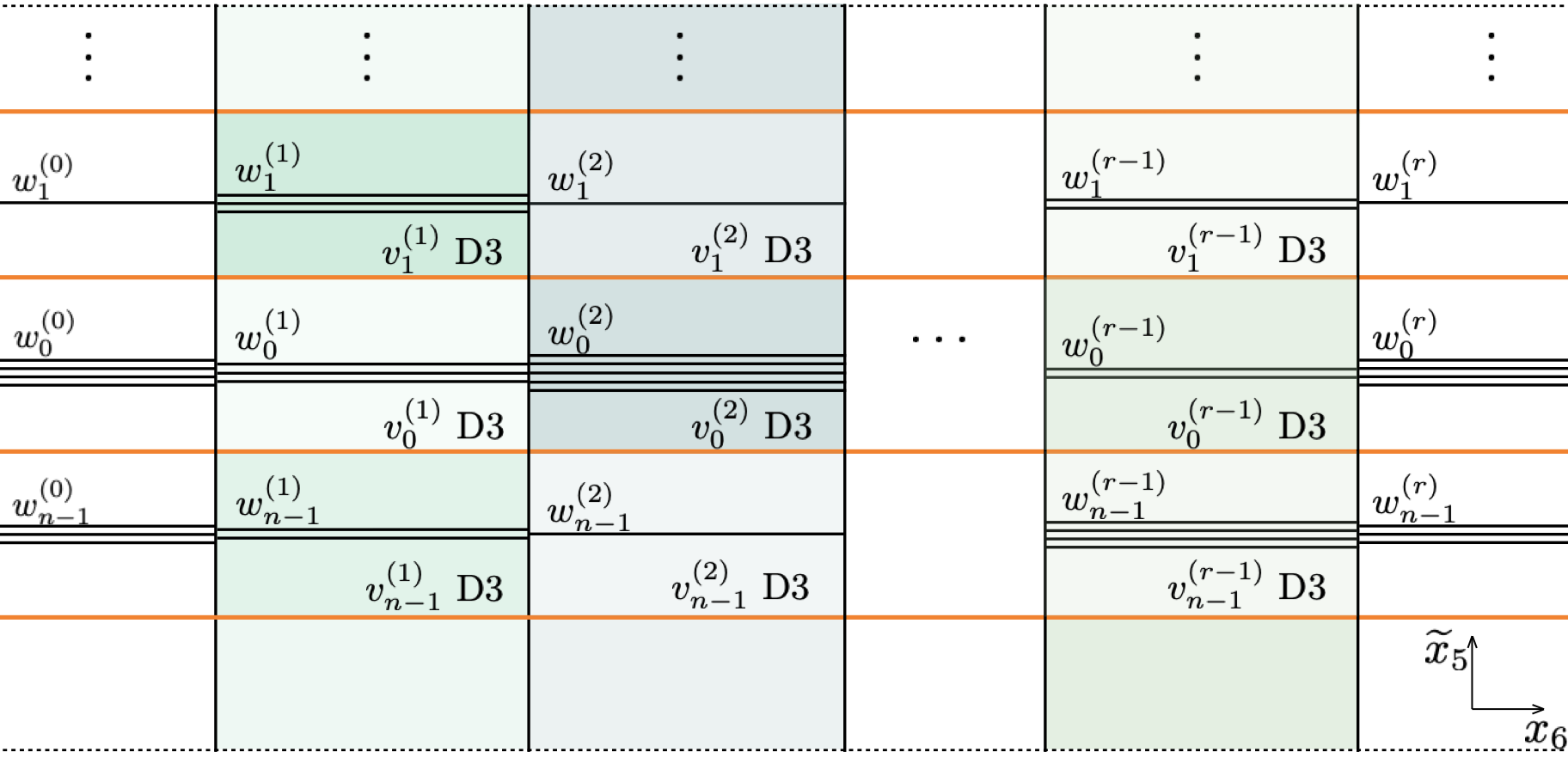}
    \caption{Simplified brane diagram corresponding to an instanton configuration of the theory $\mathcal{T}^{6d}_{r,W}$ for general monodromies $\vec{w}^{(a)}$.}
    \label{fig:IIBfrac}
\end{figure}

\paragraph{Remark.} \emph{A posteriori}, we see that it is possible to derive the quiver for theory $\mathcal{T}^{6d}_{r,W}$ for general choices of monodromies $\vec{\boldsymbol{w}}$ from the brane configuration of figure \ref{fig:beetles}, provided that we ignore the D5 brane charge of $(p,q)$ branes and treat them on the same footing as N5 branes. With this assumption, we arrive at a simplistic brane configuration for the case of unequal $\vec{w}^{(a)}$, as in figure \ref{fig:IIBfrac} from which the quiver can be read off; while this result is satisfactory, it would certainly be interesting to put the derivation of the quiver $\QQA$ for generic monodromies from the Type IIB brane configuration on a more rigorous standpoint.

\subsubsection{Cancelation of 2d gauge anomalies}
\label{sec:anom}

We now turn the issue of gauge anomalies and their cancelation. Let us for the moment focus on the nonabelian components $SU(v^{(a)}_j)$ of the gauge group of the 2d quiver gauge theory. A multiplet in the adjoint representation contributes a term $4v^{(a)}_j$ to the $ U(v^{(a)}_j) $ anomaly coefficient; both ordinary and twisted hypermultiplets, transforming in the fundamental or anti-fundamental representation of $ U(v^{(a)}_j) $, contribute $-2$ to the anomaly coefficient; finally, Fermi multipets in the fundamental or anti-fundamental representation contribute $+1$. Adding up all contributions from the fields that couple to the $U(v^{(a)}_j)$ gauge node, which are displayed in figure \ref{fig:plaqanom}, we find the following anomaly coefficient:
\be
({C}^{\widehat{A}_r}\cdot(-\vec{\boldsymbol{w}}+C^{\widehat{A}_{n-1}}\cdot \vec{\boldsymbol{v}}))^{(a)}_j
=
-({C}^{\widehat{A}_r}\cdot\boldsymbol{u}_j)^{(a)},
\ee
where $\widehat{\mathbf{C}} = C^{\widehat{A}_{r-1}}$ is the affine $SU(r)$ Cartan matrix acting on the $(a)$ label. This non-abelian gauge anomaly vanishes once the constraints \eqref{eq:constro} are imposed, which leads to the condition \eqref{eq:vsdeltas} on the gauge group ranks.\newline

On the other hand, the quivers $\QQA$  do in general suffer from abelian gauge anomalies. These are detected by the presence in the four-form anomaly polynomial $\mathcal{I}_{4,ab.}$ of the 2d quiver gauge theory of terms of the form $\Tr F_{U(v^{(a)}_j)} \Tr F_{U(v^{(b)}_k)}$. Adding the contributions from all multiplets in the quiver, we find that that these coefficients can be put in the following suggestive form:
\be
\mathcal{I}_{4,ab.} = -\sum_{a=0}^{r}\sum_{j=0}^{n-1}\left(\Tr F_{U(v^{(a)}_j)}-\Tr F_{U(v^{(a+1)}_j)}+\Tr F_{U(v^{(a+1)}_{j+1})}-\Tr F_{U(v^{(a)}_{j+1})}\right)^2,
\ee
where one sets $\Tr F_{U(v^{(0)}_j)}=\Tr F_{U(v^{(r)}_j)}$ = 0 as there are no infinitely extended D3 branes on the left and right ends of the brane diagram. The cancelation of these anomalies is achieved in exactly the same way as in \cite{DelZotto:2023rct}, by coupling the $\oplus_{a=1}^r\mathfrak{su}(n)_1^{(a)}$ currents $\vec{\boldsymbol{\mathcal{J}}}^{\mathfrak{su}(n)_1}$ to the gauge degrees of freedom of the quiver:
\be
\int_{T^2} \sum_{a=1}^r\sum_{j=1}^{n-1}\mathcal{J}_j^{\mathfrak{su}(n)_1,(a)}\left[C^{\widehat{A}_{n-1}}\cdot(\vec{\Tr A^{(a)}}-\vec{\Tr A^{(a-1)}})\right]_j.
\label{eq:gcplg}
\ee

A novel feature we encounter for the theories $\mathcal{T}^{6d}_{r,W}$ is the  existence of mixed anomalies between the abelian gauge symmetries of $\QQA$ and its global symmetries. Specifically, for the $(a,j)$ gauge node of the quiver we find the following terms:
\begin{equation}
\mathcal{I}_{4,mixed}\supset
-\Tr F_{U(v^{(a)}_j)}
\cdot
\left(\sum_{b=0}^{r}(C^{\widehat{A}_{r-1}})^{(a)(b)}\left(\Tr F_{U(w^{(b)}_j)}+(\delta^{(b)}_j+\delta^{(b)}_{j-1}) F_{U(1)_x}\right)\right).
\label{eq:abelan}
\end{equation}
It is also possible to cancel the mixed anomalies if we postulate the existence of the following additional couplings between the $\mathfrak{su}(n)^{(a)}_1$ currents and background gauge fields:
\bea
\nonumber
\int_{T^2} \sum_{a=1}^r\sum_{j=1}^{n-1}\mathcal{J}_j^{\mathfrak{su}(n)_1,(a)}
&\bigg[&
\Tr A^{U(w^{(a-1)}_j)}
-
\Tr A^{U(w^{(a)}_j)}
+
\Tr A^{U(1)_{\mathfrak{m}}}
\\
&&+(\delta^{(a)}_j+\delta^{(a)}_{j-1}-\delta^{(a-1)}_j-\delta^{(a-1)}_j)A^{U(1)_x}\bigg].
\label{eq:extracplg}
\eea
Note that the system under consideration is intrinsically strongly coupled, and in particular its Type IIB dual always necessarily involves the simultaneous presence of NS5 and D5 branes. As a consequence, a first principles derivation of these couplings is beyond the scope of the methods of \cite{Itzhaki:2005tu}, which rely on the existence of a weak coupling limit. Our choice of couplings \eqref{eq:extracplg} is the minimal choice that ensures cancelation of gauge anomalies on the BPS string worldsheet. It is quite remarkable that consistency of the BPS worldsheet theory can provide a window into such delicate aspects of this brane system.\\

To summarize, we found that the 2d (0,4) quiver describing instanton strings in the $T^2\times \mathbf{ALE}_n$ background is obtained by coupling the anomalous quiver $\QQA$ to $r$ copies of the Heisenberg algebra and $r$ copies of the $\mathfrak{su}(n)_1$ current algebra:
\be
\QQ = \mathcal{H}^r \times (\mathfrak{su}(n)_1^r \ltimes \QQA),
\ee
where the $\mathfrak{su}(n)_1^{(a)}$ currents couple to various gauge and background fields via the couplings \eqref{eq:c3cplg},\eqref{eq:gcplg}, and \eqref{eq:extracplg}. As a consequence of the coupling to the $\mathfrak{su}(n)_1$ currents, the 2d theories $\QQ$ behave as \emph{relative} theories and rather than possessing a partition function possess a vector of conformal blocks labeled by the possible choices of $\boldsymbol{\omega}^{KK}$, a fact which was already mentioned in \cite{DelZotto:2023rct} as the solitonic counterpart of the discrete choices of boundary conditions at infinity necessary to define the $T^2 \times \mathbf{ALE}_n$ partition function.
\newline

We conclude this section by pointing out that, again in parallel to the M-string case \cite{DelZotto:2023rct}, the 2d quivers $\QQ$ admit another very natural interpretation. Recall that $U(W)$ fractional instantons on $\mathbb{C}^2/\mathbb{Z}_n$ can be described in terms of a 3d $\mathcal{N}=4$ gauge theory which is encoded in terms of a Kronheimer-Nakajima quiver $\KN$, as shown in figure \ref{fig:3dKN}. The quivers $\QQ$ can be obtained by stacking along direction $x_6$ $r+1$ copies of $\KN$, with the $r$ NS5 branes giving rise to interfaces $\mathbf{NS}_n$ between them:
\bea
\resizebox{\textwidth}{!}{
$
\mathcal{KN}^{\vec{w}^{(0)}}_{\vec{0}}
\boldsymbol{\bigg{]}}
\mathbf{NS}_n
\boldsymbol{\bigg{[}}
\mathcal{KN}^{\vec{w}^{(1)}}_{\vec{v}^{(1)}}
\boldsymbol{\bigg{]}}
\mathbf{NS}_n
\boldsymbol{\bigg{[}}
\mathcal{KN}^{\vec{w}^{(2)}}_{\vec{v}^{(2)}}
\boldsymbol{\bigg{]}}
\mathbf{NS}_n
\boldsymbol{\bigg{[}}
\quad
\dots
\quad 
\boldsymbol{\bigg{]}}
\mathbf{NS}_n
\boldsymbol{\bigg{[}}
\mathcal{KN}^{\vec{w}^{(r-1)}}_{\vec{v}^{(r-1)}}
\boldsymbol{\bigg{]}}
\mathbf{NS}_n
\boldsymbol{\bigg{[}}
\mathcal{KN}^{\vec{w}^{(0)}}_{\vec{0}}
.
$
}
\nonumber
\eea
It is again the interfaces, where the $\mathfrak{su}(n)_1$ current algebras reside, that depend on a choice of monodromies $\boldsymbol{\omega}^{KK}$ for the 6d two-form fields and are responsible for the relative nature of the theories $\QQ$.
 
\subsubsection{Elliptic genus of $\QQ$}
\label{sec:quivprop}

For a given bound state of BPS strings, the supersymmetric excitations that contribute to the 6d ALE partition function of theory $\mathcal{T}^{6d}_{r,W}$ are captured by the elliptic genus
\be
\mathbb{E}^{\vec{\boldsymbol{w}},\boldsymbol{\omega}^{KK}}_{\vec{\boldsymbol{v}}}(\vec{\xi},\epsilon_+,\epsilon_-,\underline{\boldsymbol{s}},\tau)
=
\Tr (-1)^{F} q^{H_L}  \overline{q}^{H_R}e^{2\pi i \epsilon_+(J_R+J_t)}e^{2\pi i \epsilon_- J_x} e^{2\pi i \vec{\xi}\cdot\vec{J}^{\mathfrak{su}(n)}} e^{2\pi i \underline{\boldsymbol{s}}\cdot \underline{\boldsymbol{J}}},
\ee
where $\underline{J}^{(a)}$ are the Cartan generators of the global symmetry $\prod_{j=0}^{n-1}U(w_j^{(a)})$. The elliptic genus corresponding to a bound state of strings is given in terms of the following integral over the  holonomies $\vec{\boldsymbol{z}}$ of the gauge groups $\prod_{a,j} U(v^{(a)}_j)$:
\bea
\nonumber
\mathbb{E}^{\vec{\boldsymbol{w}},\boldsymbol{\omega}^{KK}}_{\vec{\boldsymbol{v}}}
\!
=
\prod_{j=0}^{n-1}
\prod_{a=1}^{r-1}
\frac{1}{v^{(a)}_j!}
\!
\int &&\!\!\!\!\!\!\!\!
\prod_{j=0}^{n-1}
\left[\!
\bigg(\prod_{a=1}^{r-1} Z_{V^{(a)}_j} Z_{X^{(a)}_j}Z_{W^{(a)}_j} Z_{\Sigma^{(a)}_j}Z_{\Theta^{(a)}_j}\bigg)
\!
\bigg(\prod_{a=1}^{r-2} Z_{Y^{(a)}_j}Z_{\Psi^{(a)}_j}Z_{\widetilde{\Psi}^{(a)}_j}\bigg)
\!\right]\\
&\times&
\bigg(\prod_{a=1}^{r}Z_{\mathcal{H}}Z^{\omega^{KK,(a)}}_{\mathfrak{su}(n)_1^{(a)}}\bigg).
\label{eq:eg}
\eea
The factors in the first row of equation \eqref{eq:eg} are the contributions from the multiplets that arise from D3 brane degrees of freedom, which are given as follows:
\bea
\nonumber
Z_{V^{(a)}_j}
&=&
\left(\prod_{k=1}^{v^{(a)}_j}\frac{\eta^2 dz^{(a)}_{j,k}}{2\pi i}\frac{\theta_1(2\epsilon_+)}{\eta}\right)
\left(\prod_{\substack{k,l=1 \\ k\neq l}}^{v^{(a)}_j}\frac{\theta_1(z^{(a)}_{j,k}-z^{(a)}_{j,l})\theta_1(2\epsilon_+ + z^{(a)}_{j,k}-z^{(a)}_{j,l})}{\eta^2}\right);\\
\nonumber
Z_{X^{(a)}_j}
&=&
\left(\prod_{k=1}^{v^{(a)}_j}\prod_{l=1}^{v^{(a)}_{j+1}}\frac{\eta^2}{\theta_1(\epsilon_++\epsilon_-+ z^{(a)}_{j+1,l}-z^{(a)}_{j,k})\theta_1(\epsilon_+-\epsilon_-+ z^{(a)}_{j,k}-z^{(a)}_{j+1,l})}\right);\\
\nonumber
Z_{Y^{(a)}_j}
&=&
\left(\prod_{k=1}^{v^{(a)}_j}\prod_{l=1}^{v^{(a+1)}_{j}}\frac{\eta^2}{\theta_1(\epsilon_++z^{(a+1)}_{j,l}- z^{(a)}_{j,k})\theta_1(\epsilon_++ z^{(a)}_{j,k}-z^{(a+1)}_{j,l})}\right);\\
\nonumber
Z_{\Psi^{(a)}_j}
&=&
\left(\prod_{k=1}^{v^{(a)}_j}\prod_{l=1}^{v^{(a+1)}_{j+1}}\frac{\theta_1(\epsilon_-+ z^{(a+1)}_{j+1,l}-z^{(a)}_{j,k})}{\eta}\right);\\
\nonumber
Z_{\widetilde{\Psi}^{(a)}_j}
&=&
\left(\prod_{k=1}^{v^{(a)}_j}\prod_{l=1}^{v^{(a+1)}_{j-1}}\frac{\theta_1(-\epsilon_-+ z^{(a+1)}_{j-1,l}-z^{(a)}_{j,k})}{\eta}\right);\\
\nonumber
Z_{W^{(a)}_j}
&=&
\left(\prod_{k=1}^{v^{(a)}_j}\prod_{K=1}^{w^{(a)}_{j}}\frac{\eta^2}{\theta_1(\epsilon_++s^{(a)}_{j,K}-z^{(a)}_{j,k})\theta_1(\epsilon_++ z^{(a)}_{j,k}-s^{(a)}_{j,K})}\right);\\
\nonumber
Z_{\Sigma^{(a)}_j}
&=&
\left(\prod_{k=1}^{v^{(a)}_j}\prod_{K=1}^{w^{(a+1)}_{j}}\frac{\theta_1(s^{(a+1)}_{j,K}-z^{(a)}_{j,k})}{\eta}\right);\\
\nonumber
Z_{\Theta^{(a)}_j}
&=&
\left(\prod_{k=1}^{v^{(a)}_j}\prod_{K=1}^{w^{(a-1)}_{j}}\frac{\theta_1(z^{(a)}_{j,k}-s^{(a-1)}_{j,K})}{\eta}\right).
\eea
In the last three expressions, the set of holonomies $\{s^{(a)}_{j,K}\}_{K=1}^{w^{(a)}_{j}}$ corresponds to the subset of the holonomies $\vec{s}^{(a)}$ associated to the $U(w^{(a)}_j)$ subgroup of $U(W)^{(a)}$.\\

In light of the discussion of section \ref{sec:anom}, the factors from the second row of equation \eqref{eq:eg} are obtained by shifting the chemical potential $\vec{\xi}$ in equation \eqref{eq:chibos} by appropriate combinations of chemical potentials for the 2d global symmetries, namely:
\be
Z_{\mathcal{H}}Z^{\omega^{KK,(a)}}_{\mathfrak{su}(n)_1^{(a)}} = \widehat{\chi}_{\mathcal{H}}\widehat{\chi}_{\omega^{KK,(a)}}^{\mathfrak{su}(n)_1}(\vec{\xi}^{(a)},\tau),
\ee
where
\be
\xi^{(a)}_j  = \xi+\mathfrak{m} + (C^{\widehat{A}_{n-1}}\cdot(Z^{(a)}-Z^{(a-1)}))_j-S^{(a)}_j+S^{(a-1)}_j+(\delta^{(a)}_j+\delta^{(a)}_{j-1}-\delta^{(a-1)}_j-\delta^{(a-1)}_{j-1})\epsilon_-,
\label{eq:xish}
\ee
where we have defined
\be
Z^{(a)}_j = \sum_{k=1}^{v^{(a)}_j}z^{(a)}_{j,k},\qquad S^{(a)}_j = \sum_{K=1}^{w^{(a)}_j}s^{(a)}_{j,K}.
\ee
Note that in the case of M-strings \cite{DelZotto:2023rct}, equation \eqref{eq:xish} reduces simply to
\be
\xi^{(a)}_j  = \xi+\mathfrak{m}-n \delta_{j,0}\mathfrak{m} +(C^{\widehat{A}_{n-1}}\cdot(Z^{(a)}-Z^{(a-1)}))_j,
\ee
and the $\mathfrak{m}$-dependent terms can be reabsorbed by a shift of the chemical potential $\vec{\xi}$ that arises from the M-theory three-form field $C_3$.  \newline

It is possible to give a combinatorial expression for the sum over the residues that contribute to the integral \eqref{eq:eg}. Recall from the discussion of section \ref{sec:anom} that our 2d quiver gauge theories arise by stacking multiple copies of the 3d $\mathcal{N}=4$ Kronheimer-Nakajima quiver gauge theory $\KN$ along direction $x_6$, with interfaces provided by NS5 branes. For each copy of the $\KN$ theory, the residues that contribute are in one-to-one correspondence with the residues that contribute to the the quiver quantum mechanics describing $SU(W)$ instantons on $\mathbb{C}^2/\mathbb{Z}_n$ \cite{Dey:2013fea}, which are given as follows. For each $a=1,\dots, r-1$, denote by $\mathcal{Y}_{\vec{v}^{(a)}}$ the set of $W$-tuples of Young diagrams $\underline{Y}^{(a)} = (Y_0^{(a)},\dots,Y_{W-1}^{(a)})$ such that the total number of boxes is 
\be
\vert \underline{Y}^{(a)}\vert
:=
\sum_{A=0}^{W-1} \vert Y^{(a)}_{A}\vert
=
\sum_{j=0}^{n-1}v^{(a)}_j.
\ee
A box $\mathfrak{b}$ appearing in a Young diagram $Y_{A(\mathfrak{b})}\in \underline{Y}$ is labeled by the index $A(\mathfrak{b})$ and coordinates $(\mathfrak{b}^1,\mathfrak{b}^2)$ inside the Young diagram. To the box $\mathfrak{b} = (\mathfrak{b}^1,\mathfrak{b}^2)$ in Young diagram $Y^{(a)}_{A(\mathfrak{b})}$, we assign the integer 
\be
\omega_\mathfrak{b} =  \omega^{(a)}_A-\mathfrak{b}^1+\mathfrak{b}^2 \qquad \text{mod } n
\label{eq:omcon}
\ee
where $\omega^{(a)}_A$ was defined in section \ref{sec:zpertc2zn} and characterizes the monodromy of $U(1)^{(a)}_A$, the $A$-th component  of the Cartan of $U(W)^{(a)}$. The residues associated to the gauge holonomies $\vec{z}^{(a)}$ are labeled by elements of a subset $\mathcal{Y}_{\vec{w}^{(a)},\vec{v}^{(a)}} $ of $\mathcal{Y}_{\vec{v}^{(a)}}$ which is defined as follows. Given a $W$-tuple $\underline{Y}^{(a)}\in \mathcal{Y}_{\vec{v}^{(a)}}$ of Young diagrams, denote by $\mathcal{B}_{\vec{w}^{(a)},j}$ the set of boxes $\mathfrak{b}$ in $\underline{Y}^{(a)}$ for which $\omega_\mathfrak{b} = j$. Then, the tuple $\underline{Y}^{(a)}$ is contained in $\mathcal{Y}_{\vec{w}^{(a)},\vec{v}^{(a)}}$ if and only if the order of the set $\mathcal{B}_{\vec{w}^{(a)},j}$ is given by
\be
\vert \mathcal{B}_{\vec{w}^{(a)},j} \vert = v^{(a)}_j
\ee
for all $j=0,\dots,n-1$. For each choice of $\underline{\boldsymbol{Y}}\in \mathcal{Y}_{\vec{w}^{(1)},\vec{v}^{(1)}}\times\dots\times\mathcal{Y}_{\vec{w}^{(1)},\vec{v}^{(1)}}$ we assign the following values to the holonomies $z^{(a)}_{j,k}$ for $a=1,\dots,r-1$, $j=0,\dots,n-1$ and $k=1,\dots,v^{(a)}_j$:
\begin{equation}
z^{(a)}_{j,k}
\to
z^*_{\underline{Y}^{(a)},j,k}
=
s^{(a)}_{A(\mathfrak{b}^{(a)}_{j,k})} + \epsilon_+ + (\epsilon_+ + \epsilon_-)\mathfrak{b}^{(a),1}_{j,k}+(\epsilon_+-\epsilon_-)\mathfrak{b}^{(a),2}_{j,k},
\label{eq:zassign}
\end{equation}
where we have denoted by $\{\mathfrak{b}^{(a)}_{j,1},\dots,\mathfrak{b}^{(a)}_{j,v^{(a)}_j}\}$ the set of boxes in $\mathcal{B}_{\vec{w}^{(a)},j}$. This involves making an arbitrary choice of ordering on the set $\mathcal{B}_{\vec{w}^{(a)},j}$, which is compensated by removing the cominatorial factor of $\prod_{j=0}^{n-1}\frac{1}{v^{(a)}_j!}$ appearing in equation \eqref{eq:eg}.

The elliptic genus is then obtained by summing over the residues evaluated at the values \eqref{eq:zassign} of the integration variables, analogously to \cite{Gadde:2015tra}. We find:
\begin{align}
\nonumber
\mathbb{E}^{\vec{\boldsymbol{w}},\boldsymbol{\omega}^{KK}}_{\vec{\boldsymbol{v}}}
=&
\hskip-.6in
\sum_{\underline{\boldsymbol{Y}}^{(1)}\in \{\mathcal{Y}_{\vec{w}^{(1)},\vec{v}^{(1)}}\times\dots\times \mathcal{Y}_{\vec{w}^{(r-1)},\vec{v}^{(r-1)}}\}}
\hskip-.1in
\left(\frac{\theta_1(2\epsilon_+)}{\eta}\right)^{\sum_{a=1}^{r-1}\sum_{j=0}^{n-1}v^{(a)}_j}
\bigg[
\bigg(\prod_{a=1}^{r-1}\!\!\phantom{\vert}^{'} Z_{X^{(a)}_j}Z_{W^{(a)}_j} Z_{\Sigma^{(a)}_j}Z_{\Theta^{(a)}_j}\bigg)
\\
&\times
\bigg(\prod_{a=1}^{r-2} Z_{Y^{(a)}_j}Z_{\Psi^{(a)}_j}Z_{\widetilde{\Psi}^{(a)}_j}\bigg)\bigg(\prod_{a=1}^{r}Z_{\mathcal{H}}Z^{\omega^{KK,(a)}}_{\mathfrak{su}(n)_1^{(a)}}\bigg)\bigg]\bigg\vert_{z^{(a)}_{j,k}\to z^*_{\underline{Y}^{(a)},j,k}}.
\label{eq:ellipticgenus}
\end{align}
As in section \ref{sec:t2c2inst},  the notation $\prod \!\phantom{\vert}^{'}$ indicates that any occurrence of $\theta_1(0)$ in the product must be replaced by a factor of $\eta$. Concrete examples of elliptic genera are provided in section \ref{sec:examples}.
\subsection{The $\mathcal{T}^{6d}_{r,W}$ partition function on $T^2\times\mathbb{C}^2/\mathbb{Z}_n$}
\label{sec:zfullc2zn}
In this section we combine the contributions from NS5 and D3 branes into the partition function of the theory $\mathcal{T}^{6d}_{r,W}$ on $T^2\times\mathbb{C}^2/\mathbb{Z}_n$. The partition function depends on the following data:
\begin{itemize}
\item[-] The monodromy of the two-form fields, specified in terms of $r$ independent integers $\omega^{KK,(1)},\dots,\omega^{KK,(r)}\in \mathbb{Z}_n$;
\item[-] The chemical potential $\vec{\xi} = \sum_{j=1}^{n-1}\xi_jc_1(\mathcal{R}_j)$ conjugate to the two-form field flux, which arises from the $C_3$ three-form field in M-theory;
\item[-] The monodromy of the background gauge field associated to the 6d global symmetry, specified in terms of an arbitrary partition $\vec{w}^{(0)}=\vec{w}^{(r)}$ of $W$;
\item[-] Monodromies of gauge fields associated to the 6d gauge symmetry, specified in terms of partitions $\vec{w}^{(1)},\dots,\vec{w}^{(r-1)}$ of $W$ belonging to the set $\mathfrak{W}_{\vec{w}^{(0)}}$ (see section \ref{sec:tdua});
\item[-] A Wilson line $\mathfrak{m}$ for the $U(1)_{\mathfrak{m}}$ isometry discussed in section \ref{sec:zper};
\item[-]  $\mathfrak{u}(W)^{r+1}$ Wilson lines $s^{(a)}_A$, for $a=0,\dots, r+1$ and $A=0,\dots, W-1$, subject to the Stuckelberg constraint \eqref{eq:stucondition}. The Wilson lines for each individual factor $\mathfrak{u}(W)^{(a)}$ are labeled so that $s^{(a)}_A \leq s^{(a)}_B$ for $A<B$;
\item[-] Tensor branch parameters $\varphi^{(a)}$ corresponding to the separation between neighboring NS5 branes, conjugate to the $\mathfrak{su}(W)^{(a)}$ instanton charge;
\item[-] The complex modulus $\tau$ of $T^2$, and equivariant parameters $\epsilon_+,\epsilon_-$ for $\mathbb{C}^2/\mathbb{Z}_n$.
\end{itemize}
The partition function is computed for fixed choices of monodromy, which we label collectively by
\be
\vec{\boldsymbol{w}} = (\vec{w}^{(0)},\vec{w}^{(1)},\dots,\vec{w}^{(r)}) \qquad \text{and}\qquad \boldsymbol{\omega}^{KK} = (\omega^{KK,(1)},\dots,\omega^{KK,(r)}).
\ee
In the partition function we sum over fluxes for the two-form fields on the M5 branes. The 6d $\mathfrak{u}(W)^{(a)}$ magnetic fluxes are all fixed by the constraint \eqref{eq:e1}. Combining the results of sections \ref{sec:zpertc2zn} and \ref{sec:d3z}, we arrive at the following expression for the partition function:
\bea
\nonumber
&&\hskip-0.75in
\mathcal{Z}^{\mathcal{T}^{6d}_{r,W},\vec{\boldsymbol{w}},\boldsymbol{\omega}^{KK}}_{T^2\times\mathbb{C}^2/\mathbb{Z}_n}(\vec{\xi},\epsilon_+,\epsilon_-,\tau,\underline{\boldsymbol{s}})
=
\\
\nonumber
q^{\frac{n\,r}{24}}
&\times&
\left(\prod_{a=1}^r\mathcal{Z}^{\text{BPS particles},\vec{w}^{(a-1)},\vec{w}^{(a)}}_{T^2\times\mathbb{C}^2/\mathbb{Z}_n}(\epsilon_+,\epsilon_-,\tau,\underline{s}^{(a-1)},\underline{s}^{(a)})\right)
\\
&\times&
\sum_{
\substack{
\boldsymbol{\kappa}=(\kappa^{(1)},\dots,\kappa^{(r-1)})\in\mathbb{Z}^{r-1}
\\
\kappa^{(a)}\geq \kappa^*_{\vec{w}^{(a)}}
}
}
e^{-\sum_{a=1}^r N_{\kappa^{(a)},\vec{w}^{(a)}}\varphi^{(a)}} \mathbb{E}^{\vec{\boldsymbol{w}},\boldsymbol{\omega}^{KK}}_{\vec{\boldsymbol{v}}_{\boldsymbol{\kappa}}}(\vec{\xi},\epsilon_+,\epsilon_-,\tau,\underline{\boldsymbol{s}}),
\label{eq:zfull}
\eea
where the BPS particles' contribution is given by equation \eqref{eq:bpsp}, while the instanton string contributions are given by equation \eqref{eq:ellipticgenus}. The instanton numbers $N_{\kappa^{(a)},\vec{w}^{(a)}}$ are given in equation \eqref{eq:instnum}.

Before turning to examples, let us comment on some basic properties of the partition function. First of all, as we have already commented in section \ref{sec:mod} the occurrence of a $(\mathfrak{su}(n)_1)^r$ current algebra gives rise to $n\cdot r$ inequivalent choices of monodromy for the tensor fields, which give rise to different partition functions that transform into each other under the modular group action. The partition function also depends on a choice of monodromy $\vec{w}^{(0)}$ for $\mathfrak{u}(W)^{(a)}$ and $\vert\mathfrak{W}_{\vec{w}^{(0)}}\vert^{r-1}$ choices of monodromy for the 6d gauge algebra. However, by a gauge transformation on the NS B-field in Type IIA (which corresponds to a rotation of the $\tilde x_5$ circle in the Type IIB frame), it is possible to simultaneously shift the monodromy for every D6 brane  by an identical factor of $e^{\frac{2\pi i}{n}\ell}$ for $\ell=1,\dots,n-1$ (see the discussion in section 3.2 of \cite{Witten:2009xu}). This is equivalent to shifting the monodromy vector $\vec{w}^{(a)}$ for all $a=0,\dots,r$ as follows:
\be
\vec{w}^{(a)} \to \vec{w}^{(a)}\langle\ell\rangle = (w^{(a)}_\ell,w^{(a)}_{\ell+1},\dots,w^{(a)}_{n-1},w^{(a)}_0,\dots,w^{(a)}_{\ell-1}),
\ee
while simultaneously shifting the D3 brane multiplicities in the same way:
\be
\vec{v}^{(a)} \to \vec{v}^{(a)}\langle\ell\rangle = (v^{(a)}_\ell,v^{(a)}_{\ell+1},\dots,v^{(a)}_{n-1},v^{(a)}_0,\dots,v^{(a)}_{\ell-1}),
\ee
At the same time, the gauge transformation leads to the following shift in the chemical potentials associated to the tensor field flux:
\be
\vec{\xi}\to \vec{\xi}\langle \ell\rangle
\ee
where we define:
\bea
\xi_j\langle \ell\rangle	&=& \xi_{j+\ell}\quad\qquad j=1,\dots,n-1-\ell\\
\xi_{n-\ell}\langle \ell\rangle	&=& \tau-\sum_{j=1}^{n-1}\xi_{j}\\
\xi_{j}\langle \ell\rangle	&=& \xi_{j+\ell-n}\qquad j=n-\ell+1,\dots,n-1.
\eea
It is straightforward to check that the partition function is invariant under the effect of the gauge transformation:
\be
\mathcal{Z}^{\mathcal{T}^{6d}_{r,W},\vec{\boldsymbol{w}}\langle\ell\rangle,\boldsymbol{\omega}^{KK}}_{T^2\times\mathbb{C}^2/\mathbb{Z}_n}(\vec{\xi}\langle\ell\rangle,\epsilon_+,\epsilon_-,\tau,\underline{\boldsymbol{s}})=\mathcal{Z}^{\mathcal{T}^{6d}_{r,W},\vec{\boldsymbol{w}},\boldsymbol{\omega}^{KK}}_{T^2\times\mathbb{C}^2/\mathbb{Z}_n}(\vec{\xi},\epsilon_+,\epsilon_-,\tau,\underline{\boldsymbol{s}}).
\ee

Finally, let us comment on the nature of the fractional instanton contributions to the partition function \eqref{eq:zfull}. We have seen in section \ref{sec:tdua} that, for each factor $\mathfrak{u}(W)^{(a)}$ of the gauge algebra, the partition function receives contributions from D3 brane bound states whose instanton charge has the same fractional part:
\be
N_{\kappa^{(a)},\vec{w}^{(a)}} = \kappa^{(a)}+\sum_{j=1}^{n-1}\frac{j(n-j)}{2n}w^{(a)}_j.
\ee
Therefore up to an overall factor of
\be
e^{-\sum_{a=1}^{r-1}\varphi^{(a)}\sum_{j=1}^{n-1}\frac{j(n-j)}{2n}w^{(a)}_j}
\ee
the partition function consists of an expansion in integer powers of the exponentiated tensor branch parameters $e^{-{\varphi}^{(a)}}$. Precisely in the case where all $w^{(a)}_j$ are identical the leading order contribution to the partition function arises from the zero instanton sector, for which
\be
\mathbb{E}^{\vec{\boldsymbol{w}},\boldsymbol{\omega}^{KK}}_{\vec{\boldsymbol{0}}}(\vec{\xi},\epsilon_+,\epsilon_-,\underline{\boldsymbol{s}},\tau)
\to
\frac{1}{\eta(\tau)^r}\prod_{a=1}^{r}\widehat{\chi}^{\mathfrak{su}(n)_1}_{\omega^{KK,(a)}}(\vec\xi,\tau),
\ee
while in other cases the leading order contributions consist of a `frozen' instanton strings which carry fractional instanton charge and are not free to move in $\mathbb{C}^2/\mathbb{Z}_n$.

\section{Properties of the BPS strings}
\label{sec:BPSstrings}
In section \ref{sec:mod} we determine the central charges and global anomalies of the 2d quiver gauge theory $\QQ$ and study its modular properties; in section \ref{sec:examples} we examine explicit examples of elliptic genera.
\subsection{Anomalies and central charges}
\label{sec:mod}
The value of the right-moving central charge $c_R$ in the infrared is given by six times the coefficient of $c_2(SU(2)_R)$ in the four-form anomaly polynomial, which can be read off from the field content of the quiver gauge theory. Based on the field content of table \ref{tab:04mult}, this coefficient receives contributions from the fermions in the vector and twisted hypermultiplets, which are charged under $SU(2)_R$, but not from the fermions belonging to the Fermi and hypermultiplets which are neutral. As a consequence, the central charge decomposes into a sum of contributions from the instanton strings for the $r-1$ $\mathfrak{u}(W)$ factors of the 6d gauge group algebra, each of which is described by a two-dimensional analog of the Kronheimer-Nakajima quiver as discussed in section \ref{sec:anom}. Adding up the contributions from these multiplets one finds:
\be
c_R = 6\sum_{a=1}^{r-1} \sum_{j=0}^{n-1}\left(v^{(a)}_jw^{(a)}_j - (v^{(a)}_j)^2+v^{(a)}_jv^{(a)}_{j+1}\right) = 6\sum_{a=1}^{r-1} \dim_{\mathbb{H}}\mathcal{M}_{\vec{w}^{(a)},\vec{v}^{(a)}},
\label{eq:hkdim}
\ee
where 
\be
\dim_{\mathbb{H}}\mathcal{M}_{\vec{w}^{(a)},\vec{v}^{(a)}} = \frac{1}{2} \,\vec{v}^{(a)}\cdot (\vec{w}^{(a)}+\vec{u}^{(a)})
\ee
is the dimension of the moduli space of $U(W)$ instantons on $\mathbb{C}^2/\mathbb{Z}_n$. Note that $c_R$ grows linearly with $\vec{v}^{(a)}$ due to the constraints \eqref{eq:constro} which fix the first Chern class of the instanton bundle in terms of the monodromy $\vec{w}^{(0)}$. The result \eqref{eq:hkdim} is precisely what one expects for a threshold bound state of instanton strings for $\prod_a \mathfrak{u}(W)$, each of which in the infrared is described by an NLSMs whose target space is the moduli space $\mathcal{M}_{\vec{w}^{(a)},\vec{v}^{(a)}}$.

We next turn to the determination of the parity anomaly which is encoded by the difference of central charges $c_L-c_R$. This can be read off from the leading $q$-power of the elliptic genus for $\omega^{KK,(1)}=\dots=\omega^{KK,(r)}=0$, which is given by $q^{\frac{2c_L-2c_R-3n\,r}{24}}$. We find that the parity anomaly can be written in a quite suggestive form:
\bea
\nonumber
c_L-c_R &=& n\, r-\frac{1}{2}\sum_{a,b=1}^{r-1}\left(C^{\widehat{A}_{r-1}}\right)^{(a)(b)}\left(\vec{v}^{(a)}\cdot {C}^{\widehat{A}_{n-1}}\cdot\vec{v}^{(b)}\right)
\\
\nonumber
&=& n\, r-\frac{1}{2}\sum_{a=0}^{r-1}\sum_{j=0}^{n-1}\left(v^{(a)}_j-v^{(a)}_{j+1}-v^{(a+1)}_j+v^{(a+1)}_{j+1}\right)^2,
\\
\nonumber
&=& \sum_{a=0}^{r-1}\sum_{j=0}^{n-1}\left[1-\frac{1}{2}\left(\delta_{\vec{w}^{(a+1)},j}-\delta_{\vec{w}^{(a)},j}\right)^2\right],
\eea
where we take $v^{(0)}_j=v^{(a)}_j=0$. The expression in the last line makes it clear that $c_L-c_R$ is fixed by the choice of the $\vec{w}^{(a)}$ and is the same for any bound state of D3 branes.\\

Turning now to the global symmetries of the CFT, a novel feature of the theories $\QQ$ is that they are \emph{relative} theories. In particular, through their coupling to the $\mathfrak{su}(n)$ current algebra they are sensitive to the choice of monodromies $\omega^{KK,(1)}, \dots, \omega^{KK,(r)},$ for the M5 branes. Under a modular $S$ transformation, the characters of $\mathfrak{su}(n)_1$ transform as follows:
\be
\widehat{\chi}^{\mathfrak{su}(n)_1}_j\left(\frac{\vec\xi}{\tau},-\frac{1}{\tau}\right) = e^{\frac{1}{2}\frac{2\pi i}{\tau}\vec{\xi}\cdot (C^{\widehat{A}_{n-1}})^{-1}\cdot\vec{\xi}}\sum_{k=0}^{n-1}\mathcal{S}_{jk} \widehat{\chi}_k^{\mathfrak{su}(n)_1}(\vec\xi,{\tau}),
\label{eq:sutr}
\ee
where
\be
\mathcal{S}_{jk} = e^{\frac{2\pi i}{n} j\, k} \qquad\text{ for } j,k=0,\dots,n-1.
\ee
As a consequence, under a modular transformation the elliptic genera for different choices of monodromy $\boldsymbol{\omega}^{KK}$ transform into one another:
\bea
\nonumber
&&\mathbb{E}^{\vec{\boldsymbol{w}},\boldsymbol{\omega}^{KK}}_{\vec{\boldsymbol{v}}}\left(\frac{\vec{\boldsymbol\xi}}{\tau},\frac{\epsilon_+}{\tau},\frac{\epsilon_-}{\tau},\frac{\underline{\boldsymbol{s}}}{\tau},-\frac{1}{\tau}\right)
=
(-i\tau)^{-\frac{r}{2}}\times
\\
\nonumber
&&
\sum_{\boldsymbol{\upsilon}^{KK}=(\upsilon^{KK,(1)},\dots,\upsilon^{KK,(r)})\in \mathbb{Z}_n^r}
\hspace{-.7in}
e^{\frac{2\pi i}{\tau}f^{\vec{\boldsymbol{w}}}_{\vec{\boldsymbol{v}}}(\vec{\boldsymbol\xi},\epsilon_+,\epsilon_-,\underline{\boldsymbol{s}})}
\left(\prod_{a=1}^r \mathcal{S}_{\omega^{KK,(a)}\upsilon^{KK,(a)}}\right)
\mathbb{E}^{\vec{\boldsymbol{w}},\boldsymbol{\upsilon}^{KK}}_{\vec{\boldsymbol{v}}}(\vec{\boldsymbol\xi},\epsilon_+,\epsilon_-,\underline{\boldsymbol{s}},\tau).\\
\eea
Note that the elliptic genus transforms with modular weight $-\frac{r}{2}$ due to the $r $ copies of the Heisenberg algebra which are decoupled from the other degrees of freedom but we have nonetheless chosen to include in the definition of the theory $\QQ$. The global symmetry anomalies on the string worldsheet give rise to a phase which is captured by the following quadratic polynomial:
\be
\resizebox{\textwidth}{!}{$\displaystyle{
f^{\vec{\boldsymbol{w}}}_{\vec{\boldsymbol{v}}}(\vec{\boldsymbol\xi},\epsilon_+,\epsilon_-,\underline{\boldsymbol{s}})= \frac{1}{2}\, \sum_{a=1}^r\vec{\xi}^{(a)} \cdot (C^{\widehat{A}_{n-1}})^{-1}\cdot\vec{\xi}^{(a)}+k_x\epsilon_-^2+k_t\epsilon_+^2+\sum_{a=0}^{r}\sum_{j=0}^{n-1}\frac{k_{\mathfrak{u}(w^{(a)}_j)}}{2}\sum_{K=1}^{w^{(a)}_j}(s^{(a)}_{j,K})^2,
}$}
\ee
where we find that the levels for $U(1)_x$ , $SU(2)_t$, and $\mathfrak{u}(w^{(a)}_j)$ are given respectively by:
\bea
k_x
&=&
-\frac{1}{2}\sum_{j=0}^{n-1}\boldsymbol{v}_j\cdot{C}^{{A}_{r-1}}\cdot\boldsymbol{v}_{j+1},
\\
\nonumber
k_t
&=&
- \sum_{a=1}^{r-1}\vec{v}^{(a)}\cdot \vec{w}^{(a)}+\frac{1}{2}\sum_{a=1}^{r-1} \vec{v}^{(a)}\cdot C^{\widehat{A}_{n-1}}\cdot \vec{v}^{(a)} + \frac{1}{2} \sum_{j=0}^{n-1} \boldsymbol{v}_j\cdot C^{{A}_{r-1}}\cdot\boldsymbol{v}_j
\\
&=&
-\sum_{a=1}^{r-1}\dim_{\mathbb{H}}\mathcal{M}_{\vec{w}^{(a)},\vec{v}^{(a)}}+ \frac{1}{2} \sum_{j=0}^{n-1} \boldsymbol{v}_j\cdot C^{{A}_{r-1}}\cdot\boldsymbol{v}_j,
\\
k_{\mathfrak{u}(w^{(a)}_j)}
&=&
\begin{cases}
-(C^{A_{r-1}}\cdot\boldsymbol{v}_j)^{(a)}\quad&\text{ for }a=1,\dots,r-1\\
v_j^{(1)}\quad&\text{ for }a=0\\
v_j^{(r-1)}\quad&\text{ for }a=r
\end{cases}
.
\eea
\newline
\subsection{Examples}
\label{sec:examples}
In this section we work out explicitly the expressions for the elliptic genus of BPS strings for various classes of examples.

\subsubsection{One string on $\mathbb{C}^2/\mathbb{Z}_n$, for $\mathcal{T}_{2,W}$.}
Let us first consider the case where the 6d gauge and global symmetries possess trivial monodromy: $\vec{w}^{(0)}=\vec{w}^{(1)}=\vec{w}^{(2)}=(W,0,\dots,0)$. 
The lowest possible instanton charge is zero, corresponding to $\vec{v}^{(1)}=0$. In this case,
\be
\mathbb{E}^{\vec{\boldsymbol{w}},\boldsymbol{\omega}^{KK}}_{\vec{\boldsymbol{0}}}(\vec{\xi},\epsilon_+,\epsilon_-,\underline{\boldsymbol{s}},\tau)
=
\frac{1}{\eta(\tau)^2}
\widehat{\chi}^{\mathfrak{su}(n)_2}_{\omega^{KK,(2)}}(\vec\xi^{(1)},\tau)
\widehat{\chi}^{\mathfrak{su}(n)_2}_{\omega^{KK,(2)}}(\vec\xi^{(2)},\tau),
\ee
where
\be
\xi_j^{(1)}=\xi_j^{(2)}=\xi_j+(1-n\delta_{j,0})\mathfrak{m}
\ee
due to the Stuckelberg constraint \eqref{eq:stucondition}. The instanton charge 1 configuration corresponds to $\vec{v}^{(1)}=(1,\dots,1)$. In this case, by imposing the condition \eqref{eq:omcon} one finds that the elements of $\mathcal{Y}_{\vec{w}^{(1)},\vec{v}^{(1)}}$ are $W$-tuples of Young diagrams of the form
\be
\left(\emptyset,\dots,\emptyset,
\resizebox{.1\textwidth}{!}{
\begin{ytableau}
       ~ & \none & \none & \none\\
       \vdots & \none & \none & \none\\
        & \none & \none & \none \\
       & \cdots  &
\end{ytableau}
}
\!\!\!\!\!\!
,\emptyset,\dots,\emptyset\right),
\ee
where only one Young diagram (the $A$-th, say) is nonempty and has nonzero components 
\bea
\nonumber
Y_{A,0}&=&\ell,\\
Y_{A,i}&=&1\quad\text{for } i=1,\dots,n-\ell.
\eea
Keeping $A$ fixed and summing over $\ell$ gives rise to the following terms in the elliptic genus:
\be
\frac{\displaystyle{\prod_{K=1}^{W}\theta_1(s^{(2)}_{0,K}-s^{(1)}_{0,A}-\epsilon_+)\theta_1(s^{(1)}_{0,A}-s^{(0)}_{0,K}+\epsilon_+)}}{\displaystyle{\prod_{\substack{K=1\\K\neq A}}^{W}\theta_1(s^{(1)}_{0,K}-s^{(1)}_{0,A})\theta_1(2\epsilon_++s^{(1)}_{0,A}-s^{(1)}_{0,K})}}
\mathbb{E}_{\mathbf{ALE}_n}(\vec\xi,\epsilon_+,\epsilon_-,m,\tau).
\ee
Here,
\be
\mathbb{E}_{\mathbf{ALE}_n}(\vec\xi,\epsilon_+,\epsilon_-,m,\tau)=
-
\sum_{j=1}^{n}
\frac{
\widehat{\chi}^{\mathfrak{su}(n)_1}_{\omega^{KK,(a)}}(\vec{\xi}^{(1)}[j])
\widehat{\chi}^{\mathfrak{su}(n)_1}_{\omega^{KK,(a+1)}}(\vec{\xi}^{(2)}[j])}
{\eta^2\theta_1(\epsilon_+[j]+\epsilon_-[j])\theta_1(\epsilon_+[j]-\epsilon_-[j])}.
\ee
where
\bea
\xi^{(1)}_{k}[j] &=& \xi_k+(n\delta_{k,0}-1)\mathfrak{m}+\delta_{k,j}\epsilon_1[j]+\delta_{k,j-1}\epsilon_2[j],\\
\xi^{(2)}_{k}[j] &=& \xi_k+(n\delta_{k,0}-1)\mathfrak{m}-\delta_{k,j}\epsilon_1[j]-\delta_{k,j-1}\epsilon_2[j],
\eea
and
\be
\epsilon_+[j]=\epsilon_+,\qquad \epsilon_-[j] = n\,\epsilon_-+(n+1-2j).
\label{eq:epj}
\ee
Summing over different $A$, we get:
\be
\mathbb{E}^{\vec{\boldsymbol{w}},\boldsymbol{\omega}^{KK}}_{\vec{\boldsymbol{v}}}(\vec{\xi},\epsilon_+,\epsilon_-,\underline{\boldsymbol{s}},\tau)
=
\widetilde{\mathbb{E}}_{\boldsymbol{{v}}}(\epsilon_+,\epsilon_-,\underline{\boldsymbol{s}},\tau)\mathbb{E}_{\mathbf{ALE}_n}(\vec\xi,\epsilon_+,\epsilon_-,m,\tau),
\ee
where 
\bea
\nonumber
\widetilde{\mathbb{E}}_{\boldsymbol{{v}}}(\epsilon_+,\epsilon_-,\underline{\boldsymbol{s}},\tau)
&=&
\sum_{A=1}^{W}\frac{\displaystyle{\prod_{K=1}^{W}\theta_1(s^{(2)}_{0,K}-s^{(1)}_{0,A}-\epsilon_+)\theta_1(s^{(1)}_{0,A}-s^{(0)}_{0,K}+\epsilon_+)}}{\eta^2\displaystyle{\prod_{\substack{K=1\\K\neq A}}^{W}\theta_1(s^{(1)}_{0,K}-s^{(1)}_{0,A})\theta_1(2\epsilon_++s^{(1)}_{0,A}-s^{(1)}_{0,K})}}
\\
&=&\left(\frac{\eta^2}{\theta_1(\epsilon_++\epsilon_-)\theta_1(\epsilon_+-\epsilon_-)}\right)^{-1}\mathbb{E}_{\boldsymbol{{v}}}(\epsilon_+,\epsilon_-,\underline{\boldsymbol{s}},\tau)
\eea
is the elliptic genus of one $\mathfrak{su}(W)$ instanton string of theory $\mathcal{T}_{2,W}$ on $\mathbb{C}^2$, stripped of its center of mass degrees of freedom. On the other hand, $\mathbb{E}_{\mathbf{ALE}_n}(\vec\xi,\epsilon_+,\epsilon_-,m,\tau)$ encodes the current algebra and center-of-mass degrees of freedom on $\mathbf{ALE}_n$.\newline

Let us now consider an example with a more general choices of monodromies for the 6d gauge symmetry. We take the theory $\mathcal{T}^{6d}_{2,2}$ with monodromies:
\be
\vec{w}^{(0)}=\vec{w}^{(1)} =\vec{w}^{(2)} = (1,1),
\ee
and consider the one-string sector with $v^{(1)}= (1,1)$. The set $\mathcal{Y}_{\vec{w}^{(a)},\vec{v}^{(a)}}$ has five components:
\be
\left(
\resizebox{.067\textwidth}{!}{
\begin{ytableau}
       ~ & ~\\
\end{ytableau}
}
,\emptyset\right),
\quad
\left(
\resizebox{.033\textwidth}{!}{
\begin{ytableau}
       ~ \\
       ~ \\
\end{ytableau}
}
,\emptyset\right),
\quad
\left(\emptyset,
\resizebox{.067\textwidth}{!}{
\begin{ytableau}
       ~ & ~\\
\end{ytableau}
}
\right),
\quad
\left(\emptyset,
\resizebox{.033\textwidth}{!}{
\begin{ytableau}
       ~ \\
       ~ \\
\end{ytableau}
}
\right),
\quad
\left(
\resizebox{.033\textwidth}{!}{
\begin{ytableau}
       ~ \\
\end{ytableau}
},
\resizebox{.033\textwidth}{!}{
\begin{ytableau}
       ~ \\
\end{ytableau}
}
\right).
\ee
It is convenient to parametrize the chemical potentials in the following way: $s^{(a)}_{j,1} = \frac{1}{2}S^{(0)}+a\, \mathfrak{m}+(2j-1)\zeta^{(a)}$. We find that the elliptic genus is given by:
\bea
\nonumber
&&\mathbb{E}^{\vec{\boldsymbol{w}},\boldsymbol{\omega}^{KK}}_{\vec{\boldsymbol{v}}}(\vec{\xi},\epsilon_+,\epsilon_-,\underline{\boldsymbol{s}},\tau)
=\eta^{-2}\bigg[
\\
\nonumber
&&
\sum_{\substack{s_1=\pm\\s_2=\pm}}
\frac{
\theta_1(\mathfrak{m}+\epsilon_++s_2(\zeta^{(1)}-\zeta^{(0)}))
\theta_1(\mathfrak{m}-\epsilon_++s_2(\zeta^{(2)}-\zeta^{(1)}))
}
{\theta_1(2s_1\epsilon_-)
\theta_1(2\epsilon_+ + 2s_1\epsilon_-)}
\\
\nonumber
&&
\times
\frac{
\theta_1(\mathfrak{m}+2\epsilon_++s_1\epsilon_-+s_2(\zeta^{(0)}+\zeta^{(1)}))
\theta_1(\mathfrak{m}-2\epsilon_+-s_1\epsilon_--s_2(\zeta^{(1)}+\zeta^{(2)}))
}{
\theta_1(\epsilon_+ + s_1\epsilon_-+2s_2\zeta^{(1)})
\theta_1(3\epsilon_+ + s_1\epsilon_-+2s_2\zeta^{(1)})
}
\\
\nonumber
&&
\times
\widehat{\chi}^{\mathfrak{su}(2)_1}_{\omega^{KK,(1)}}(\xi_1\!-\!2\epsilon_+\!-\!2s_1\epsilon_-\!+\!s_2(\zeta^{(0)}\!-\!\zeta^{(1)}))
\widehat{\chi}^{\mathfrak{su}(2)_1}_{\omega^{KK,(2)}}(\xi_1\!+\!2\epsilon_+\!+\!2s_1\epsilon_-\!+\!s_2(\zeta^{(1)}\!-\!\zeta^{(2)}))
\\
&&
\nonumber
+\prod_{s_1=\pm}
\frac{
\theta_1(\mathfrak{m}+\epsilon_++s_1(\zeta^{(0)}-\zeta^{(1)}))
\theta_1(\mathfrak{m}-\epsilon_++s_1(\zeta^{(1)}-\zeta^{(2)}))}
{
\theta_1(\epsilon_++\epsilon_-+2s_1\zeta^{(1)}))
\theta_1(\epsilon_+-\epsilon_-+2s_1\zeta^{(1)}))
}
\\
&&
\times
\widehat{\chi}^{\mathfrak{su}(2)_1}_{\omega^{KK,(1)}}(\xi_1+\zeta^{(0)}+3\zeta^{(1)})
\widehat{\chi}^{\mathfrak{su}(2)_1}_{\omega^{KK,(2)}}(\xi_1-\zeta^{(2)}-3\zeta^{(1)})\bigg].
\eea

\subsubsection{Frozen strings on $\mathbb{C}^2/\mathbb{Z}_n$}
One of the noteworthy features of the theories $\mathcal{T}^{6d}_{r,W}$ is the existence of BPS strings which are pinned at the $\mathbf{ALE}_n$ singularity. This is reflected in the absence of a center-of-mass factor in the elliptic genus. Such a factor would comprise bosonic terms in the denominator, which are charged with respect to $SU(2)_t\times U(1)_x$ but not with respect to the 6d gauge symmetries. From the explicit form of the integrand of the elliptic genus, equation \eqref{eq:eg}, one can see that such terms cannot arise unless all $v^{(a)}_j \geq 1$ for some $a$, in which case a bound state of $(1,1,\dots,1)$ fractional D3 branes may be moved away from the singularity.

\begin{figure}
    \centering
    \includegraphics[scale=0.65]{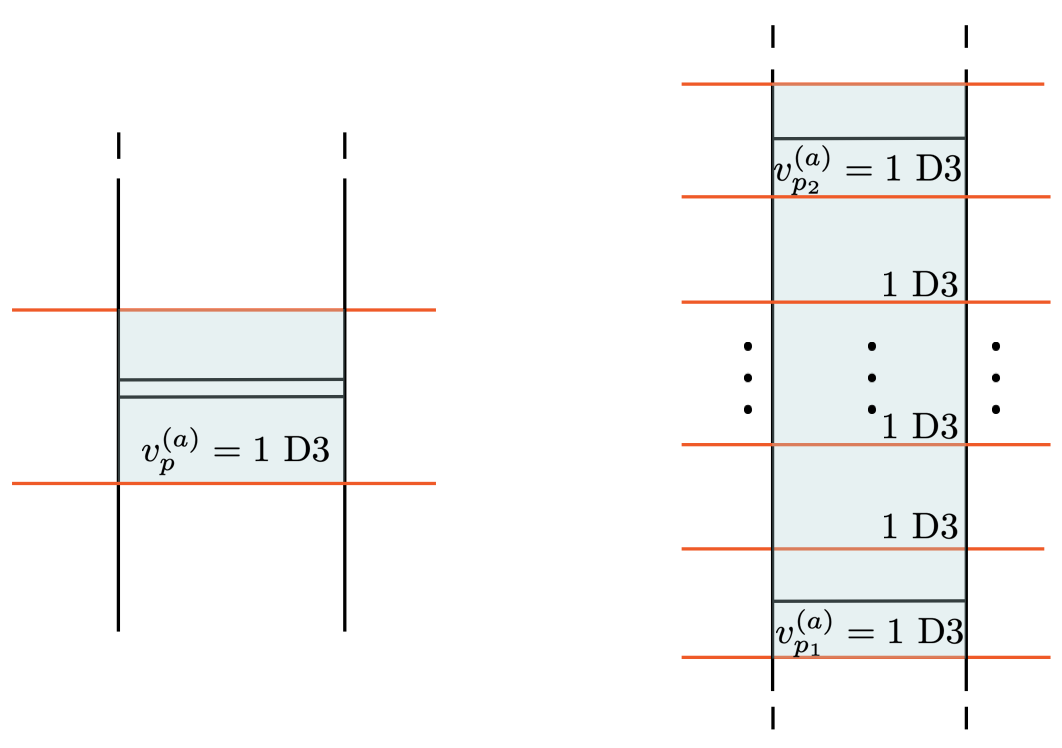}
    \caption{Type IIB realization of the classes of frozen strings with $\dim_{\mathbb{H}}\mathcal{M}_{\vec{w}^{(a)},\vec{v}^{(a)}}=1$. The only plaquettes supporting D3 branes are the ones shown in the figure. Additional horizontal D5-brane segments whose worldvolume does not intersect that of the D3 branes are not shown in the figure.}
    \label{fig:mod1}
\end{figure}

Even if the degrees of freedom associated to moving in $\mathbf{ALE}_n$ are absent, frozen strings do retain internal degrees of freedom. In this section we look at a few simple examples. Recall that the right-moving central charge of a string is given by
\be
c_R = 6\dim_{\mathbb{H}}\mathcal{M}_{\vec{w}^{(a)},\vec{v}^{(a)}} = 3 \,\vec{v}^{(a)}\cdot (\vec{w}^{(a)}+\vec{u}^{(a)}).
\ee
\paragraph{Elliptic genera for frozen BPS string instantons with $c_R = 6$}. The simplest nontrivial classes of frozen strings are those for which $c_R = 6$, corresponding to an instanton moduli space of quaternionic dimension 1. In fact, in quaternionic dimension one, the moduli space must be diffeomorphic to an ALE space, although this cannot be the geometric $\mathbf{ALE}_n$ space for a frozen string. We will confirm this picture below. One can have $\dim_{\mathbb{H}}\mathcal{M}_{\vec{w}^{(a)},\vec{v}^{(a)}} =1 $ if a given vector $\vec{v}^{(a)}$ is non-vanishing while all other vectors are identically zero. It is easy to see that only the following two configurations are possible:
\be
\vec{v}^{(a)}_j = \delta_{j,p},\qquad w^{(a)}_p=2,\qquad w^{(a\pm1)}_p=0,
\ee
and 
\be
\vec{v}^{(a)}_{j\text{ mod }n}
=
\begin{cases}
1 & p_1\leq j\leq p_2\\
0 & \text{otherwise}
\end{cases}
,
\ee
where $0\leq p_1\leq n-1$, $p_1< p_2 <p_1+n$, and the monodromies are given by $w^{(a)}_{p_1}=w^{(a)}_{p_2\text{ mod }n}=1$, $w^{(a)}_{j\text{ mod }n}=0$ for $p_1<j<p_2$, and $w^{(a\pm1)}_{j\text{ mod }n}=0$ for $p_1\leq j\leq p_2$.
The Type IIB dual picture for these two cases are depicted respectively on the left- and right-hand side of figure \ref{fig:mod1}.

In the first case, the elliptic genus is given simply by:
\be
\mathbb{E}^{\vec{\boldsymbol{w}},\boldsymbol{\omega}^{KK}}_{\vec{\boldsymbol{v}}}=
\sum_{s_1=\pm}\frac{1}{\theta_1(s_1(s^{(a)}_{j,1}-s^{(a)}_{j,2}))\theta_1(2\epsilon_++s_1(s^{(a)}_{j,1}-s^{(a)}_{j,2}))}\widehat{\chi}^{\mathfrak{su}(n)_1}_{\omega^{KK,(1)}}(\vec{\xi}^{(1)}_{s_1})\widehat{\chi}^{\mathfrak{su}(n)_1}_{\omega^{KK,(2)}}(\vec{\xi}^{(2)}_{s_1}),
\ee
where the arguments $\vec{\xi}^{(a)}_{\pm}$ of the $\mathfrak{su}(n)_1$ characters depend on the specific choices of monodromies $\vec{\boldsymbol{w}}$ and of chemical potentials. If we take $\omega^{KK,(1)}=\omega^{KK,(2)}=0$ and focus on the lowest energy states in the elliptic genus, we find:
\be
\mathbb{E}^{\vec{\boldsymbol{w}},\boldsymbol{0}}_{\vec{\boldsymbol{v}}}=q^{-1/6}t \mathcal{H}_{\mathbf{ALE}_2}\left(t,s^{(a)}_{j,2}/s^{(a)}_{j,1}\right)+\mathcal{O}(q^{5/6}),
\ee
where
\be
\mathcal{H}_{\mathbf{ALE}_2}\left(t,x\right)=
\frac{1+t^2}{\left(1-t^2x^2\right)\left(1-t^2x^{-2}\right)}
\ee
denotes the Hilbert series of $\mathbf{ALE}_2$. Notice that the $SU(2)$ isometry of the $\mathbf{ALE}_2$ space is distinct from the geometric $\mathbf{ALE}_n$, and indeed is $SU(2)$ isometry couples to the chemical potentials for the 6d gauge symmetries.

In the second case, the elliptic genus picks residues labeled by $W$-tuples of Young diagrams $\underline{Y}^{(a)}$ with two non-zero entries $Y^{(a)}_A$ and $Y^{(a)}_B$ corresponding to $\omega_A^{(a)} = p_1 $ and $\omega_B^{(a)} = p_2 $:
\be
Y^{(a)}_A =
\begin{ytableau}
       ~ &\cdots &~\\
\end{ytableau},
\qquad
Y^{(a)}_B
=
\begin{ytableau}
       ~ \\
       \vdots\\
       \\
\end{ytableau}
\ee
The residues are labeled by the number of boxes $\ell\in\{0,\dots,p_2-p_1+1\}$ in diagram $Y^{(a)}_A$, while $Y^{(a)}_B$ contains $p_2-p_1+1-\ell$ boxes. The elliptic genus for this configuration is given by:
\be
\mathbb{E}^{\vec{\boldsymbol{w}},\boldsymbol{\omega}^{KK}}_{\vec{\boldsymbol{v}}}=
\sum_{j=1}^{\widetilde{n}}\frac{1}{\theta_1(\epsilon_+[j]+\widetilde{\epsilon}_-[j])\theta_1(\epsilon_+[j]-\widetilde{\epsilon}_-[j])}\widehat{\chi}^{\mathfrak{su}(n)_1}_{\omega^{KK,(1)}}(\vec{\xi}^{(1)}_{j})\widehat{\chi}^{\mathfrak{su}(n)_1}_{\omega^{KK,(2)}}(\vec{\xi}^{(2)}_{j}),
\ee
where $\widetilde{n}=p_2-p_1+2$, $\epsilon_+[j]$ is given in equation \eqref{eq:epj} and
\be
\widetilde{\epsilon}_-[j] = \widetilde{n}\,\widetilde{\epsilon}_- +(\widetilde{n}+1-2j)\epsilon_+.
\ee
Here, we define
\be
\widetilde{\epsilon}_- = \epsilon_-+\frac{s^{(a)}_{p_2,1}-s^{(a)}_{p_2,1}-2\epsilon_-}{\widetilde{n}}.
\label{eq:epti}
\ee
The $\vec{\xi}^{(a)}_{\ell}$ of the $\mathfrak{su}(n)_1$ characters depend on the specific choices of monodromies $\vec{\boldsymbol{w}}$ and of chemical potentials $s^{(a)}_{j,k}$. We find that the frozen string configuration has a moduli space of vacua which is diffeomorphic to the space $\mathbf{ALE}_{\widetilde{n}}$, where the $SU(2)$ isometry of the ALE space couples to the combination \eqref{eq:epti} of chemical potentials. Indeed, taking trivial monodromies $\omega^{KK,(a)}=0$ and focusing on the ground states, we find:
\be
\mathbb{E}^{\vec{\boldsymbol{w}},\boldsymbol{0}}_{\vec{\boldsymbol{v}}} = q^{-1/6}t\mathcal{H}_{\mathbf{ALE}_{\widetilde{n}}}(\epsilon_+,\widetilde\epsilon_-)+\mathcal{O}(q^{5/6}),
\ee
where 
\be
\mathcal{H}_{\mathbf{ALE}_{n}}(\epsilon_+,\epsilon_-) = \frac{1-t^{2n}}{(1-t^2)(1-t^n x^n)(1-t^nx^{-n})}
\ee
is the Hilbert series of $\mathbf{ALE}_{n}$. This is similar to what was found in \cite{Dey:2013fea} for certain related classes of instantons in pure $SU(n)$ supersymmetric gauge theory. 

\paragraph{Elliptic genera for frozen BPS string instantons with $c_R = 12$.} We conclude this section by giving more nontrivial examples of frozen strings, corresponding to the case $\dim_{\mathbb{H}}\mathcal{M}_{\vec{w}^{(a)},\vec{v}^{(a)}} = 2$. In our first example, we consider the theory $\mathcal{T}^{6d}_{2,W}$ and take $\vec{w}^{(0)}=\vec{w}^{(2)}=(2,W-2),$ $\vec{w}^{(1)}=(0,W)$, and  $\vec{v}^{(1)}=(0,1)$. We find a very simple formula for the elliptic genus:
\be
\mathbb{E}^{\vec{\boldsymbol{w}},\boldsymbol{\omega}^{KK}}_{\vec{\boldsymbol{v}}}
\!=\!
-\!
\sum_{K=1}^{W}
\frac{
\displaystyle{
\prod_{L=1}^{W-2}
\theta_1(\epsilon_+\!+\!s^{(1)}_{1,K}\!-\!s^{(0)}_{1,L})
\theta_1(-\epsilon_+\!+\!s^{(1)}_{1,K}\!-\!s^{(2)}_{1,L})}}
{
\displaystyle{
\prod_{\substack{L=1\\L\neq K}}^W
\theta_1(s^{(1)}_{1,K}-s^{(1)}_{1,L})
\theta_1(2\epsilon_++s^{(1)}_{1,K}-s^{(1)}_{1,L})}
}
\widehat{\chi}^{\mathfrak{su}(2)_1}_{\omega^{KK,(1)}}(\vec{\xi}^{(1)}_K)
\widehat{\chi}^{\mathfrak{su}(2)_1}_{\omega^{KK,(1)}}(\vec{\xi}^{(2)}_K),
\label{eq:egex}
\ee
where
\bea
\vec{\xi}^{(1)}_K &=& (\xi_1+\mathfrak{m}+2\epsilon_++2s^{(1)}_{1,K}+S^{(0)}_{1}-S^{(1)}_{1}),
\\
\vec{\xi}^{(2)}_K &=& (\xi_1+\mathfrak{m}-2\epsilon_+-2s^{(1)}_{1,K}+S^{(1)}_{1}-S^{(2)}_{1}).
\eea
Although it is not immediately obvious from the form of equation \eqref{eq:egex}, the poles at $s^{(1)}_{K}=s^{(1)}_L$  are fictitious due to cancelations between the different summands.

In our final example, we consider a bound state between two strings coupled to different tensor multiplets in the theory $\mathcal{T}^{6d}_{3,2}$ on $\mathbb{C}^2/\mathbb{Z}_n$. We choose
\bea
\vec{w}^{(0)}&=&\vec{w}^{(0)}=\vec{w}^{(1)}=\vec{w}^{(3)}=(2,0)\\
\vec{w}^{(2)}&=&(0,2)
\eea
and
\bea
\vec{v}^{(1)}&=&(1,1)\\
\vec{v}^{(0)}&=&(0,1).
\eea
We find the following expression for the elliptic genus:
\bea
\nonumber
&&
\mathbb{E}^{\vec{\boldsymbol{w}},\boldsymbol{\omega}^{KK}}_{\vec{\boldsymbol{v}}}
=
\sum_{s_1,s_2,s_3=\pm}\frac{\widehat\chi^{\mathfrak{su}(2)_1}_{\omega^{KK,(1)}}(\xi_1+\mathfrak{m}+2\epsilon_++2s_1\epsilon_-)}{\eta\,\theta_1(2s_1\epsilon_-)\theta_1(2\epsilon_++2s_1\epsilon_-)}
\\
\nonumber
&&
\times
\widehat\chi^{\mathfrak{su}(2)_1}_{\omega^{KK,(2)}}(\xi_1+\mathfrak{m}-2s_1\epsilon_- +2s_3\zeta^{(2)})
\widehat\chi^{\mathfrak{su}(2)_1}_{\omega^{KK,(3)}}(\xi_1+\mathfrak{m}-2\epsilon_+ -2s_3\zeta^{(2)})
\\
\nonumber
&&
\times
\frac{
\theta_1(\mathfrak{m}+\epsilon_++(2+s_2)\zeta^{(1)}-\zeta^{(0)})
\theta_1(\mathfrak{m}+\epsilon_++(2+s_2)\zeta^{(1)}-3\zeta^{(0)})
}{
\theta_1(2s_2\zeta^{(1)})
\theta_1(2s_3\zeta^{(2)})
}
\\
\nonumber
&&
\times
\frac{
\theta_1(\mathfrak{m}\!+\!s_1\epsilon_-\!-\!(2\!+\!s_2)\zeta^{(1)}\!\!+\!(2\!-\!s_2s_3)\zeta^{(2)})
\theta_1(\mathfrak{m}\!-\!2\epsilon_+\!-\!s_1\epsilon_-\!-\!(2\!+\!s_2)\zeta^{(1)}\!\!+\!(2\!-\!s_3)\zeta^{(2)})
}{
\theta_1(2\epsilon_++2s_2\zeta^{(1)})
\theta_1(2\epsilon_++2s_3\zeta^{(2)})
}
.
\\
\eea
Here, we have expressed the chemical potentials for the 6d gauge symmetries as:
\bea
s^{(a)}_{0,k} &=& \frac{1}{2}S^{(0)}+\mathfrak{m}^a+(2k-1)\zeta^{(a)}\qquad\text{for }a=0,1,3;\\
s^{(a)}_{1,k} &=& \frac{1}{2}S^{(0)}+\mathfrak{m}^a+(2k-1)\zeta^{(a)}\qquad\text{for }a=2.
\eea
These results are the starting point to analyze the physics of the resulting BPS string CFTs along the lines of \cite{DelZotto:2018tcj}. This type of detailed study is beyond the scope of this paper and is left to future work.

\section*{Acknowledgments}
We are grateful to Du Pei, Vivek Shende, Kaiwen Sun, Richard Szabo, and especially Wei Gu and Albrecht Klemm for helpful discussions.  We would like to thank the Institut Mittag-Leffler for hospitality during the workshop ``Enumerative Invariants, Quantum Fields and String Theory Correspondences'' in June 2022 where part of this research was conducted. The work of MDZ and GL has received funding from the European Research Council (ERC) under the European Union’s Horizon 2020 (grant agreement No. 851931) and Horizon Europe (grant agreement No. 101078365) research and innovation programs. MDZ also acknowledges support from the Simons Foundation (grant \#888984, Simons Collaboration on Global Categorical Symmetries).

\appendix

\section{Derivation of the bound on $v^{(a)}_j$}\label{sec:TXapp}

In this appendix we derive the bound in equation \eqref{eq:deltacondition}. Using
\be
\vert\vec\delta^{(a)}\vert=\sum_j \delta^{(a)}_j = 0
\label{eq:traceless}
\ee
and defining
\be
\norm{\vec{\delta}^{(a)}}^2=\sum_{j=0}^{n-1}(\delta^{(a)}_j)^2,
\ee
one has
\be
2n \norm{\vec{\delta}^{(a)}}^2 = \sum_{j,k=0}^{n-1}(\delta^{(a)}_j-\delta^{(a)}_k)^2 = \sum_{j=1}^{n-1}\sum_{k=j}^{n-1}(\eta^{(a)}_j+\dots+\eta^{(a)}_k)^2,
\ee
where $\eta^{(a)}_j = -(\widehat{C}\cdot v^{(a)})_j$. By the Cauchy-Schwarz inequality, one then finds
\be
2 \norm{\vec{\delta}^{(a)}}^2 \leq\sum_{j=1}^{n-1}\sum_{k=j}^{n-1}((\eta^{(a)}_j)^2+\dots+(\eta^{(a)}_k)^2)=\sum_{j=1}^{n-1}j(n-j)(\eta^{(a)}_j)^2.
\ee
Now, since $w^{(0)}_j-W\leq \eta^{(a)}_j\leq w^{(0)}_j$ one has $(\eta^{(a)}_j)^2\leq \max\left((w^{(0)}_j-W)^2,(w^{(0)}_j)^2\right)$,
which gives the desired bound
\be
\norm{\vec{\delta}^{(a)}}^2 \leq \frac{1}{2} \sum_{j=1}^{n-1}j(n-j)\max\left((W-w^{(0)}_j)^2,(w^{(0)}_j)^2\right).
\ee
which is satisfied by a finite set $\mathfrak{D}_{\vec{w}^{(0)}}$ of $n$-tuples $\vec{\delta}^{(a)} = (\delta^{(a)}_0,\dots,\delta^{(a)}_{n-1})\in \mathbb{Z}^{n}$ (with $\vert\vec{\delta}^{(a)}\vert=0$).\footnote{ \  Alternatively, using $w^{(0)}_j\leq W$ leads to the simpler but weaker bound $\norm{\vec\delta^{(a)}}^2\leq \frac{1}{12}W^2n(n^2-1)$.}

\section{A BPS particle on $S^1\times \mathbb{C}^2/\mathbb{Z}_n$}
\label{sec:bps}

In this appendix we determine the partition function on $S^1\times \mathbb{C}^2/\mathbb{Z}_n$ of a 5d BPS multiplet carrying charges $(0, k_t)$ with respect to the Cartan of $U(1)_x\times SU(2)_t$. We take this particle to carry charges $(-1,+1)$ with respect to a pair of $U(1)$ (possibly background) gauge fields $U(1)^{(a)}_A$ and $U(1)^{(b)}_B$, with Wilson lines $s^{(a)}_A$ and $s^{(b)}_B$. We take the connection for $U(1)^{(a)}_A$ to have monodromy $e^{\frac{2\pi i}{n} \omega^{(a)}_A}$ on the asymptotic boundary of $\mathbb{C}^2/\mathbb{Z}_n$, where $\omega^{(a)}_A\in\mathbb{Z}_n$, and similarly for $U(1)^{(b)}_B$. There are two possible ways to proceed, which we will show lead to equivalent results. One approach is to start from the partition function on $S^1\times\mathbb{C}^2$ with charges $(0,k_t)$ with respect to the Cartan of the $SO(4)\simeq SU(2)_x\times SU(2)_t$ isometry of $\mathbb{C}^2$, which is given by
\be
Z^{BPS,U(1)_A^{(a)},U(1)_B^{(b)}}_{S^1\times\mathbb{C}^2}(\epsilon_+,\epsilon_-,m,k_t)
=
\prod_{i,j=0}^\infty (1-e^{2\pi i(m+s^{(b)}_B-s^{(a)}_A)} t^{i-j+2k_t}x^{i+j+1})^{(-1)^{2k_t}},
\label{eq:zbpsa}
\ee
and project onto states with the appropriate $SU(2)_x$ charge. Recall that $\mathbb{Z}_n$ is embedded into the Cartan of $SU(2)_x$, such that the generator of $\mathbb{Z}_n$ maps to $e^{\frac{2\pi i}{n}}$. As the BPS particle is charged under gauge fields that transform with monodromy $e^{\frac{2\pi i}{n}\omega_{1,2}}$, orbifolding projects onto fields that carry charge $\omega_{AB}^{(a)(b)}=\omega_{A}^{(a)}-\omega_{B}^{(b)} $ mod $n$, leading to:
\be
Z^{BPS,U(1)_A^{(a)},U(1)_B^{(b)}}_{S^1\times{\mathbb{C}^2/\mathbb{Z}_n}}(\epsilon_+,\epsilon_-,m,k_t)
=
\hspace{-.4in}
\prod_{\substack{i,j=0\\i+j+1=\omega_{AB}^{(a)(b)}\text{ mod }n}}^\infty
\hspace{-.4in}
(1-e^{2\pi i(m+s^{(b)}_B-s^{(a)}_A)} t^{i-j+2k_t}x^{i+j+1})^{(-1)^{2k_t}}.
\label{eq:zbpsproj}
\ee

Alternatively, we may resort to the factorization property \cite{Nekrasov:2003vi} of the partition function on the resolved space $S^1\times \widetilde{\mathbb{C}^2/\mathbb{Z}_n}$, which in our case leads to the following product of contributions from the $n$ fixed points of the $\mathbb{Z}_n$ action:
\bea
\nonumber
&&Z^{BPS,U(1)_A^{(a)},U(1)_B^{(b)}}_{S^1\times\widetilde{\mathbb{C}^2/\mathbb{Z}_n}}(\epsilon_+,\epsilon_-,m,k_t)\\
&&
\hskip.5in=
\prod_{\ell=1}^{n}Z^{BPS,U(1)_A^{(a)},U(1)_B^{(b)}}_{S^1\times\mathbb{C}^2}(\epsilon_+[\ell],\epsilon_-[\ell],s^{(a)}_A[\ell],s^{(b)}_B[\ell],m,k_t).
\hskip.5in
\label{eq:blup}
\eea
The equivariant parameters associated by the $\ell$-th fixed point are given by:
\be
\epsilon_+[\ell] = \epsilon_+,\qquad \epsilon_-[\ell] = n\epsilon_- +(n+1-2\ell)\epsilon_+.
\ee
The shifted Coulomb branch parameters, on the other hand, are given by
\bea
\nonumber
s^{(a),(\ell)}_A
&=&
s^{(a)}_A +\epsilon_+[\ell](h^{(a)}_{A,\ell}+h^{(a)}_{A,\ell-1})+\epsilon_-[\ell](h^{(a)}_{A,\ell}-h^{(a)}_{A,\ell-1})\\
&=&
s^{(a)}_A +\epsilon_+((n+2-2\ell)h^{(a)}_{A,\ell}-(n-2\ell)h^{(a)}_{A,\ell-1})+n\,\epsilon_-(h^{(a)}_{A,\ell} - h^{(a)}_{A,\ell-1}),
\eea
where $h^{(a)}_{A,0}=h^{(a)}_{A,n}=0$ and
\be
h^{(a)}_{A,\ell}
=
\sum_{s=1}^{n-1} (C^{A_{n-1}})^{-1}_{\ell\, s} u^{(a)}_{A,s},\qquad \ell=1,\dots,n-1.
\ee
We also define
\be
\delta h^{(b)(a)}_{BA,\ell} = h^{(b)}_{B,\ell}-h^{(a)}_{A,\ell}\in\frac{1}{n}\mathbb{Z}^{n-1}
\ee
and
\be
\delta u^{(b)(a)}_{BA,\ell} = u^{(b)}_{B,\ell}-u^{(a)}_{A,\ell}\in\mathbb{Z}^{n-1}.
\ee

Plugging in these explicit expressions, the right hand side of equation \eqref{eq:blup} can be written as:
\be
\prod_{\ell=1}^{n}
\prod_{i,j=0}^\infty
(1-e^{2\pi i(m+s^{(b)}_B-s^{(a)}_A)} t^{n_1(i,j,\ell,k_t,\delta h^{(b)(a)}_{BA,\ell})}x^{n_2(i,j,\ell,k_t,\delta h^{(b)(a)}_{BA,\ell})})^{(-1)^{2k_t}}.
\label{eq:zbpsorbi}
\ee
where
\bea
\nonumber
n_1(i,j,\ell,k_t,\delta h^{(b)(a)}_{BA,\ell})
&=&
i-j+2k_t+(n+1-2\ell)(i+j+1)\\
\nonumber
&+&(n+2-2\ell)\delta h^{(b)(a)}_{BA,\ell}-(n-2\ell)\delta h^{(b)(a)}_{BA,\ell-1}\\
\nonumber
&=&
i-j+2k_t+(n+1-2\ell)(i+j+1)-\omega^{(b)(a)}_{BA}\\
&+&
n\sum_{k=\ell}^{n-1}u^{(b)(a)}_{BA,k}+2\sum_{k=1}^{\ell-1}k\,u^{(b)(a)}_{BA,k}
\label{eq:n1}
\eea
and
\bea
\nonumber
n_2(i,j,\ell,k_t,\delta h^{(b)(a)}_{BA,\ell})
&=&
n(i+j+1+\delta h^{(b)(a)}_{BA,\ell}-\delta h^{(b)(a)}_{BA,\ell-1})\\
&=&
n\left(i+j+1+\sum_{k=\ell}^{n-1}u^{(b)(a)}_{BA,k}\right) -\omega^{(b)(a)}_{BA}.
\label{eq:n2}
\eea
\newline

In fact, the two expressions \eqref{eq:zbpsproj} and \eqref{eq:zbpsorbi} coincide. We will prove this by showing that there exists a bijection between the factors in the infinite products of the two expressions. Let us start by making several observations: first of all, each factor is associated to a BPS particle of definite charges under the Cartan of $U(1)_x\times SU(2)_t$; note also that in equation \eqref{eq:zbpsproj} there exists at most one choice of $(i,j)$ which corresponds to a given value of the charges; furthermore, in both equations \eqref{eq:zbpsproj} and \eqref{eq:zbpsorbi} the $U(1)_x$ charge automatically equals $-\omega^{(b)(a)}_{BA}$ mod $n$. Let us now restrict to states with a definite $U(1)_x$ charge given by 
\be
n\alpha -\omega^{(b)(a)}_{BA},
\label{eq:nalpha}
\ee
where $\alpha\in\mathbb{Z}_{\geq0}$ or $\alpha\in\mathbb{Z}_{>0}$ depending on whether $\omega^{(b)(a)}_{BA}$ equals $0$ or not. There are exactly  $n\alpha -\omega^{(b)(a)}_{BA}$ factors in equation \eqref{eq:zbpsproj} that carry this $U(1)_x$ charge; the corresponding $U(1)_t$ charges are given by:
\be
{\Large\{}-n \alpha+\omega^{(b)(a)}_{BA}+1+2k_t,\quad -n \alpha+\omega^{(b)(a)}_{BA}+3+2k_t,\quad\dots,\quad n \alpha-\omega^{(b)(a)}_{BA}-1+2k_t{\Large\}}.
\label{eq:projch}
\ee
We will show that for these choices of charges the same factor appears exactly once in equation \eqref{eq:zbpsorbi}. Indeed, the $U(1)_x$ charges appearing in equation \eqref{eq:zbpsorbi} satisfy
\be
n_2(i,j,\ell,k_t,\delta h^{(b)(a)}_{BA,\ell}) = \omega^{(b)(a)}_{BA}\qquad\text{ mod }n.
\ee
Setting
\be
n_2(i,j,\ell,k_t,\delta h^{(b)(a)}_{BA,\ell}) = n\alpha-\omega^{(b)(a)}_{BA}
\ee
corresponds to imposing the constraint
\be
i+j+1 = \alpha-\sum_{k=\ell}^{n-1}u^{(b)(a)}_{BA,k}
\label{eq:ij1}
\ee
on the set of allowed $(i,j,\ell)$, which in particular, for fixed $\ell$, restricts the range of $i$ to 
\be
0\leq i \leq \alpha-\sum_{k=\ell}^{n-1}u^{(b)(a)}_{BA,k}-1,
\label{eq:irange}
\ee
and similarly for $j$. Plugging \eqref{eq:ij1} into equation \eqref{eq:n2}, we see that the corresponding $U(1)_t$ charge is given by:
\be
i-j+2k_t+(n+1-2\ell)\alpha+(2\ell-1)\sum_{k=\ell}^{n-1}u^{(b)(a)}_{BA,k}-\omega^{(b)(a)}_{BA}+2\sum_{k=1}^{\ell-1}k\,u^{(b)(a)}_{BA,k},
\ee
which within the range \eqref{eq:irange} takes each of the values
\bea
\nonumber
{\Large\{}\text{min}_\ell:=&&(n-2\ell)\alpha+2\ell\sum_{k=\ell}^{n-1}u^{(b)(a)}_{BA,k}-\omega^{(b)(a)}_{BA}+2\sum_{k=1}^{\ell-1}k\,u^{(b)(a)}_{BA,k}+2k_t+1,\\
\nonumber
&&(n-2\ell)\alpha+2\ell\sum_{k=\ell}^{n-1}u^{(b)(a)}_{BA,k}-\omega^{(b)(a)}_{BA}+2\sum_{k=1}^{\ell-1}k\,u^{(b)(a)}_{BA,k}+2k_t+3,\\
\nonumber
&&\dots\\
\nonumber
\text{max}_\ell:=&&(n+2-2\ell)\alpha+(2\ell-2)\sum_{k=\ell}^{n-1}u^{(b)(a)}_{BA,k}-\omega^{(b)(a)}_{BA}+2\sum_{k=1}^{\ell-1}k\,u^{(b)(a)}_{BA,k}+2k_t-1{\Large\}},\\
&&
\label{eq:lset}
\eea
exactly once. We have denoted  by $\text{min}_{\ell}$ and $ \text{max}_{\ell}$, respectively, the lowest and highest $U(1)_R$ charge that appear at fixed $\ell$. It is straightforward to check explicitly that
\bea
\nonumber
\text{min}_{n} &=& -n \alpha+\omega^{(b)(a)}_{BA}+1+2k_t,\\
\text{max}_{1} &=& n \alpha-\omega^{(b)(a)}_{BA}-1+2k_t,\\
\text{min}_{\ell-1} &=&\text{max}_{\ell} +2.
\eea
From this it follows that the union of the sets \eqref{eq:lset} for $\ell=1,\dots,n$ coincides with the set of all possible charges \eqref{eq:projch} that appear in equation \eqref{eq:zbpsproj}.
In other words, the factors that appears in equation \eqref{eq:zbpsproj} are in one-to-one correspondence with the factors that appear in equation \eqref{eq:zbpsorbi}, which proves equality between the two expressions for the partition function.

\section{Partition functions of the 5d abelian quiver gauge theory}
\label{sec:5dabapp}
In this appendix we derive expressions for the $S^1\times \mathbb{C}^2$ and $S^1\times\mathbb{C}^2/\mathbb{Z}_n$ partition functions of the 5d abelian quiver gauge theory of figure \ref{fig:5dquiv}, which arises from the circle compactification of one M5 brane probing the $\mathbf{TN}_{W}$ space. Conjecturally, these partition functions coincide respectively with the M5 brane's partition functions on $T^2\times X$, for $X = \mathbb{C}^2$ or $\mathbb{C}^2/\mathbb{Z}_n$:
\be
\mathcal{Z}^{M5^{(a)}}_{T^2\times X} =\mathcal{Z}^{\mathcal{T}^{KK,(a)}}_{S^1\times X}.
\ee

\subsection{Partition function on $S^1\times\mathbb{C}^2$}
\label{sec:5dc2app}
The K-theoretic Nekrasov partition function factorizes into a product of classical, perturbative, and instanton factors:
\be
\mathcal{Z}_{S^1\times\mathbb{C}^2}^{\mathcal{T}^{KK,(a)}} = \mathcal{Z}_{S^1\times\mathbb{C}^2}^{\mathcal{T}^{KK,(a)},\text{ class}}\mathcal{Z}_{S^1\times\mathbb{C}^2}^{\mathcal{T}^{KK,(a)}\text{, pert}}\mathcal{Z}_{S^1\times\mathbb{C}^2}^{\mathcal{T}^{KK,(a)}\text{, inst}},
\label{eq:zfullm5a}
\ee
and we focus on the latter two factors. The perturbative part of the partition function can be written in the following closed form:
\be
\mathcal{Z}_{S^1\times\mathbb{C}^2}^{\mathcal{T}^{KK,(a)}\text{, pert}}(\epsilon_+,\epsilon_-,\underline{\mu}^{(a)},\underline{a}^{(a)})
= 
\prod_{i,j=0}^\infty
\prod_{A=0}^{W-1}
\frac{
(1- \tilde\mu^{(a)}_{A} t^{i-j} x^{i+j+1})}
{(1- t^{i-j+1} x^{i+j+1})},
\label{eq:zper}
\ee
where 
\be
\tilde\mu^{(a)}_{A} = \mu^{(a)}_{A}\, e^{2\pi i(a^{(a)}_{A}-a^{(a)}_{A-1})} = e^{2\pi i \nu^{(a)}_A}.
\ee

The instanton part of the partition function can be expressed as the following sum\footnote{ \  This expression, as well as equation \eqref{eq:zper} above, are obtained by uplifting to five dimensions the results of \cite{Bruzzo:2014jza} for four-dimensional quiver gauge theories, replacing rational factors by trigonometric ones. We are cavalier about overall exponential prefactors which would require careful regularization.}:
\be
\resizebox{\textwidth}{!}{$\displaystyle{
\mathcal{Z}_{S^1\times\mathbb{C}^2}^{\mathcal{T}^{KK,(a)}\text{, inst}}(\epsilon_+,\epsilon_-,\underline{\mu}^{(a)},\underline{a}^{(a)},\underline{q}^{(a)})
=
\sum_{\underline{{Y}}^{(a)}}
\left(\prod_{A=0}^{W-1} (q^{(a)}_{A})^{\vert Y^{(a)}_{A} \vert}\right)
\prod_{A=0}^{W-1}\frac{m_{Y^{(a)}_{A},Y^{(a)}_{A+1}}(t,x,t^{-1}\tilde\mu^{(a)}_{A})}{m_{Y^{(a)}_{A},Y^{(a)}_{A}}(t,x,1)},
}$}
\ee
where $\underline{{Y}}^{(a)}$ denotes a $W$-tuple $(Y^{(a)}_{0},\dots,Y^{(a)}_{W-1})$ of Young diagrams, and
\be
\resizebox{\textwidth}{!}{$
\displaystyle{m_{Y,Y'}(\epsilon_+,\epsilon_-,\mu)
=
\prod_{i=1}^{\vert Y\vert}
\prod_{j=1}^{Y_i}
\frac{1-\mu(t x)^{i-Y'^{tr}_j}(t/x)^{Y^{tr}_i-j+1}}{\sqrt{\mu(t x)^{i-Y'^{tr}_j}(t/x)^{Y^{tr}_i-j+1}}}
\prod_{i=1}^{\vert Y'\vert}
\prod_{j=1}^{Y'_i}
\frac{1-\mu(t x)^{Y^{tr}_j-i+1}(t/x)^{j-Y'_i}}{\sqrt{\mu(t x)^{Y^{tr}_j-i+1}(t/x)^{j-Y'_i}}}.}
$}
\ee
The coefficient of $\prod_{A=0}^{W-1}(q^{\vert Y^{(a)}_A \vert}_A)$ captures the contribution of states with instanton charges $(\vert Y^{(a)}_{0} \vert,\dots,\vert Y^{(a)}_{W-1} \vert)$ with respect to the gauge groups $(U(1)^{(a)}_{0},\dots,U(1)^{(a)}_{W-1})$ in the quiver.

\begin{figure}
\begin{center}
\includegraphics[width=.55\textwidth]{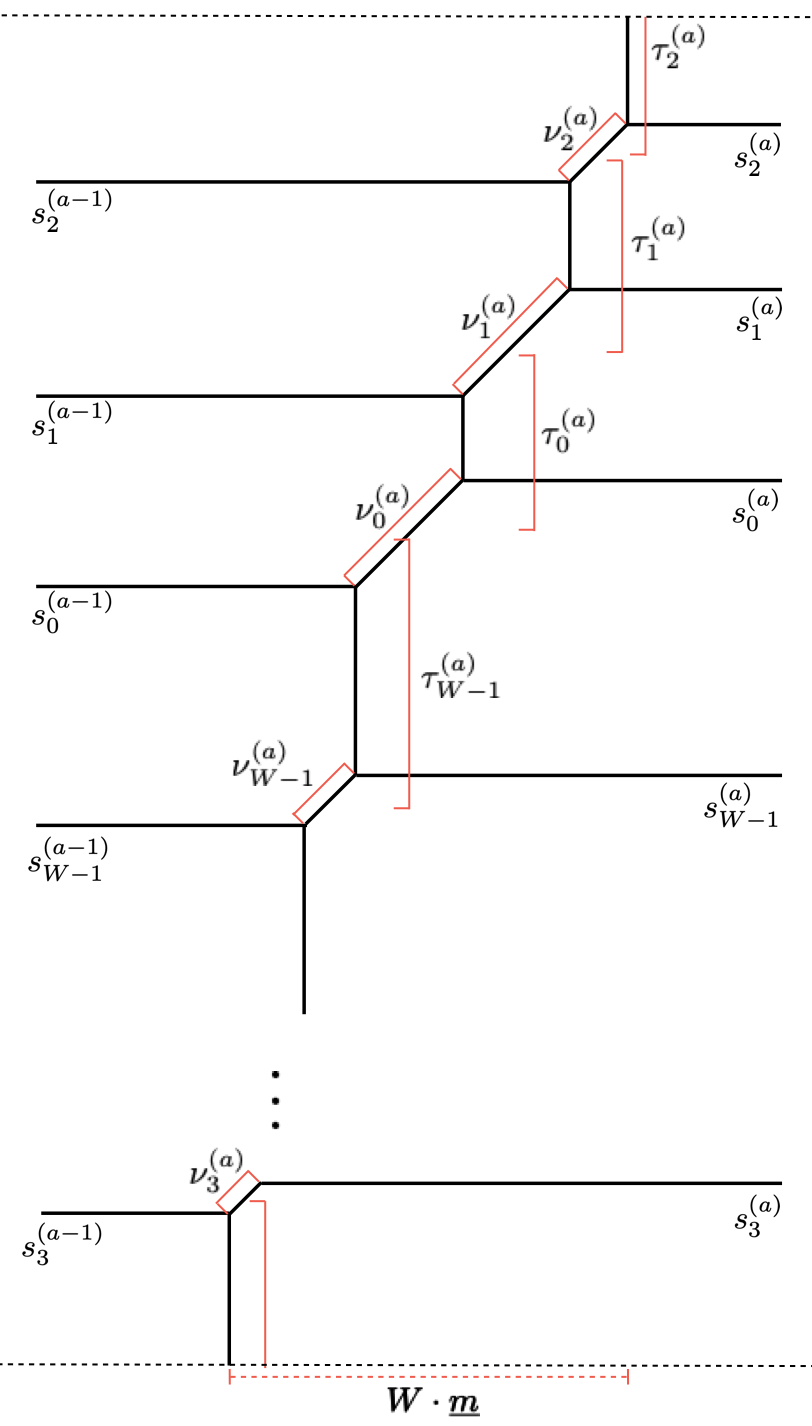}
\caption{A $(p,q)$ web giving rise to the abelian quiver gauge theory of figure \ref{fig:5dquiv}.}
\label{fig:pqa}
\end{center}
\end{figure}

With M-theory geometric engineering in mind, we find that the instanton part can be rewritten in the following form:\footnote{ \  For $W\leq 4$ we have checked the validity of this conjectural identity by expanding up to monomials of a given degree $\lambda$ in the variables $q^{(a)}_0,\dots, q^{(a)}_{A-1}$.}
%
\bea
\nonumber
&&
\resizebox{.45\textwidth}{!}{$
\displaystyle{
\mathcal{Z}_{S^1\times\mathbb{C}^2}^{\mathcal{T}^{KK,(a)}\text{, inst}}(\epsilon_+,\epsilon_-,\underline{\mu}^{(a)},\underline{a}^{(a)},\underline{q}^{(a)})
=
q^{\frac{1}{24}}\widehat{\chi}_{\mathcal{H}}(\tau)
}$}
\\
\nonumber
&&
\resizebox{.62\textwidth}{!}{$
\displaystyle{
\times
\prod_{A=0}^{W-1}
\prod_{i,j,k=0}^\infty
\frac{
(1- q^{k+1}{\tilde\mu}^{(a)}_{A} t^{i-j} x^{i+j+1})
(1- q^{k+1}\frac{1}{\tilde\mu^{(a)}_{A}}t^{i-j} x^{i+j+1})}
{(1- q^{k+1} t^{i-j+1} x^{i+j+1})
(1- q^{k+1}t^{i-j-1} x^{i+j+1})}
}$}
\\
\nonumber
&&
\resizebox{.95\textwidth}{!}{$
\displaystyle{
\times
\prod_{\substack{A,B=0\\B> A}}^{W-1}
\prod_{i,j,k=0}^\infty
\frac{
(1- q^{k} \sqrt{{\tilde\mu}^{(a)}_{A} {\tilde\mu}^{(a)}_{B}}(\prod_{L=A}^{B-1}q^{(a)}_L)t^{i-j} x^{i+j+1})
(1- q^{k+1} \frac{1}{\sqrt{{\tilde\mu}^{(a)}_{A}{\tilde\mu}^{(a)}_{B}}}(\prod_{L=A}^{B-1}\frac{1}{q^{(a)}_L})t^{i-j} x^{i+j+1})}
{(1- q^{k}\sqrt{\frac{{\tilde\mu}^{(a)}_{B}}{{\tilde\mu}^{(a)}_{A}}}(\prod_{L=A}^{B-1}q^{(a)}_L)t^{i-j+1} x^{i+j+1})
(1- q^{k+1}\sqrt{\frac{{\tilde\mu}^{(a)}_{A}}{{\tilde\mu}^{(a)}_{B}}}(\prod_{L=A}^{B-1}\frac{1}{q^{(a)}_L})t^{i-j+1} x^{i+j+1})}
}$}
\\
\nonumber
&&
\resizebox{.95\textwidth}{!}{$
\displaystyle{
\times
\prod_{\substack{A,B=0\\B> A}}^{W-1}
\prod_{i,j,k=0}^\infty
\frac{
(1- q^{k}\frac{1}{\sqrt{{\tilde\mu}^{(a)}_{A} {\tilde\mu}^{(a)}_{B}}}(\prod_{L=A}^{B-1}q^{(a)}_L)t^{i-j} x^{i+j+1})
(1- q^{k+1}\sqrt{{\tilde\mu}^{(a)}_{A} {\tilde\mu}^{(a)}_{B}}(\prod_{L=A}^{B-1}\frac{1}{q^{(a)}_L})t^{i-j} x^{i+j+1})}
{(1- q^{k} \sqrt{\frac{{\tilde\mu}^{(a)}_{A}}{{\tilde\mu}^{(a)}_{B}}}(\prod_{L=A}^{B-1}q^{(a)}_L)t^{i-j-1} x^{i+j+1})
(1- q^{k+1}\sqrt{\frac{{\tilde\mu}^{(a)}_{B}}{{\tilde\mu}^{(a)}_{A}}}(\prod_{L=A}^{B-1}\frac{1}{q^{(a)}_L})t^{i-j-1} x^{i+j+1})},
}$}
\\
\label{eq:Zm5inst}
\eea
which has already appeared in topological string computations in \cite{Haghighat:2013tka} (see also \cite{Hohenegger:2013ala}), up to the factor of the Heisenberg algebra character
\be
\widehat{\chi}_{\mathcal{H}}(\tau)=\frac{1}{\eta(\tau)}= \frac{1}{q^{\frac{1}{24}}\prod_{k=1}^\infty(1-q^k)}
\ee
and up to parameter redefinitions.  The infinite product factors in equations \eqref{eq:zper} and \eqref{eq:Zm5inst} can be interpreted as contributions from BPS M2 branes wrapped on holomorphic two-cycles in an elliptic Calabi-Yau threefold whose toric diagram (up to birational equivalence) corresponds to the brane configuration of figure \ref{fig:pqa}.
It is useful to rewrite \eqref{eq:zfullm5a} in a way that makes the interpretation of the quiver as $U(W)^{(a-1)}\times U(W)^{(a)}$ bifundamental matter manifest. To do this, let $s^{(a-1)}_A$ and $s^{(a)}_A$ for $A=0,\dots W{-1}$ denote the positions of the D5 brane segments in the $(p,q)$ web of figure \ref{fig:pqa}. A given $s^{(a)}_A$ corresponds in Type IIA to a Wilson lines for the factor $U(1)_A^{(a)}$ in the Cartan subgroup of $U(W)^{(a)}$, and we have picked a Weyl chamber in which $s^{(a-1)}_A\leq s^{(a-1)}_B$ and $s^{(a)}_A\leq s^{(a)}_B$ for $A< B$.  Note that the parameters $\nu^{(a)}_A$ and $\tau^{(a)}_A$ appearing in the M5 brane partition function \eqref{eq:Zm5a} are determined in terms of the Wilson line parameters $\underline{\boldsymbol{s}}$. Namely, one has:
\be
\nu^{(a)}_A = s^{(a)}_A-s^{(a-1)}_A,\qquad \tau^{(a)}_A = \frac{1}{2}\left(s^{(a)}_{A+1}-s^{(a)}_{A}+s^{(a-1)}_{A+1}-s^{(a-1)}_{A}\right).
\ee
Let $\mathfrak{S}^{(a)}$ denote the set of index pairs $(A,B)$ such that $s^{(a-1)}_A\leq s^{(a)}_B $, and $\overline{\mathfrak{S}}^{(a)}$ its complement among the set of all pairs $(A,B)$. Then, the full partition function (up to classical prefactors) can be written as follows:
\bea
\nonumber
&&
\mathcal{Z}_{S^1\times\mathbb{C}^2}^{\mathcal{T}^{KK,(a)}}
=
q^{\frac{1}{24}}
\widehat{\chi}_{\mathcal{H}}(\tau)\\
\nonumber
&\times&
\prod_{k=0}^{\infty}
\prod_{\substack{A,B=0\\A\leq B}}^{W-1} 
Z^{BPS,U(1)_A^{(a-1)},U(1)_B^{(a-1)}}_{S^1\times{\mathbb{C}^2}}(k\tau,1)
Z^{BPS,U(1)_A^{(a-1)},U(1)_B^{(a-1)}}_{S^1\times{\mathbb{C}^2}}((k+1)\tau,-1)\\
\nonumber
&\times&
\prod_{k=0}^{\infty}
\prod_{\substack{A,B=0\\A < B}}^{W-1}
Z^{BPS,U(1)_A^{(a)},U(1)_B^{(a)}}_{S^1\times{\mathbb{C}^2}}(k\tau,1)
Z^{BPS,U(1)_A^{(a)},U(1)_B^{(a)}}_{S^1\times{\mathbb{C}^2}}((k+1)\tau,-1)\\
\nonumber
&\times&
\prod_{k=0}^{\infty}
\prod_{(A,B)\in\mathfrak{S}^{(a)}}
Z^{BPS,U(1)_A^{(a-1)},U(1)_B^{(a)}}_{S^1\times\mathbb{C}^2}(k\tau,0)
Z^{BPS,U(1)_B^{(a)},U(1)_A^{(a-1)}}_{S^1\times\mathbb{C}^2}((k+1)\tau,0)\\
\nonumber
&\times&
\prod_{k=0}^{\infty}
\prod_{(A,B)\in\overline{\mathfrak{S}}^{(a)}}
Z^{BPS,U(1)_B^{(a)},U(1)_A^{(a-1)}}_{S^1\times\mathbb{C}^2}(k\tau,0)
Z^{BPS,U(1)_A^{(a-1)},U(1)_B^{(a)}}_{S^1\times\mathbb{C}^2}((k+1)\tau,0).
\label{eq:zzfull}
\eea
The toric diagram of figure \ref{fig:pqa} corresponds to an ordering of D5 brane Wilson line parameters for which $\overline{\mathfrak{S}}^{(a)}$ is empty. Other orderings can be engineered by performing sequences of flops of $\mathcal{O}(-1)\oplus\mathcal{O}(-1)\to\mathbb{P}^1$ curves in the toric diagram, in which case the same functional form \eqref{eq:zzfull} still holds for the $S^1\times\mathbb{C}^2$ partition function.

\subsection{Partition function on $S^1\times\mathbb{C}^2/\mathbb{Z}_n$}
\label{sec:5dpf}
We now place the abelian quiver gauge theory on the background $S^1\times\mathbb{C}^2/\mathbb{Z}_n$ and compute the partition function based on the results of \cite{Bruzzo:2014jza}. This requires specifying a monodromy $e^{\frac{2\pi i}{n}\omega_A}$ for each gauge node, which is encoded in terms of a $W$-tuple of integers $\underline{\omega} = (\omega_0,\dots,\omega_{W-1})\in \mathbb{Z}_n^W$. Equivalently, we may encode the information about the gauge field monodromies in terms of a $W$-tuple of $n$-dimensional vectors $\vec{w}_A$ such that
\be
(\vec{w}_A)_j = \delta_{j,\omega_A}.
\ee
Note that while in our brane engineering setup is that we can also turn on monodromies for the background gauge fields associated to the global symmetry of the quiver, this possibility was not considered in \cite{Bruzzo:2014jza}. As a consequence, we are only able to compare our results with the gauge theory computations of \cite{Bruzzo:2014jza} for trivial global symmetry monodromies. This corresponds for us to turning on a monodromy $\omega^{KK}$ for the $U(1)^{KK}$ gauge field that arises from the two-form field, but not for the 6d one-form fields. Since in the setup of \cite{Bruzzo:2014jza} $U(1)^{KK}$ corresponds to the diagonal combination of all abelian gauge fields in the circular quiver, to compare with their result we must set the monodromies for all quiver gauge nodes to be identical: $\omega_0=\dots=\omega_{W-1}= \omega^{KK}$. Other possible choices of monodromy in the abelian quiver gauge theory do not arise straightforwardly in our brane setup. Nevertheless for the moment we will allow for generic monodromies $\underline{\omega}$ and specialize to our case of interest at a later moment.\newline

Instantons of the quiver gauge theory are labeled by the first Chern classes of a $W$-tuple of self-dual gauge bundles $\underline{\mathcal{U}}$, that is, by a set of $W$ vectors $\vec{u}_A\in\mathbb{Z}^{n-1}$ such that 
\be
c_1(\mathcal{U}_A) = \sum_{j=1}^{n-1} (u_A)_{j} c_1(\mathcal{R}_j),
\ee
where $\mathcal{R}_j$ is the flat line bundle on $\mathbb{C}^2/\mathbb{Z}_n$ with monodromy $e^{2\pi i \frac{j}{n}}$ at infinity.

 The instanton number $n_A = \int ch_2(\mathcal{U}_A)$ is determined in terms of the first Chern classes as \cite{Douglas:1996sw}:
\be
n_A = \sum_j \frac{j(n-j)}{2 n}(u_A)_j,
\ee
where the $\vec{u}_A=(u_{A,1},\dots,u_{A,n-1})\in\mathbb{Z}^{n-1}$ are subject to the constraints
\be
\sum_{j=1}^{n-1} j u_{A,j} = \omega_{A}\qquad\text{mod }n.
\label{eq:c1hol}
\ee
For each node of the quiver we turn on chemical potentials $\vec{\xi}_A$ conjugate to the magnetic flux. According to \cite{Bruzzo:2014jza}, conformal symmetry of the 5d quiver gauge theory leads to constraints on the first Chern classes of the gauge fields, which for the abelian quiver gauge theory translates to the vanishing of the following quantities:
 \bea
 \nonumber
 d_A(\vec{\underline{u}},\vec{\underline{w}}) &=& \frac{1}{4}\left(\lambda_A(\vec{\underline{w}})+\lambda_{A+1}(\vec{\underline{w}})\right)-\frac{1}{2}\vec{u}_A \cdot (C^{{A}_{n-1}})^{-1}\cdot (2\vec{u}_A-\vec{u}_{A-1}-\vec{u}_{A+1})\\
 \label{eq:dv}
 \eea
 for $A=0,\dots,W-1$. Here we set $\vec{\underline{u}}= (\vec{u}_0,\dots,\vec{u}_{W-1})$, $\vec{\underline{w}} = (\vec{w}_0,\dots,\vec{w}_{W-1})$, and 
\be
\lambda_A(\underline{w}) = \frac{1}{n}\vert \omega_A-\omega_{A-1}\vert (n-\vert \omega_A-\omega_{A-1}\vert).
\ee
Vanishing of the quantities $d_A(\vec{\underline{u}},\vec{\underline{w}})$ given in equation \eqref{eq:dv} is quite constraining. To see this, define $\delta \vec{u}_A = \vec{u}_A-\vec{u}_{A-1}$ for $A\in \mathbb{Z}_W$. Summing over $A$, equation \eqref{eq:dv} implies that
\be
\sum_{A=0}^{W-1} \lambda_A(\vec{\underline{w}}) =\sum_{A=0}^{W-1}\delta \vec{u}_A\cdot (C^{{A}_{n-1}})^{-1}\cdot \delta \vec{u}_A.
\label{eq:dsum}
\ee
Since the matrix $(C^{{A}_{n-1}})^{-1}$ is positive definite, equation \eqref{eq:dsum} is satisfied by at most a finite number of choices of $(\delta \vec{u}_{1},\dots,\delta \vec{u}_{W-1})$. For a given solution to \eqref{eq:dsum}, it is straightforward to see that either \eqref{eq:c1hol} and \eqref{eq:dv} do not admit any solution, or they admit an infinite number of solutions. Indeed, suppose one has found a solution

\bea
\nonumber		&&\vec{u}_0^\star,\\
\nonumber		&&\vec{u}_1^\star=\vec{u}_0^\star+\delta u_1,\\
\nonumber		&&\dots\\
				&&\vec{u}_{W-1}^\star = \vec{u}_0^\star+\sum_{A=1}^{W-1}\delta \vec{u}_A.
\eea
Then any other solution can be obtained by shifting $\vec{u}_0^\star$ by a vector $\vec{\gamma}\in\mathbb{Z}^{n-1}$ subject to the constraint $\sum_{j=1}^{n-1} j \gamma_j = 0$ mod $n$ and to the set of linear conditions 
\be
\vec{\gamma} \cdot (C^{{A}_{n-1}})^{-1}\cdot (\delta \vec{u}_{A+1}-\delta \vec{u}_{A}) = 0
\ee
 for $A=0,\dots,W-1$, which admit infinitely many solutions. We will denote by $\mathfrak{U}(\vec{\underline{w}})$ the set of distinct families of solutions to equations \eqref{eq:c1hol} and \eqref{eq:dv}, which we label by a choice of representative $\underline{u}^*=(\vec{u}_0^\star,\vec{u}_1^\star,\dots,\vec{u}_{W-1}^\star)$ and by the set $\Gamma$ of all corresponding $\vec{\gamma}$.\newline

Note that when $\vec{w}_0=\dots=\vec{w}_{W-1}$, which corresponds to our case of interest, the only solution to \eqref{eq:dsum} is the vector $(\delta \vec{u}_1,\dots,\delta \vec{u}_{W-1})=(0,\dots,0)$, and there is a single family of  solutions to constraints \eqref{eq:dv} corresponding to  $\vec{u}^\star_0=\dots=\vec{u}^\star_{W-1}=\vec{w}_0$ and $\gamma\in\mathbb{Z}^{n-1}$, such that $\sum_j j (\vec{u}^\star_{0})_j =\omega_0 $ and $\sum_j j \gamma_j = 0$ mod $n$. On the other hand, solutions corresponding unequal monodromies do not always exist. For example, when $n=2$, it is straightforward to show that the set of solutions $\mathfrak{U}(\underline{{w}})$ is nonempty if and only if $\vec{w}_0=\dots=\vec{w}_{W-1}$. For $n>2$ it is in certain cases possible to find nontrivial solutions. For instance, let us take $n=3, W=2$ and 
\be
\vec{\underline{w}}^{(a)} = ((0,1,0),(0,0,1)),
\label{eq:wexample}
\ee
or equivalently $(\omega_0,\omega_1) = (1,2)$. Then, the set $\mathfrak{U}(\vec{\underline{w}}^{(a)})$ contains three families of solutions corresponding to 
\be
(\vec{u}_0^\star,\vec{u}_1^\star) \in \{((1,0),(0,1)), ((1,0),(1,-1)),((-1,1),(0,1))\}.
\ee
In these three cases $\vec{\gamma}$ is given respectively by $(n,n)$, $(2n,-n)$, and $(-n,2n)$, with $n\in \mathbb{Z}$.\newline

Due to the Nekrasov master formula \cite{Nekrasov:2003vi,Nakajima:2003pg,Gasparim:2009sns,DelZotto:2021gzy}, the partition function on $S^1\times \mathbb{C}^2/\mathbb{Z}_n$ of the 5d abelian quiver gauge theory can be expressed in terms of $n$ copies of the partition function on $S^1\times\mathbb{C}^2$. Explicitly, combining the results of \cite{Bruzzo:2014jza} with the factorized formula \eqref{eq:zzfull} for the $S^1\times \mathbb{C}^2$ partition function, we find (up to the classical factor):
\bea
\nonumber
\mathcal{Z}_{S^1\times\mathbb{C}^2/\mathbb{Z}_n}^{\mathcal{T}^{KK,(a)},\vec{\underline{w}}^{(a)}}
&=&
\sum_{(\vec{u}^\star,\Gamma)\in\mathfrak{U}(\underline{\vec{w}}^{(a)})}
\sum_{\gamma\in\Gamma}
\prod_{A=0}^{W-1}\left(
q_A^{\frac{1}{2}(\vec{u}^\star_A+\vec\gamma)\cdot (C^{A_{n-1}})^{-1}\cdot (\vec{u}^\star_A+\vec\gamma)}
e^{2\pi i\vec\xi_{A}^{(a)}\cdot(C^{A_{n-1}})^{-1}\cdot(\vec{u}^\star_{A}+\vec\gamma)}
			    \right)\\
&\times&
q^{\frac{n}{24}}\widehat\chi_{\mathcal{H}}(\tau)^n\prod_{j=1}^{n} \mathcal{Z}_{S^1\times\mathbb{C}^2}^{M5^{(a)}}(\epsilon_+[j],\epsilon_-[j],\underline{\mu}^{(a)},\underline{a}^{(a)}[j],\underline{q}^{(a)}),
\eea
where
\be
\epsilon_+[j]=\epsilon_+,\qquad \epsilon_-[j] = n\,\epsilon_-+(n+1-2j),
\ee
and $\mathcal{Z}_{S^1\times\mathbb{C}^2}^{BPS}(\epsilon_+,\epsilon_-,\underline{\mu}^{(a)},\underline{a}^{(a)},\underline{q}^{(a)})$ denotes the product over BPS particle contributions appearing in equation \eqref{eq:Zm5inst}.
The partition function also depends on shifted Coulomb branch parameters:
\be
a_A^{(a)}[j] = a_A^{(a)}+h_{A,j}(\epsilon_+[j]+\epsilon_-[j])+h_{A,j-1}(\epsilon_+[j]-\epsilon_-[j]]),
\ee
where
\be
h_{A,j}= ((C^{A_{n-1}})^{-1} u_A)_j \qquad\text{ for } j=1,\dots,n-1
\ee
and $h_{A,0}=h_{A,n}=0.$ Crucially, the perturbative and instanton parts of the partition function depend on the ${a}^{(a)}_A[j]$ and $\mu_A^{(a)}$ only through the combinations
\be
\tilde\mu^{(a)}_A[j](\vec{u}_A) = \mu_A^{(a)} e^{2\pi i(a_A^{(a)}[j] - a_{A-1}^{(a)}[j])} =\tilde\mu_A^{(a)} (t x)^{((C^{A_{n-1}})^{-1}\delta \vec{u}_{A})_j}(t/x)^{((C^{A_{n-1}})^{-1}\delta \vec{u}_{A})_{j-1}},
\ee 
which are independent of $\gamma$. As a consequence, we can rewrite the partition function as:
\bea
\nonumber
\mathcal{Z}_{S^1\times\mathbb{C}^2/\mathbb{Z}_n}^{\mathcal{T}^{KK,(a)},\underline{\vec{w}}^{(a)}}
&=&
\sum_{(\vec{\underline{u}}^\star,\Gamma)\in\mathfrak{U}(\underline{\vec{w}})}
\prod_{j=1}^{n} \mathcal{Z}_{S^1\times\mathbb{C}^2}^{\mathcal{T}^{KK,(a)}}(\epsilon_+,\epsilon_-[j],\underline{\tilde\mu}^{(a)}[j](\underline{\vec{u}}),\underline{q}^{(a)})\\
&\times&
\frac{q^{\frac{n}{24}}}{\eta(\tau)^n}
\sum_{\gamma\in\Gamma}
\prod_{A=0}^{W-1}
q_A^{\frac{1}{2}(\vec{u}^\star_{A}+\vec{\gamma})\cdot (C^{A_{n-1}})^{-1}\cdot (\vec{u}^\star_{A}+\vec{\gamma})}
e^{2\pi i\vec{\xi}_{A}^{(a)}\cdot(C^{A_{n-1}})^{-1}\cdot(\vec{u}^\star_{A}+\vec{\gamma})}
.
\nonumber\\
 \label{eq:zabe}
\eea
We can further manipulate the second line in this expression to get:
\bea
&&
\sum_{\gamma\in\Gamma}
\prod_{A=0}^{W-1}\left(
q_A^{\frac{1}{2}(\vec{u}^\star_{A}+\vec{\gamma})\cdot (C^{A_{n-1}})^{-1}\cdot (\vec{u}^\star_{A}+\vec{\gamma})}
e^{2\pi i\vec{\xi}_{A}^{(a)}\cdot(C^{A_{n-1}})^{-1}\cdot(\vec{u}^\star_{A}+\vec{\gamma})}
\right)
\\
\nonumber
&=&
\prod_{A=1}^{W-1} q_A^{\frac{1}{2}(\vec{u}^\star_A-\vec{u}^\star_0)\cdot (C^{A_{n-1}})^{-1}\cdot (\vec{u}^\star_A-\vec{u}^\star_0)}
e^{2\pi i\vec{\xi}_{A}^{(a)}\cdot(C^{A_{n-1}})^{-1}\cdot(\vec{u}^\star_{A}-\vec{u}^\star_{0})}\\
&\times&
\sum_{\vec{\gamma}\in\Gamma}
q^{\frac{1}{2}(\vec{u}_0^\star+\vec{\gamma})\cdot (C^{A_{n-1}})^{-1}\cdot (\vec{u}_0^\star+\vec{\gamma})}
e^{2\pi i(\vec{u}^\star_0+\vec{\gamma})\cdot(C^{A_{n-1}})^{-1}\cdot\left(\sum_{A=0}^{W-1}(\vec{\xi}^{(a)}_A+\tau_A^{(a)} (\vec{u}^\star_A-\vec{u}^\star_0))\right)},
\eea
where $q=e^{2\pi i \tau}=\prod_{A=0}^{W-1}q_A$ and $e^{2\pi i \tau_A^{(a)}}=q_A^{(a)}$.\newline

Let us now specialize to the case of interest in the paper, that is when all monodromies are taken to be identical:
\be
\omega_0=\dots=\omega_{W-1}=\omega^{KK,(a)},
\ee
so that $w_{A,j}^{(a)} = \delta_{j,\omega^{KK,(a)}}$ for all $A$. In this case, the expression for the partition function, equation \eqref{eq:zabe}, simplifies. In particular all $\underline{\tilde\mu}^{(a)}[j]$ are identical and coincide with $\underline{\tilde\mu}^{(a)}$; moreover, all $\vec{u}^\star_0=\dots=\vec{u}^\star_{W-1}\equiv\vec{w}_0^{(a)}$, so we obtain simply:
\bea
\nonumber
\mathcal{Z}_{S^1\times\mathbb{C}^2/\mathbb{Z}_n}^{\mathcal{T}^{KK,(a)},\underline{\vec{w}}^{(a)}}
&=&
\frac{q^{\frac{n}{24}}}{\eta(\tau)}
\widehat{\chi}^{\mathfrak{su}(n)_1}_{\omega^{KK,(a)}}(\sum_{A=0}^{W-1}\vec{\xi}_A^{(a)},\tau)
\times
\left(\prod_{j=1}^{n}
\mathcal{Z}_{S^1\times\mathbb{C}^2}^{\mathcal{T}^{KK,(a)}}(\epsilon_+,\epsilon_-[j],\underline{\tilde\mu}^{(a)},\underline{q}^{(a)})\right).\\
\label{eq:zns5fact}
\eea
Note that only the linear combination $\vec\xi^{(a)}=\sum_{A=0}^{W-1} \vec{\xi}^{(a)}_A$ of chemical potentials, which couples to the magnetic flux of the diagonal $U(1)$ gauge group that arises from the 6d two-form field, appears in the character. Finally, using the identity between the orbifold projection formula \eqref{eq:zbpsproj} and the localization formula for the partition functions of BPS particles, we see that the gauge theoretical expression \eqref{eq:zns5fact} for the partition function of the abelian quiver gauge theory coincides with the partition function of one NS5 brane on $T^2\times \mathbb{C}^2/\mathbb{Z}_n$, equation \eqref{eq:zns5full}.\newline

Although not directly relevant to this paper, it is also interesting to consider examples where the monodromies $\vec{w}_A$ are not all identical and examine the structure of the partition function. For definiteness, let us consider the case $W=2,n=3$ with $\vec{\underline{w}}^{(a)}$ given by equation \eqref{eq:wexample}. Then, summing over the three different families of allowed magnetic fluxes gives rise to the following partition function:
\bea
\nonumber
&&\hskip-.17in\mathcal{Z}_{S^1\times\mathbb{C}^2/\mathbb{Z}_n}^{\mathcal{T}^{KK,(a)},\underline{\vec{w}}^{(a)}}
=
\frac{q^{\frac{5}{24}}}{\eta(\tau)^2}
e^{\frac{\pi i}{3}(\xi^{(a)}_{0,0}-\xi^{(a)}_{1,0}-\xi^{(a)}_{0,1}+\xi^{(a)}_{1,1})}
\\
\nonumber
&&
\times\bigg[
\widehat{\chi}^{\mathfrak{su}(2)_1}_1\left(\xi^{(a)}_{0,0}+\xi^{(a)}_{1,0}+\xi^{(a)}_{0,1}+\xi^{(a)}_{1,1},\tau\right)
\prod_{j=1}^{3} \mathcal{Z}_{S^1\times\mathbb{C}^2}^{M5^{(a)}}(\epsilon_+,\epsilon_-[j],\underline{\tilde\mu}[j](((1,0),(0,1)),\underline{q})
\\
\nonumber
&&+
e^{\pi i(\xi^{(a)}_{0,1}-\xi^{(a)}_{1,1})}
\widehat{\chi}^{\mathfrak{su}(2)_1}_1\left(\xi^{(a)}_{0,0}+\xi^{(a)}_{1,0},\tau\right)
\prod_{j=1}^{3} \mathcal{Z}_{S^1\times\mathbb{C}^2}^{M5^{(a)}}(\epsilon_+,\epsilon_-[j],\underline{\tilde\mu}[j](((1,0),(1,-1))),\underline{q})
\\
\nonumber
&&+
e^{\pi i(-\xi^{(a)}_{0,0}+\xi^{(a)}_{1,0})}
\widehat{\chi}^{\mathfrak{su}(2)_1}_1\left(\xi^{(a)}_{0,1}+\xi^{(a)}_{1,1},\tau\right)
\prod_{j=1}^{3} \mathcal{Z}_{S^1\times\mathbb{C}^2}^{M5^{(a)}}(\epsilon_+,\epsilon_-[j],\underline{\tilde\mu}[j](((-1,1),(0,1))),\underline{q})\bigg].\\
\eea
Interestingly the partition function retains good modular properties, although the current algebra symmetry is now broken to $\mathfrak{su}(2)_1$ in different ways in the three different sectors. It would be interesting to reproduce this choice of monodromies and understand this phenomenon within our string-theoretic setup.

\bibliography{untitled}
\bibliographystyle{utphys}

\end{document}